\documentclass[twocolumn,tighten,times]{aastex63}

\shorttitle{Identification of single emission lines through supervised machine learning}
\shortauthors{Baronchelli I.}

\usepackage{graphicx}
\usepackage{CJK} 

\newcommand{\mic}{$\mu$m}
\newcommand{\ha}{H$\alpha$}

\newcommand{\hb}{H$\beta$}
\newcommand{\oiii}{[O\thinspace{\sc iii}]}
\newcommand{\oii}{[O\thinspace{\sc ii}]}
\newcommand{\sii}{[S\thinspace{\sc ii}]}
\newcommand{\siii}{[S\thinspace{\sc iii}]}

\newcommand{\pdfn}{$\mathrm{PDF}^{*}$}

\DeclareGraphicsExtensions{.pdf,.png,.jpg}

\begin{document}
\pdfoutput=1"
\begin{CJK*}{UTF8}{gbsn} 


\title{Identification of single spectral lines through supervised machine learning in a large HST survey (WISP): a pilot study for Euclid and WFIRST}

\correspondingauthor{I. Baronchelli}
\email{ivano.baronchelli@unipd.it}

\author{I. Baronchelli}
\affiliation{Dipartimento di Fisica e Astronomia, Universit${\grave{\mathrm{a}}}$ di Padova, vicolo Osservatorio, 3, 35122 Padova, Italy.}

\author{C. M. Scarlata}
\affiliation{MN Institute for Astrophysics, University of Minnesota, 116 Church St. SE,  Minneapolis, MN 55455, USA.}

\author{G. Rodighiero}
\affiliation{Dipartimento di Fisica e Astronomia, Universit${\grave{\mathrm{a}}}$ di Padova, vicolo Osservatorio, 3, 35122 Padova, Italy.}

\author{L. Rodr\'{\i}guez-Mu\~noz}
\affiliation{Dipartimento di Fisica e Astronomia, Universit${\grave{\mathrm{a}}}$ di Padova, vicolo Osservatorio, 3, 35122 Padova, Italy.}

\author{M. Bonato}
\affiliation{$INAF-$Istituto di Radioastronomia and Italian ALMA Regional Centre, Via Gobetti 101, I-40129, Bologna, Italy}
 \affiliation{$INAF-$Osservatorio Astronomico di Padova, Vicolo dell'Osservatorio 5, I-35122, Padova, Italy}

\author{M. Bagley}
\affiliation{College of Natural Sciences, The University of Texas at Austin, 2515 Speedway, Austin, TX 78712, USA}

\author{A. Henry}
\affiliation{Space Telescope Science Institute, 3700 San Martin Drive, Baltimore, MD, 21218, USA}

\author{M. Rafelski}
\affiliation{Space Telescope Science Institute, 3700 San Martin Drive, Baltimore, MD 21218, USA}
\affiliation{Department of Physics and Astronomy, Johns Hopkins University, Baltimore, MD 21218, USA}

\author{M. Malkan}
\affiliation{Department of Physics and Astronomy, UCLA, Physics and Astronomy Bldg., 3-714, LA CA 90095-1547, USA}

\author{J. Colbert}
\affiliation{IPAC, Mail Code 314-6, Caltech, 1200 E. California Blvd., Pasadena, CA 91125, USA.}

\author{Y. S. Dai(戴昱)}
\affiliation{Chinese Academy of Sciences South America Center for Astronomy (CASSACA)/NAOC, 20A Datun Road, Beijing 100101, China}

\author{H. Dickinson}
\affiliation{School of Physical Sciences, The Open University, Walton Hall, Milton Keynes, MK7 6AA, UK}
\affiliation{MN Institute for Astrophysics, University of Minnesota, 116 Church St. SE,  Minneapolis, MN 55455, USA.}

\author{C. Mancini}
\affiliation{Dipartimento di Fisica e Astronomia, Universit${\grave{\mathrm{a}}}$ di Padova, vicolo Osservatorio, 3, 35122 Padova, Italy.}

\author{V. Mehta}
\affiliation{MN Institute for Astrophysics, University of Minnesota, 116 Church St. SE,  Minneapolis, MN 55455, USA.}

\author{L. Morselli}
\affiliation{Dipartimento di Fisica e Astronomia, Universit${\grave{\mathrm{a}}}$ di Padova, vicolo Osservatorio, 3, 35122 Padova, Italy.}

\author{H. I. Teplitz}
\affiliation{IPAC, Mail Code 314-6, Caltech, 1200 E. California Blvd., Pasadena, CA 91125, USA.}

\begin{abstract}

Future surveys focusing on understanding the nature of dark energy (e.g., Euclid and WFIRST) will cover large fractions of the extragalactic sky in near-IR slitless spectroscopy. These surveys will detect a large number of galaxies that will have only one emission line in the covered spectral range. In order to maximize the scientific return of these missions, it is imperative that single emission lines are correctly identified. 
Using a supervised machine-learning approach, we classified a sample of single emission lines extracted from the WFC3 IR Spectroscopic Parallel survey (WISP), one of the closest existing analogs to future slitless surveys. Our automatic software integrates a SED fitting strategy with additional independent sources of information. We calibrated it and tested it on a ``\emph{gold}'' sample of securely identified objects with multiple lines detected. 
The algorithm correctly classifies real emission lines with an accuracy of 82.6\%, whereas the accuracy of the SED fitting technique alone is low ($\sim$50\%) due to the limited amount of photometric data available ($\leq6$ bands).
While not specifically designed for the Euclid and WFIRST surveys, the algorithm represents an important precursor of similar algorithms to be used in these future missions.

\end{abstract}


\section{Introduction}
\label{introduction}

A \emph{spectroscopic} measurement provides the most precise estimate of the redshift of a given source. 
Even the most precise spectroscopic surveys, however,  have a fraction of spectroscopic failures due to incorrectly identified spectral features \citep[see, e.g.,][]{2018ApJ...858...77H}.

The typical approach adopted in a spectroscopic survey exploits a flagging scheme used to characterize the reliability of a galaxy's redshift measurement. This approach is well exemplified by the VIMOS VLT deep survey \citep{2005A&A...439..845L}, where the quality of the emission lines' classification is expressed through a four step scale (or flag). The probability that a classification is not correct can be estimated for each flag value using different methods. In the VIMOS survey,  even  the best quality sample shows a residual $\sim$0.5-1.0\% chance of lines being incorrectly identified. This fraction rises to up $\sim$50\% for the lowest-quality sample. The effect of a misidentified line identification results in what are commonly referred to as catastrophic redshift failures, or ``\emph{outliers}'' in the real versus measured redshift plot. 

The danger of line misidentification is that the Gaussian uncertainty associated with the $\lambda$ position is directly translated to the redshift uncertainty. However, given the line misidentification problem, the actual redshift uncertainty is often not Gaussian and it is characterized by the presence of multiple peaks. 
The Gaussian assumption, therefore, can lead to a large underestimation of the final uncertainty associated with any physical quantity derived from line fluxes and redshifts. 

In an alternative to the spectroscopic approach, many color-based criteria allow us to estimate redshifts, albeit with a lower precision than when spectra are available. Photometric redshifts have  historically been used to build samples of high-$z$ galaxies by exploiting the shape and features of the UV/optical part of the spectra, in particular the presence of the \emph{Balmer} and \emph{Lyman breaks} (at $\lambda\sim 4000$\AA\ and 1216\AA, respectively) as, for example, in the initial works in the Hubble deep field \cite[e.g.][]{1996MNRAS.280L..43C,1996MNRAS.283.1388M,1998hdf..symp..219D,2001MNRAS.325..897S}. 
More refined techniques allow us to extrapolate information from photometric measurements by fitting the spectral energy distributions (SEDs) with theoretical and/or empirical models \citep[for a recent review of this topic see, e.g.,][]{2018NatAs.tmp...68S}. Photometric analyses may also result in outliers, as a consequence, e.g., of the misidentification of the Balmer/Lyman break.



Various strategies can be adopted to reduce the number of catastrophic failures in redshift estimates, e.g., by  combining spectroscopic and photometric analysis. 
First of all, instead of a simplistic Gaussian approximation for each source's redshift, one can use the information in the full redshift  probability distribution function (PDF). PDFs are typically created by most common software that performs fitting to spectral energy distributions (SEDs), such as, e.g., \emph{Hyperz} \citep{2000A&A...363..476B}. Besides allowing for a more formally correct treatment of the uncertainty and its propagation, the redshift PDF\footnote{Multiple peaks are commonly observed in a typical PDF. These peaks are due to the degeneracy existing among different SED models when fitted to the available photometric data.} provides a way to compute the probability of a redshift being an outlier. Using the full redshift PDF allows for the selection of samples with different degrees of purity. In any case, the correct computation of PDFs from SED fitting strategies crucially depends on the availability of template galaxy models able to fit the photometric data.

The outlier problem can  also be mitigated using photometric priors. These priors may include galaxy colors (as it is the case in photometric redshift estimates via SED fitting) or observed galaxy fluxes.
To first approximation, brighter sources are more likely located at lower redshift, so this  information can help to disentangle, for example, bright local objects with high metallicity (i.e., steeper optical continuum) from  high redshift sources. In this context, some publicly available SED-fitting software provide the user with the option of applying  optical priors \citep[e.g. \emph{EAZY}:][]{2008ApJ...686.1503B}. Additional empirical techniques can be exploited to improve the precision of photometric redshift estimates and to correct outliers in specific cases. For example, \cite{2018ApJ...857...64B} demonstrate how additional optical priors, in combination with a far-IR detections, can help improve the accuracy.

The approaches described above can be used to identify single emission lines in galaxy spectra.
However, while the photometric redshift estimate can be refined for every source using these techniques, the wavelength position of a spectral line remains the primary source of information in the case of a spectroscopic determination. 
In other words, these techniques can help in identifying an emission line, but they do not affect the spectroscopic redshift of a source when the emission lines are already unequivocally identified.


In this paper, we address the problem of correctly identifying single emission lines detected in grism spectra of the HST-WISP survey  
\citep[WFC3 Infra-red Spectroscopic Parallel survey, ][ Baronchelli, I. et al. in preparation]{2010ApJ...723..104A}, 
by combining different sources of information. 

Another goal of this paper is to provide a testing ground for the definition of similar algorithms to be used in the context of the future ESA's Euclid \citep{2011arXiv1110.3193L} and NASA's WFIRST \citep{2011arXiv1108.1374G} missions, in order to maximize the scientific return of their near-IR spectroscopic surveys. It is worth noting that the spectroscopic coverage of the grisms and the photometric bands available in the WISP survey are both very similar to those that are planned to be employed by Euclid and WFIRST. These similarities, together with the wide sky area covered by WISP, make this survey one of the most important proxies for future space-based spectroscopic missions. The Euclid and WFIRST surveys will probably benefit from a large amount of ancillary data. In this sense, focussing on the WISP survey, our analyses represent a \emph{pilot} study of these future missions. In any case, the modular structure of our algorithm is specifically designed to easily include and remove additional modules and sources of information.

The paper is organized as follows: in Section~\ref{SEC:DATA} we present the WISP survey and the samples we use to calibrate and test the algorithm. Section~\ref{sec:The_algorithm} describes the algorithm and its modular structure. 
In Section~\ref{SECT:PRECISION} we compute the precision of the algorithm, also in terms of completeness and contamination of differently selected samples. In Section~\ref{SEC:NEW_CLASSIF} we report  the new classification of WISP sources obtained using our software, while in Section~\ref{SECT:Discussion} we discuss the implications of our work in the context of future dark energy missions. The main results and future perspectives are finally summarized in Section~\ref{SECT:Conclusion}.

\section{Data}
\label{SEC:DATA}
We tested our algorithm on the second data release of the WISP survey 
\citep[WFC3 Infra-red Spectroscopic parallel survey:][Baronchelli, I. et al. in preparation]{2010ApJ...723..104A}\footnote{https://archive.stsci.edu/prepds/wisp/}.

\subsection{The WISP survey}
\label{SEC:WISP}


WISP is a pure-parallel, near-infrared slitless grism spectroscopic survey  that efficiently collects WFC3 data while other HST instruments are in use \citep[P.I. M. Malkan,][]{2010ApJ...723..104A}. The spectral coverage of the WFC3's grisms, G102 (0.8-1.1\mic, R$\sim$210) and G141 (1.07-1.7\mic, R$\sim$130), enables the detection of the H$\alpha$ line from $z\sim$0.3 to $z\sim$1.5  and the \oiii\ emission lines from $z\sim$0.7 to $z\sim$2.3. To aid in extracting the 1D spectra from the dispersed images and to enable the wavelength calibration,  the WISP fields were also observed in direct imaging mode with filters chosen to match the grisms' spectral coverage: F110W for G102 and either F140W or F160W for G141. A relevant fraction of the WISP fields ($\sim$ one third) are also observed with the WFC3 UVIS camera with a subset of the available filters (F475X, F600LP, F606W, and F814W). Finally, about half of the fields are also covered by {\it Spitzer}/IRAC observations in channel 1 and/or 2 (3.6\mic\ and 4.5\mic, respectively). To date, the WISP survey has observed 483 fields, collectively covering an area of more than 2000 arcmin$^{2}$ (the actual available data in the WISP spectroscopic catalog refer to a total area of 1520 arcmin$^{2}$).

Being a parallel survey, the observing strategy of WISP depends on the details of the primary observations. This means that the depth of the coverage varies from field to field. Accordingly, we divided visit opportunities into two categories: ``short'' and ``long'' visits, corresponding to fewer than four, and four or more orbits, respectively. During the short opportunities, only the G141 grism and the F140W (or F160W) filter are used, while observations in the G102 grism and in the F110W filter are added during long opportunities. In the latter case, the relative integration time between the G102 and G141 grisms is chosen to balance the sensitivity reached by the two grisms. Given these premises, the median 5$\sigma$ depth reached in both grisms is $5\times 10^{-17}$ erg s$^{-1}$ cm$^{-2}$, with a factor of approximately two  field-to-field variation.

For the line detection, we apply a wavelet convolution and a SExtractor-type threshold 
through a custom line-finding software. This approach dramatically reduces the number of false positive detections. For the detection, we require at least three contiguous pixels with S/N$\geq$2.3. This translates into S/N$\geq$4 integrated over the full emission line (i.e. the sum over the pixels involved). However, this value does not directly correspond to the S/N ratio reported in the final catalog. In that case, the flux and flux uncertainty are measured after the lines and the full spectrum are fit (meaning that the signal becomes the area under the Gaussian curve rather than the sum of the individual pixel values).

Once detected, each emission line is identified through a visual inspection streamlined with a Python-based interactive line fitting and measuring code. Every object is analyzed by at least two separate reviewers, and a total of 10 reviewers were involved in this process. During this phase, false identifications due to contamination by overlapping spectra or to specific noise features are  excluded from the sample. All identified real emission lines are then fitted with a Gaussian profile $+$ continuum, using a line fitting algorithm that required little input from the reviewer. When an emission line is detected in a spectrum, additional emission lines, initially undetected, can also be fit even if their flux was below the original pixel-based detection threshold.

The quantities measured for each line (total flux, equivalent width (EW), full-width-half-maximum FWHM) are then merged into a catalog that has a unique entry for each galaxy with all the spectroscopic information. Differences among reviewers' classifications and line fits are used to obtain average solutions, when possible, and to define the quality flag associated with each spectrum. In particular, when no agreement is found between the identifications of two reviewers, this information is included in the quality flag. In this case, in the final catalog, only the solution associated with the line fit that minimizes the $\chi^{2}$ parameter is included. More details on the creation of the WISP emission line catalog are presented in Bagley, M. et al. (2020, in preparation).

\subsection{Calibration and test samples}
\label{SEC:cal_test_samples}


The algorithm is calibrated (Section~\ref{sec:ALGORITHM_INPUTS}) and tested (Section~\ref{SECT:PRECISION}) on two not independent (almost completely overlapping) subsamples extracted from  the WISP spectroscopic catalog of secure identifications. The choice of using two overlapping samples is justified in Section~\ref{SECT:NOT_INDIP_SAMPLES}. For these samples, the standard method used to identify the spectral lines can be applied, i.e., two or more spectral lines are detected above a 2$\sigma$ threshold. Additionally, we require that the reviewers agree on the identification of the lines. We only include sources detected in fields covered by both WFC3 grisms. This selection reduces the WISP area usable for calibration and testing to $\sim$900 arcmin$^{2}$. Finally, we exclude all the emission lines located at the low sensitivity ends of the grisms' wavelength range ($\lambda<$8500\AA, $\lambda>$16700\AA\ and 11000\AA$<\lambda<$11400\AA)\footnote{The high level of noise makes it difficult to clearly identify the strongest line in this spectral region.}. 
We call these two subsamples the \emph{calibration} (2128 sources) and the \emph{test} (2283 sources) ``\emph{gold}'' sample. The only difference between the two samples is that in the \emph{calibration} sample we consider only sources detected in the F110W photometric band, while the same requirement is relaxed when testing the algorithm (measuring accuracy, completeness, and contamination). This selection is due to the fact that the F110W magnitude is required to calibrate one of the modules of the algorithm (the magnitude prior described in Section~\ref{SEC:mag_prior}), but the same measurement is not necessary when running the software on unidentified spectral lines. 
Consequently, the two samples almost fully overlap, with the entire \emph{calibration} sample included in the \emph{test} sample (only $\sim$7\% of the sources used for the test are not used for the calibration).

After calibrating and testing the algorithm, we run the software on a WISP subsample of sources covered by both grisms but with only one emission line detected above a 2$\sigma$ threshold (Section~\ref{SEC:NEW_CLASSIF}).
We highlight the fact that the algorithm is designed to identify the brightest line detected in a spectrum while it is blind to the possible presence of additional lines besides the strongest one. Consequently, their presence in a spectrum does not influence the calibration and test of the algorithm itself.

 \section{The algorithm}
\label{sec:The_algorithm}
\subsection{Rationale}
The purpose of the algorithm is to provide a probabilistic identification of the strongest emission line observed in a spectrum, by combining various sources of information.

In principle, when only one line is detected in a spectrum,  an effective method to identify the line is to compare the spectroscopic redshift expected from different species/ions (e.g., $z_{\mathrm{H}\alpha}$, $z_{\mathrm{[OIII]}}$, $z_{\mathrm{[OII]}}$, etc.), with the independent redshift solution suggested by an SED fit of the available photometric data. The reliability of this approach, however, strongly depends on the number of available photometric measurements and their wavelength coverage. The precision of such a method rapidly declines when only a few photometric measurements are available. In this case, the photometric redshift PDF  will not  clearly distinguish between, e.g., the H$\alpha$ and the \oiii\ redshift solutions. The presence of multiple peaks in the PDF can further complicate the issue, generating photometric redshift outliers.

\begin{figure*}[!ht]
\centering
\includegraphics[width=17.5cm]{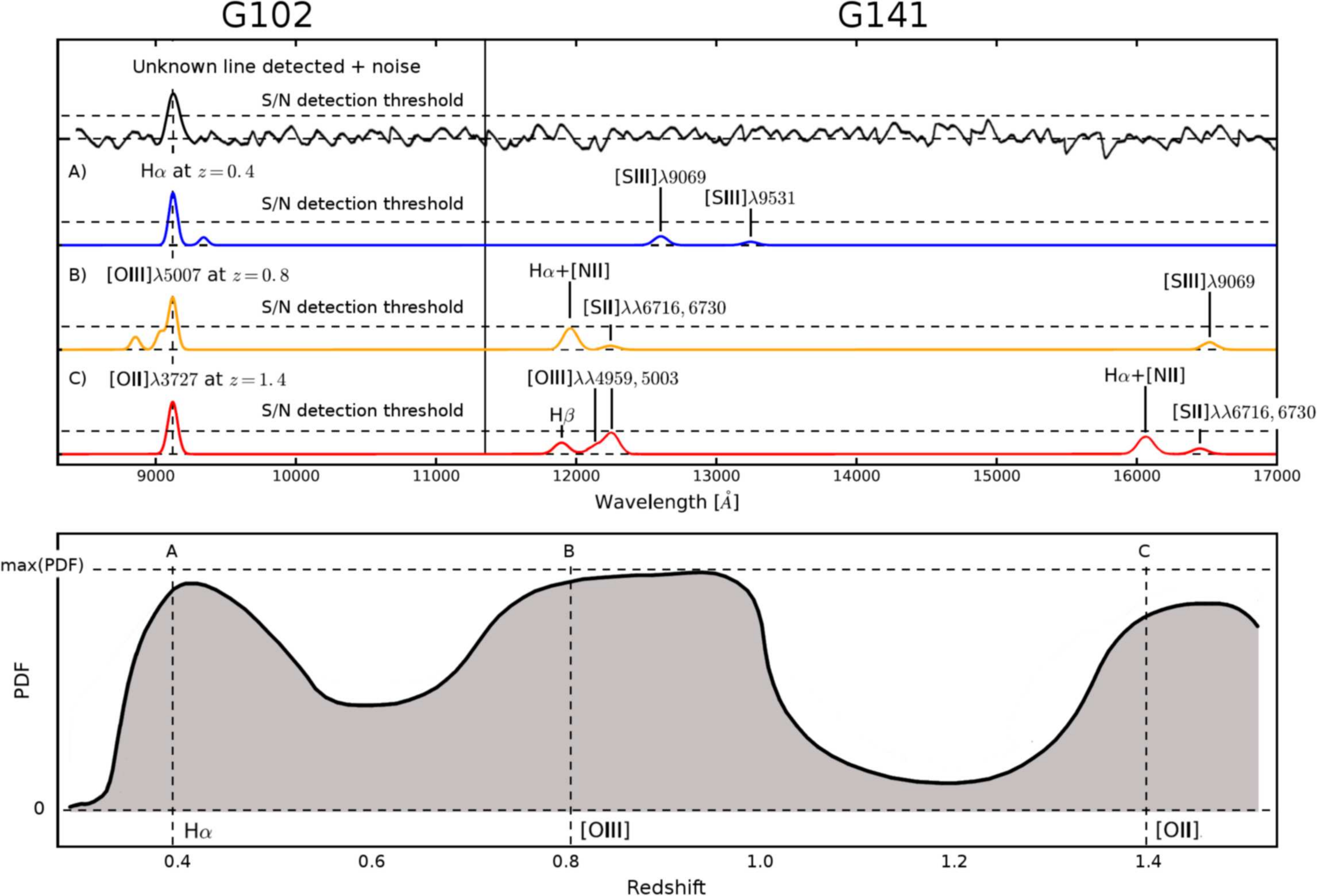}
\caption{{\bf Top panel:} In the upper of the four plots, the schematic representation of a WISP grism spectrum with only one line detected above the S/N threshold (uppermost plot). The lower three plots of the top panel represent three different possible solutions: \ha\ (A), \oiii\ (B), and \oii\ (C). {\bf Bottom panel:} a poorly constrained PDF with multiple probability peaks. This kind of PDF would not be able to discriminate among the three possibilities shown in the top panel (A, B, C). This Figure is a modified version of Figure 3 of \cite{2017ApJ...837...11B}. }
\label{img:Bad_PDF_ex}
\end{figure*}

Figure~\ref{img:Bad_PDF_ex} (top panel) shows the schematic representation of a WISP grism spectrum (G102 and G141) with only one line detected. The same observation can be equally well described as being due to \ha, \oiii\ or \oii\ (cases A, B, and C in the Figure), if the emission due to the other additional lines is below the S/N detection threshold. In the same Figure, the bottom panel illustrates the effects of a poorly constraining multi-peak PDF, on the redshift determination.

For a Gaussian PDF, the \emph{width} of the curve (its value of $\sigma$) univocally indicates the uncertainty associated with the value (on the x-axes) corresponding to the peak of the PDF itself.
In the more general case, when the shape of the PDF is not Gaussian (e.g.,  bottom panel of Figure~\ref{img:Bad_PDF_ex}), it is difficult to unequivocally describe the width of the probability distribution function with one number.
In this case, the redshift range $\Delta z(p)$ corresponding to a given total probability (e.g., p=95\%) is a more correct way to represent the uncertainty associated with the most likely value.
Alternatively, one could also consider the integral of \pdfn, defined as:
\begin{equation}
\label{EQ:PDF*_DEF}
\int{\mathrm{PDF}^{*}(z)\ \mathrm{d}z}=\int{\frac{\mathrm{PDF}(z)}{\max(\mathrm{PDF}(z))}\ \mathrm{d}z}
\end{equation}
Lower \pdfn\ integrals or smaller $\Delta z(p)$, however, are not always associated with more precise solutions, such as when a PDF shows multiple peaks. This is true even if each of the peaks can singularly be described by a Gaussian function with a small value of $\sigma$. This case represents the typical outlier problem. While each Gaussian peak is characterized by a small $\sigma$ (high precision), the real uncertainty could be badly underestimated (low accuracy).

On the other hand, some correlations such as the magnitude (or size) versus redshift relation are not particularly tight. Using these quantities as priors to determine the redshift itself thus generates wide PDFs (low precision). Because of this characteristic, however, these methods are less prone to the problem of outliers. In other words, while the uncertainty is high, the measure of the uncertainty is more accurate.

Thus, the most effective way to preserve the precision while maximizing the accuracy (limiting the number of outliers) is to exploit the broader PDFs to identify the most probable peaks in the narrower PDFs. This can be obtained by simply multiplying all the PDFs to each other. Besides allowing the PDFs to be computed, the same parameters can provide estimates of the expected flux ratios. These ratios can be used by the algorithm to obtain an independent prediction of the most prominent line observed in a spectrum.

In this section, we analyze in detail all the different methods that we eventually combine in our algorithm. Every method is treated as an independent module of the algorithm. In particular, each module represents one single method to obtain information on one only type from one or more input parameters. This approach allows us to more clearly describe the different types of information that the same input parameter can provide, the relations existing between input parameters and outputs, and the limits of each method.


The output of the algorithm we describe below is, for the strongest emission line in each spectrum, a set of indices $P_{\mathrm{transition}}$. These indices are directly related to the probability that the observed emission line is correctly identified to each of the transitions considered (\ha, H$\beta$, H$\gamma$, \oiii, \oii, \sii)\footnote{These indices do not correspond to the actual probabilities for the following reason: first, the algorithm considers only the listed species, without taking into account different possibilities. Moreover, the algorithm does not compute the probability that a detected line is not a real emission, but the result of noise or contamination. For these reasons, the indices are normalized so that the highest probability index $P_{\mathrm{transition}}$ is always equal to 1.}. The transition with the highest value of $P_{\mathrm{transition}}$ will provide the (probabilistically) best redshift solution. We note that, during this process, the algorithm always assumes that the observed line is a real spectral feature and not a spurious detection due to noise or contamination.


\subsection{Structure of the algorithm}
\label{sec:ALGORITHM_INPUTS}
\begin{figure*}[!ht]
\centering
\includegraphics[width=17.5cm]{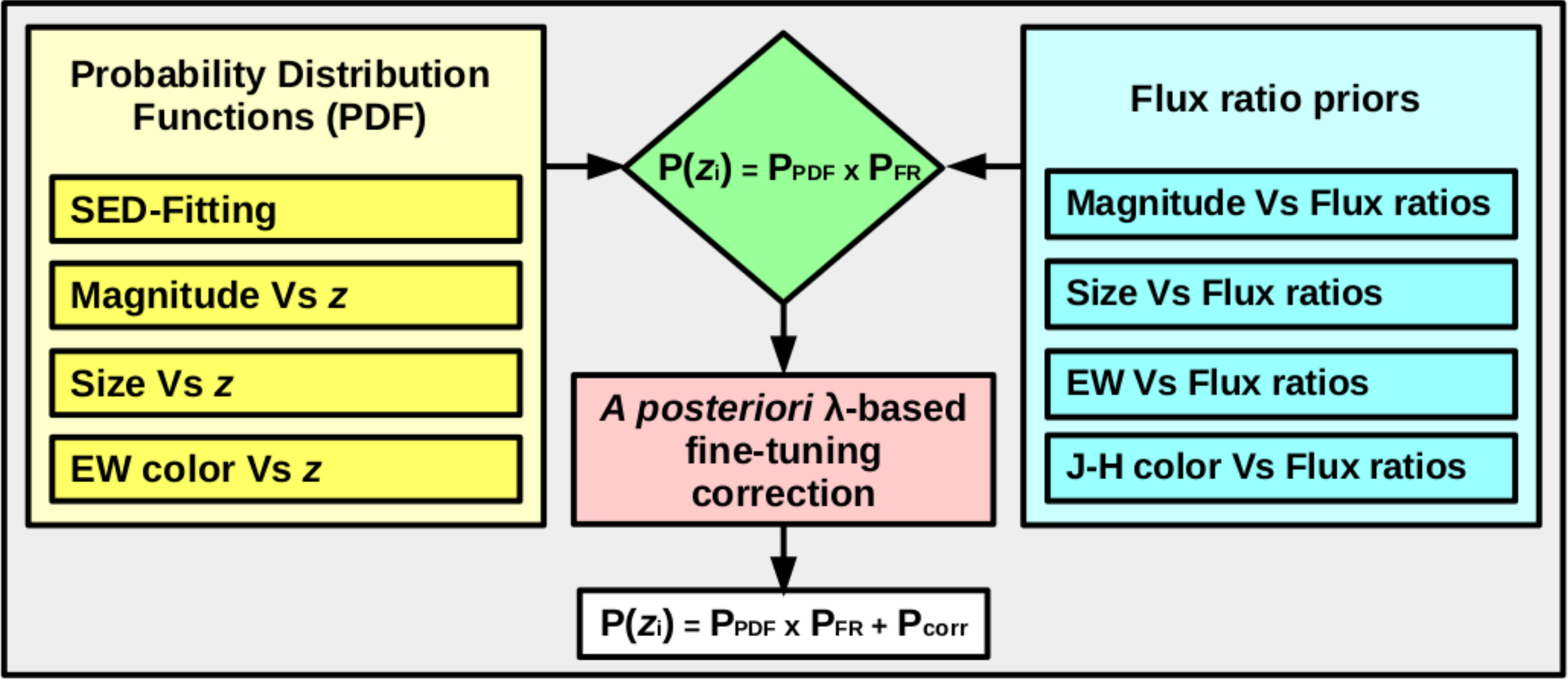}
\caption{Organization of the modules in the algorithm. The modules can be divided into three categories, or blocks, depending on the kind of information supplied. For each spectrum (for each source), the probability distribution functions PDF($z$) provide a continue probability estimation (regression), between $z=0$ and $z=3.3$. The line flux ratios can be used to predict which, among the considered species/transitions, is more likely responsible for the strongest line observed (classification). Finally, the probability ratios can be fine-tuned using a $\lambda$-dependent \emph{a posteriori} correction (optimization).  }
\label{img:scheme1}
\end{figure*}

The modular structure of the algorithm is organized as shown in Figure~\ref{img:scheme1}. In brief, the modules can be divided in three main categories, or blocks, depending on the kind of information provided. Using PDFs, the algorithm can estimate the probability associated with each value of $z$ (regression) in the redshift range considered ($0<z<3.3$). The flux ratios allow us to forecast which species/transition is most probably responsible for the strongest emission (classification). Finally, comparing the results with the input test sample, the algorithm can automatically fine-tune the final function, in order to increase the accuracy of the entire process (optimization). In a scheme:

\begin{itemize}

\item{\thinspace{\sc  Regression block:} photo-$z$ PDFs estimated from SED fitting (Section~\ref{SEC:photoz}), J band apparent magnitude (Section~\ref{SEC:mag_prior}), apparent size (Section~\ref{SEC:size_prior}), and equivalent width of the strongest line (Section~\ref{SEC:EW_PDF});}
\item{\thinspace{\sc  Classification block:} a set of probability ratios estimated using existing relations between flux ratios and J band apparent magnitude, size, line EW, and J-H color index (Section~\ref{SEC:lratios_prior});}
\item{\thinspace{\sc Optimization block:} a posteriori fine-tuning of the probability ratios based on the observed wavelength (Section~\ref{SEC:empirical_corr}).}

\end{itemize}

In this section we describe the single modules and how we calibrated them, while in Section~\ref{sect:op_descr}, we detail on how the modules are combined with each other to output the final probability estimates.

\subsubsection{Photo-z PDF from SED fitting}
\label{SEC:photoz}
\begin{figure}[!ht]
\centering
\includegraphics[width=8.5cm]{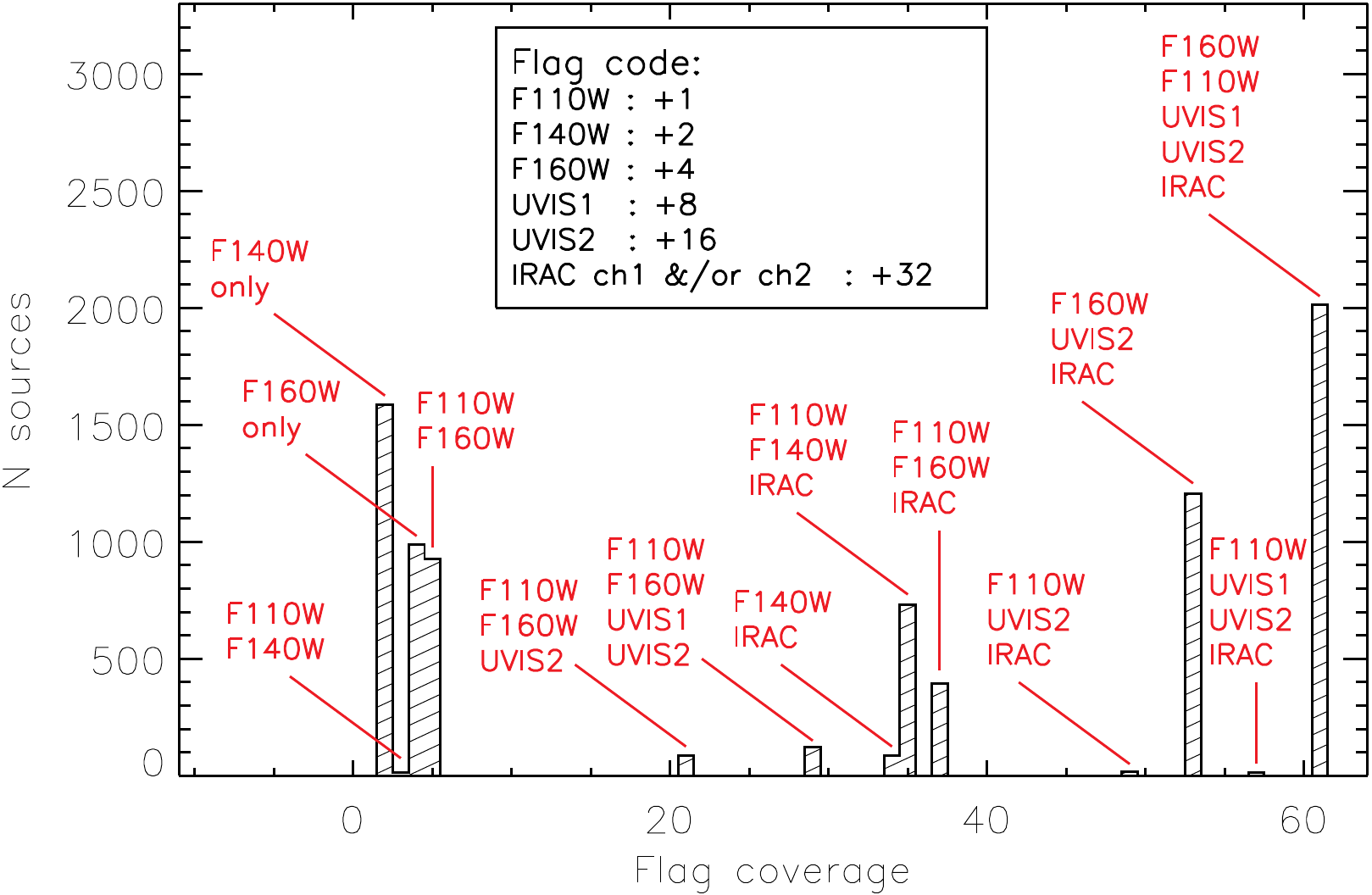}
\caption{Number of sources versus coverage in the available photometric bands, for the original WISP spectroscopic catalog (at least one spectral line measured). No selection is applied. The coverage flag (x axes) is given by the sum of: +1 when the source is observed through the F110W filter, +2 for F140W, +4 for F160W, +8 for UVIS1, +16 for UVIS2, and +32 for IRAC channel 1 and/or channel 2 coverage. The naming convention ``UVIS1'' and ``UVIS2'' is used to represent the bluer and the redder UVIS filters used in a single observation, while a combination of either F475X or F606W and F600LP or F814W were actually used in the WISP survey. }
\label{img:histogr_cover}
\end{figure}

We compute  photometric redshifts using the \emph{Hyperz} software \citep{2000A&A...363..476B}. For each WISP field and for each source, we considered all the measurements in the available photometric bands among the WFC3/IR F110W, F140W, F160W, WFC3/UVIS F606W, F600LP, F814W, and IRAC 3.6 and 4.5 \mic\ filters. Figure~\ref{img:histogr_cover} shows the number of sources covered by every combination of photometric bands. Given the limited number of bands used in the photometric redshift calculation (larger than 2 only for approximately 55\% of the sample), the resulting photometric redshift estimate will not be precise. However, we are not interested in the absolute value of the  photometric redshift, typically assumed to correspond to the highest peak of the PDF, but rather in the full shape of the PDF itself. This PDF will be combined with additional probability functions as described in Section~\ref{sect:op_descr}. Therefore, even a photo-$z$ estimation derived from two bands only (corresponding to one single color index) adds information on the redshift of a source. 
For example, the poorly constrained PDF shown in the bottom panel of Figure~\ref{img:Bad_PDF_ex} does not allow for a reliable estimate of the photometric redshift. However, it does allow us to safely say that the source considered \emph{is not} located below $z\sim$0.3 or in the range $1.1 \lesssim z \lesssim 1.3$. This kind of information can be particularly relevant if combined with additional information from other independent sources \footnote{For example, the upper panel of Figure~\ref{img:PDF_example} shows three poorly constrained PDFs. Their combination is shown in the bottom panel of the same Figure.}. 


For the \emph{Hyperz} run, we considered a combination of template models from \cite{2003MNRAS.344.1000B}, characterized by an exponentially declining star formation rate (SFR) with $\mathrm{SFR}\propto\exp(-t/\tau)$, where $\tau=$0.3, 1, 2, 3, 5, 10, 15, and 30 Gyr. We consider  solar metallicity Z=Z$_{\odot}$ and an extinction law in the \cite{2000ApJ...533..682C} form, with A$_{V}$ ranging from 0.0 to 3.0.

We computed photometric redshifts for all the sources in the sample, regardless of the number of photometric bands available. 
The typical output PDF, rarely shows well defined and unique probability peaks. More commonly, multiple peaks are present. In some cases, their presence is due to the degeneracy of the input parameters, generating solutions at different redshifts with similar probabilities\footnote{Often, few photometric optical data can equivalently well be fitted by both the spectrum of an obscured low redshift source or by the spectrum of an high redshift galaxy with low/normal extinction.}. In other cases, they are a consequence of the discretized grid of models used for the fit, especially when the maxima are close to each other (in redshift). To correct for the  effect of model discretization, we smooth the PDFs using a Gaussian filter. Although a good compromise between precision and mitigation of the discretization effect can be obtained with a constant value of $\sigma_{z}$ ($\sim$0.1), we decided to vary $\sigma_z$ as a function of the integral of the \pdfn\ (see Equation~\ref{EQ:PDF*_DEF}), for the reasons explained next.

As we previously discussed, narrow  photometric redshift PDFs  may still be centered on the wrong redshifts, with resulting uncertainties that are underestimated (resulting in outliers). In general, the overconfidence in the estimation of the $z$--PDF is a well known problem, especially for what concerns their low--probability tails \citep[see, for example,][]{2016MNRAS.457.4005W}. Because of the small (underestimated) uncertainties, the (possibly wrong) photometric redshift solution would prevail over any other redshift indicator. Thus, in these cases, also the emission line identification would also be compromised. To limit the effects of this problem, we smooth the $z$--PDFs, a similar solution to that adopted by, e.g., \cite{2019MNRAS.485..586R}. Differently from that work, where a constant $\sigma=0.2$ is considered, in our Gaussian smoothing, $\sigma$ is proportional to the inverse of the \pdfn\ integrals and asymptotically converging to $\sigma_{z}=$0.1:
\begin{equation}
\label{EQ:smooth_fact}
\sigma_{z}=0.1\times\left( 1+ \frac{P^{*}_{z}}{ \int^{3.3}_{0.0}{\mathrm{PDF}^{*}(z)\ \mathrm{d}z}} \right).
\end{equation}

\begin{figure}[!ht]
\centering
\includegraphics[width=8.5cm]{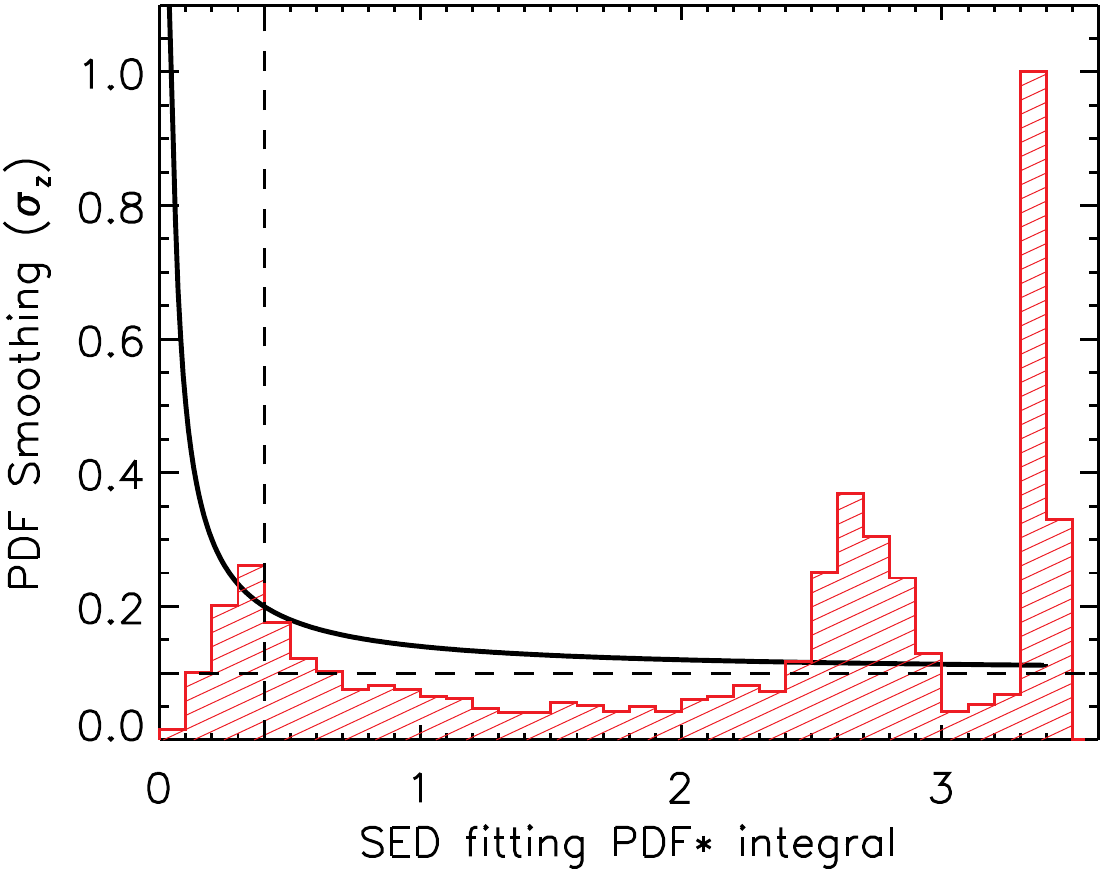}
\caption{The black curve represents the width of the Gaussian filter used to smooth the photo-$z$ PDF ($\sigma_{z}$ in Equation~\ref{EQ:smooth_fact}), as a function of the \pdfn\ integral $\int^{3.3}_{0.0}{\mathrm{PDF}^{*}(z)\ \mathrm{d}z}$. 
The red histogram shows the distribution of the \pdfn\ integral for the WISP galaxies. Generally speaking, high values of the \pdfn\ integral indicate poorly constrained photometric redshifts (a value of 3.4 corresponds to a \pdfn\ equal to 1.0 everywhere between $z$=0 and $z$=3.3). However, given the possible presence of multiple probability peaks (outliers) and the intrinsic discretization of the grid of models used for the SED fit, the precision of the algorithm is not guaranteed by narrower PDFs.
The horizontal dashed line represents the best average smoothing factor able to mitigate the effects of the use of a discretized grid of models without limiting the precision of the final PDFs. In Equation~\ref{EQ:smooth_fact} this value corresponds to an asymptotic limit for $\sigma_{z}$ ($\sigma_{z}\rightarrow$ 0.1 for high values of the \pdfn\ integral). The vertical dashed line represents the best empirically derived value of $P^{*}_{z}$, for which the smoothing factor $\sigma_{z}$ is doubled with respect to its asymptotic value. }
\label{img:PDF_photoz_integral}
\end{figure}

Both the best average smoothing factor ($\sigma_{z}=$0.1) and the constant value $P^{*}_{z}=0.4$ are empirically determined by maximizing the accuracy of the algorithm when run on the test ``\emph{gold}'' sample. Figure~\ref{img:PDF_photoz_integral}, shows the value of the variable smoothing factor $\sigma_{z}$ as a function of the original photo-$z$ \pdfn\ integral. 

The SED fitting method relies on the combination of color indices in different photometric bands. As a consequence, if two different sources are characterized by very different apparent magnitudes or sizes, but identical color indices for all the bands considered, there will be no difference between their output redshifts. Thus, to improve the method, we include the magnitude and size priors described in Sections~\ref{SEC:mag_prior} and \ref{SEC:size_prior}.

\subsubsection{Photo-z PDF from F110W (J band) magnitude}
\label{SEC:mag_prior}


Galaxies can not be arbitrarily bright \citep{1976ApJ...203..297S}. The consequence is that, when surveying a fixed solid angle $\Omega$, galaxies that \textit{look} very bright are preferentially located at low redshifts, 
while faint sources more likely correspond to high redshift galaxies.

\begin{figure}[!ht]
 \centering
 \includegraphics[width=8.5cm]{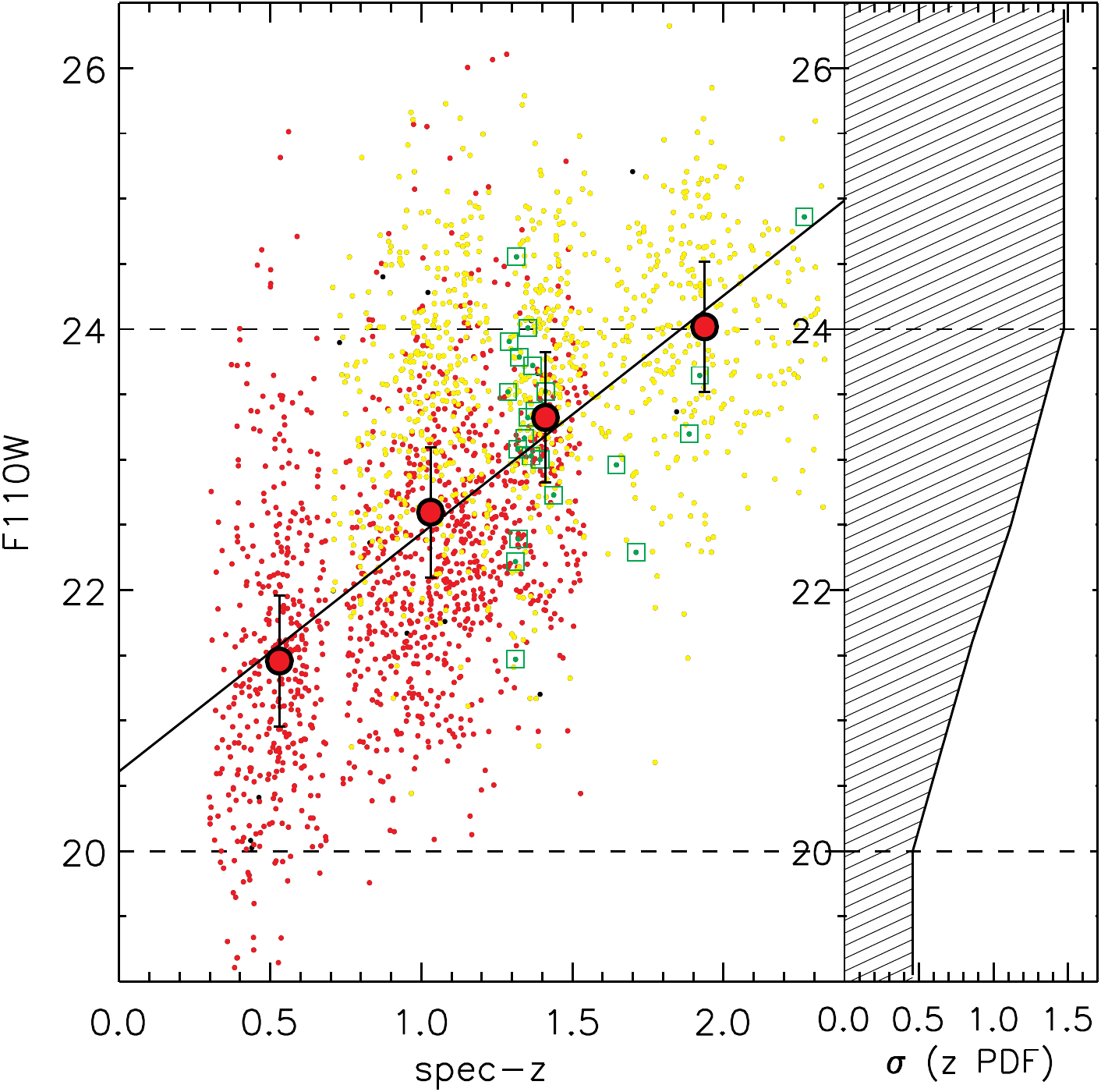}
 \caption{{\bf Left panel:} Relation between AB magnitude measured in the F110W band (J) and spectroscopic redshift for the calibration ``\emph{gold}'' sample. All the data are considered for the linear fit, that we use as a prior in the algorithm. However, the prior does not consider values of magnitude J$<$J$_{\mathrm{min}}$ or J$>$J$_{\mathrm{max}}$ (dashed lines): in these cases, the prior assumes J$=$J$_{\mathrm{min}}$=20 and J$=$J$_{\mathrm{max}}$=24 respectively. This method limits the contamination of the numerically small sample of sources dominated by the \oii\ emission, from \oiii\ dominated sources. The actual nature of the strongest emission line is represented with yellow circles for \oiii, red circles for H$\alpha$ and green circles surrounded by squares for \oii. {\bf Right panel:} given the J magnitude of a source, the PDF is assumed to be a Gaussian function, with $\sigma$ equivalent to the horizontal dispersion of the data around that specific value of magnitude ($\pm$0.5 mag). }
 \label{img:JMAG_PDF}
 \end{figure}

Figure~\ref{img:JMAG_PDF} 
shows the observed F110W magnitude (hereafter J) as a function of the spectroscopic redshift for WISP sources in the calibration ``\emph{gold}'' sample.
The algorithm uses the linear fit to these data (solid line in the Figure) as photo-z prior.
Specifically, given the J magnitude of a galaxy, the algorithm assumes a Gaussian shaped PDF(z), centered at the redshift indicated by the linear fit, and with a $\sigma$ equivalent to the horizontal dispersion of the data in the plot of Figure~\ref{img:JMAG_PDF} (right panel).

While we computed the linear fit using all the data, when applying the prior we limit the range of validity to: J$_{\mathrm{min}}$$<$J$<$J$_{\mathrm{max}}$, with J$_{\mathrm{min}}$=20 and J$_{\mathrm{max}}$=24. Sources with J$<$J$_{\mathrm{min}}$ or J$>$J$_{\mathrm{max}}$ are set to J$=$J$_{\mathrm{min}}$ and J$=$J$_{\mathrm{max}}$ respectively. This empirical approach allows us to limit the contamination of the numerically small sample of sources dominated by the \oii\ emission, from sources leaking from the tails of the numerically more consistent  \oiii\ and \ha\ distributions.

\begin{figure*}[!ht]
 \centering
 \includegraphics[width=8.5cm]{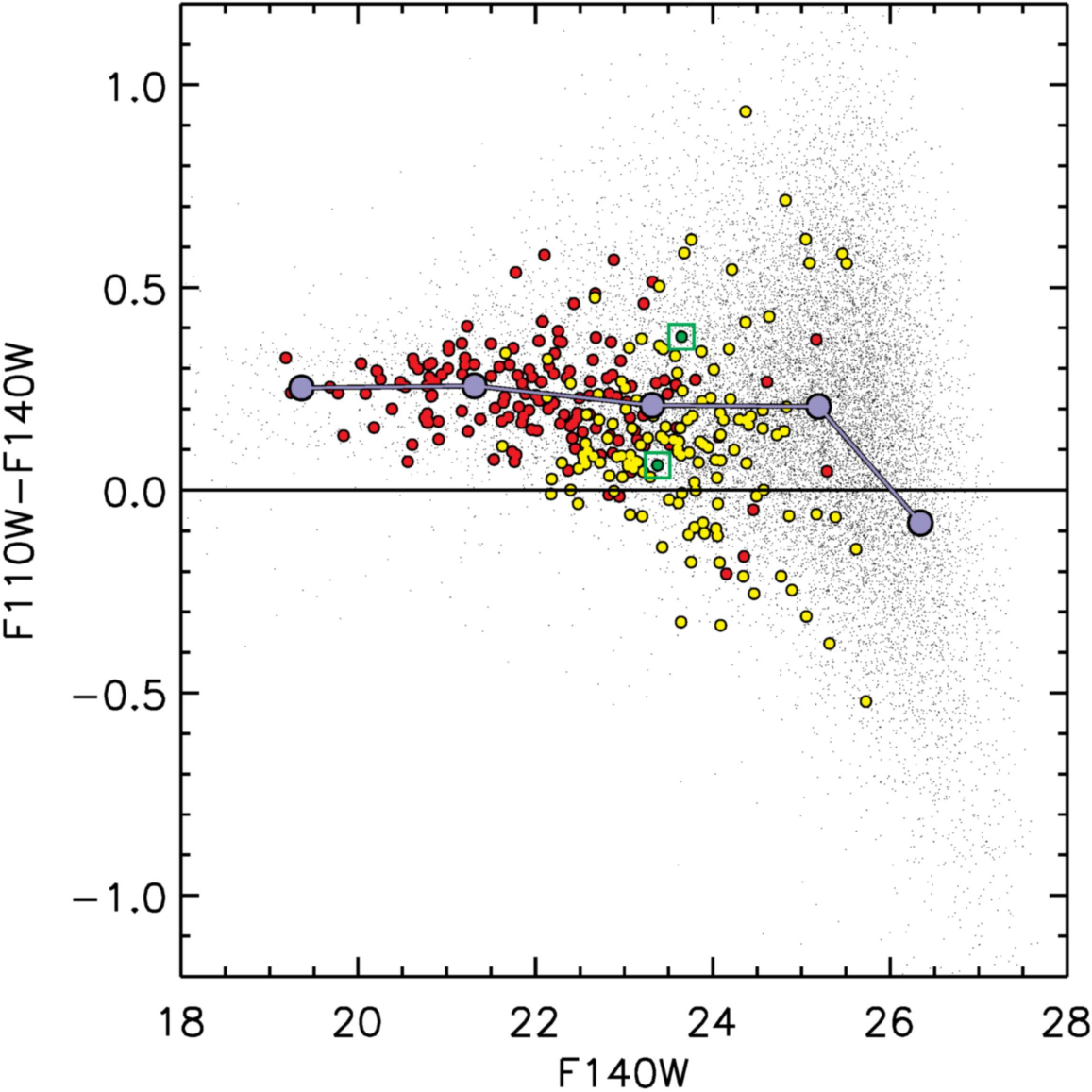}
 \includegraphics[width=8.5cm]{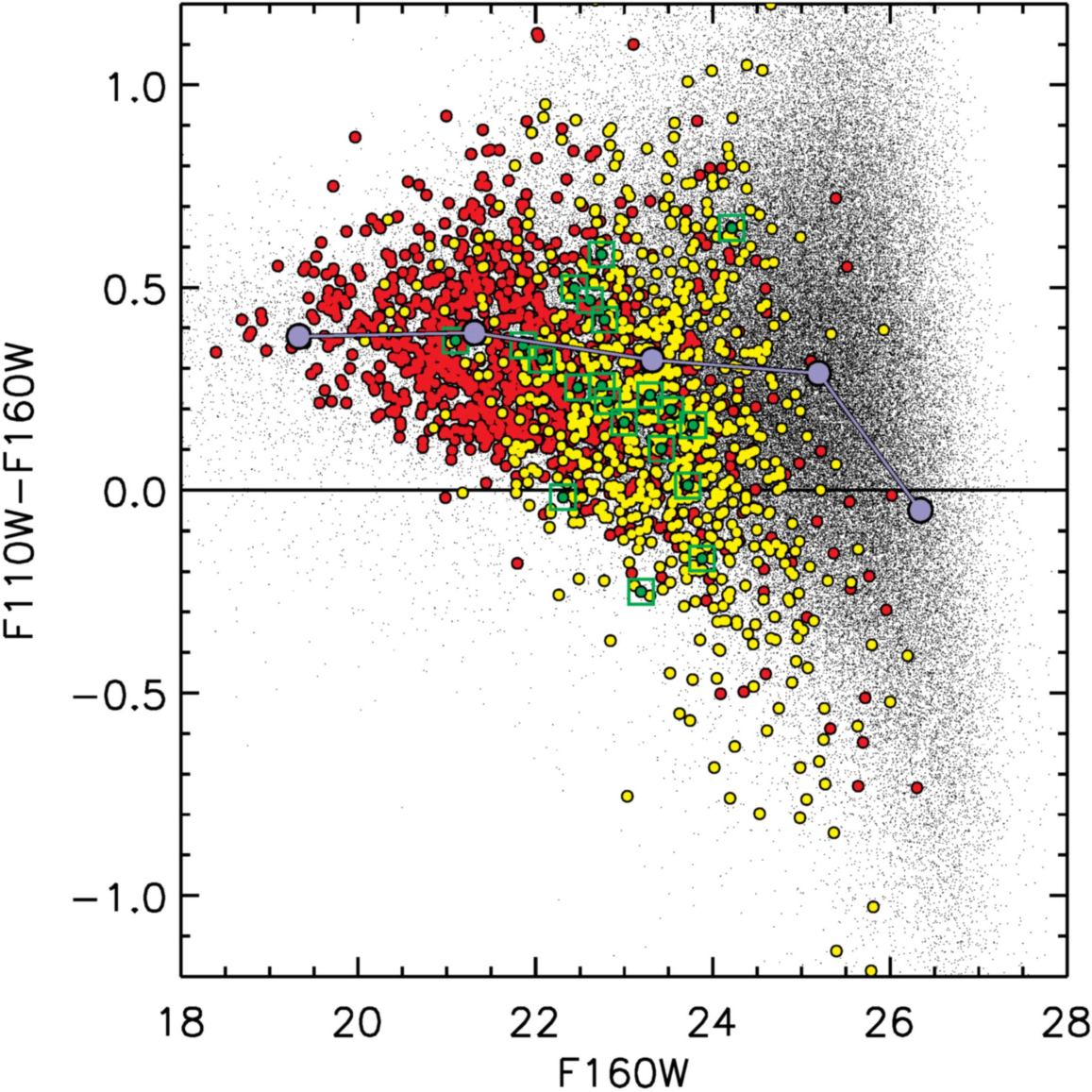}
 \caption{{\bf Left panel:} J-H (F110W-F140W) color index as a function of the F140W magnitude. We use this color index to infer the expected F110W  magnitude from the F140W (see Section \ref{SEC:mag_prior}). The correction is weakly dependent on magnitude, and does not  depend on the identification of the strongest emission. Red circles show H$\alpha$, yellow circles  \oiii, green circles surrounded by a square  \oii, black dots for no lines detected. The average value of the color index is represented using light purple lines and filled circles, in 5 bins of magnitude. {\bf Right panel:} J-H color index computed for the F110W-F160W combination, as a function of F160W.}
 \label{img:MAG_corr}
 \end{figure*}

As shown in Figure~\ref{img:histogr_cover}, many sources are not covered by observations in the F110W band. However, to first approximation,
we can convert between the H  (F140W or F160W)  and the J band magnitudes, as shown in Figure~\ref{img:MAG_corr}. In this figure we plot the magnitude-dependent correction that can be applied to the measurements obtained using the F140W and F160W filters, to recover the magnitude expected in the F110W band. The same figure also shows that the magnitude correction does not depend on which emission line is the strongest in the spectrum. 

\subsubsection{Photo-z PDF from apparent size}
\label{SEC:size_prior}

The mass-size relation \citep[e.g.][]{2013ApJ...762...77P,2014ApJ...788...28V} and the shape of the stellar mass function \citep[see e.g.][and Figure 4 therein]{2017ApJ...847...13L} indicate that the same arguments used for the apparent magnitude can be applied to the apparent size as well: 
apparently larger sources are more likely located at lower rather than higher redshifts. 
Following the Mattig equation \citep{1958AN....284..109M}, the apparent size of a galaxy decreases with the distance, up to $z\sim$1.5.  

\begin{figure}[!ht]
 \centering
 \includegraphics[width=8.5cm]{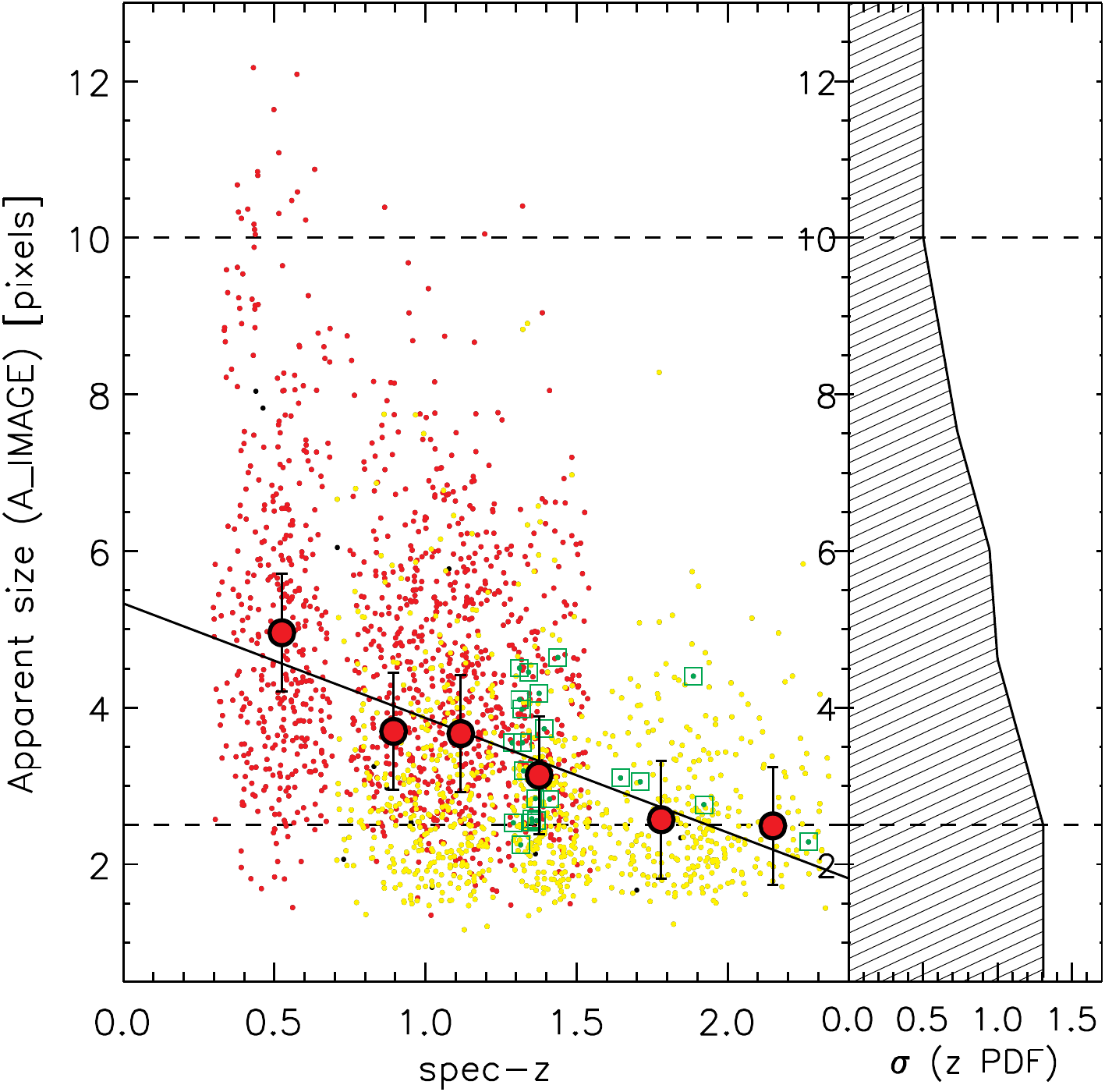}
 \caption{{\bf Left panel:} Relation between apparent size (\emph{SExtractor} A\_IMAGE parameter) measured in the F110W band (J) and spectroscopic redshift, for the calibration ``\emph{gold}'' sample. All the data are used to compute the linear fit, that we use as a prior in the algorithm. However, the prior does not consider values of A\_IMAGE A$<$A$_{\mathrm{min}}=$2.5 pixels or A$>$A$_{\mathrm{max}}=$10 pixels (dashed lines): in these cases, the prior assumes A$=$A$_{\mathrm{min}}$ and A$=$A$_{\mathrm{max}}$ respectively. The actual nature of the strongest emission line is represented with yellow circles for \oiii, red circles for H$\alpha$ and green circles surrounded by squares for \oii. {\bf Right panel:} given the value of A\_IMAGE of a source, the PDF is assumed to be a Gaussian function, with $\sigma$ equivalent to the horizontal dispersion of the data around that specific value of A\_IMAGE ($\pm$0.75). }
 \label{img:A_IMAGE_PDF}
 \end{figure}

 In Figure~\ref{img:A_IMAGE_PDF}, we show the galaxies' apparent size (A\_IMAGE\footnote{The A\_IMAGE parameter is an output of the \emph{SExtractor} software corresponding to the RMS of the luminosity distribution of a galaxy, measured in pixels, along the semi-major axes.}) computed in the J band (F110W), as a function of the spectroscopic redshift. We include only sources in the WISP calibration ``\emph{gold}'' sample. The linear fit to the size-magnitude relation is used as an additional redshift prior in our algorithm. In particular, given the apparent size of a source, we assume a Gaussian PDF in redshift, centered at the redshift indicated by the linear fit, and with a $\sigma$ equivalent to the local horizontal dispersion of the data in the plot of Figure~\ref{img:A_IMAGE_PDF} (right panel).

Also in this case, we computed the linear fit using all the data, but when applying the prior we limit the range to: A\_IMAGE$_{\mathrm{min}}$$<$A\_IMAGE$<$A\_IMAGE$_{\mathrm{max}}$, with A\_IMAGE$_{\mathrm{min}}$=2.5 pixels and A\_IMAGE$_{\mathrm{max}}$=10 pixels. Sources with sizes larger or smaller than these limits are set to A\_IMAGE=A\_IMAGE$_{\mathrm{max}}$ and A\_IMAGE=A\_IMAGE$_{\mathrm{min}}$, respectively.
When no images are available in the J band, the A\_IMAGE parameter is computed using the H band image. Given the negligible differences in the J and H PSFs and sampled stellar populations, no corrections are considered in these cases.

\subsubsection{Photo-z PDF from apparent equivalent width}
\label{SEC:EW_PDF}

\begin{figure}[!ht]
 \centering
 \includegraphics[width=8.5cm]{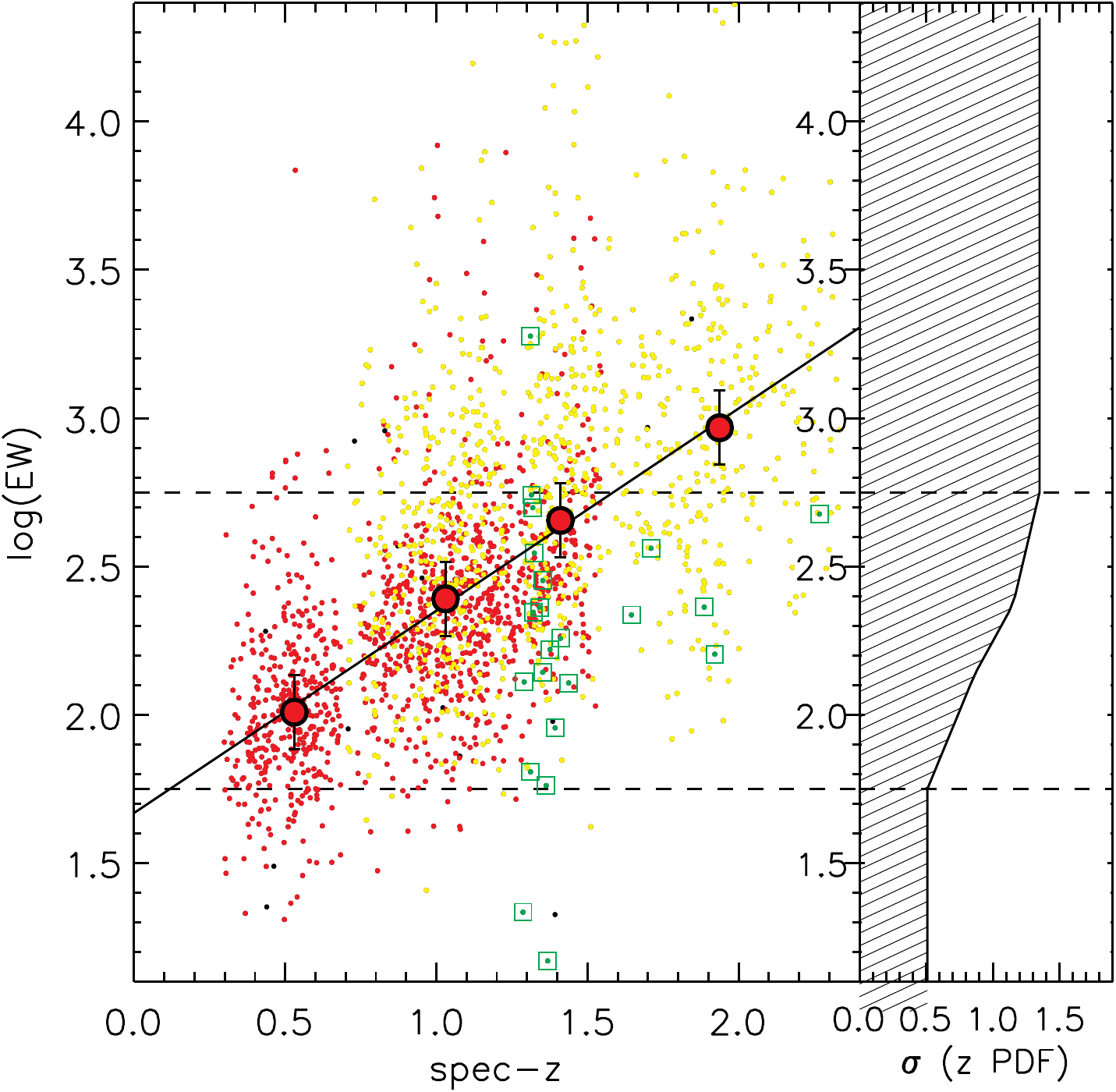}
 \caption{{\bf Left panel:} Relation between apparent equivalent width and spectroscopic redshift, for the calibration ``\emph{gold}'' sample. All the data are used to compute the linear fit, that we use as a prior in the algorithm. However, the prior does not consider values of equivalent width $\log(\mathrm{EW})<\log(\mathrm{EW}_{\mathrm{min}})=$1.75 or $\log(\mathrm{EW})>\log(\mathrm{EW}_{\mathrm{max}})=$2.75 (dashed lines): in these cases, the prior assumes $\log(\mathrm{EW})=\log(\mathrm{EW}_{\mathrm{min}})$ and $\log(\mathrm{EW})=\log(\mathrm{EW}_{\mathrm{max}})$ respectively. The nature of the strongest emission line is represented with yellow circles for \oiii, red circles for H$\alpha$ and green circles surrounded by squares for \oii.  {\bf Right panel:} given the value of EW, the PDF is assumed to be a Gaussian function, with $\sigma$ equivalent to the horizontal dispersion of the data around that specific value of EW ($\pm$0.125). }
 \label{img:EW_PDF}
 \end{figure}

Figure~\ref{img:EW_PDF}, shows the relation existing between the apparent equivalent width of the measured lines and the spectroscopic redshift of the sources. As we do for the apparent J magnitude and size priors (Sections~\ref{SEC:mag_prior} and \ref{SEC:size_prior} respectively), we compute a linear fit to the observed relation (solid line in the Figure). The best fitting relation is used as a redshift prior by our algorithm. In particular, given the EW of an observed emission line, we consider a Gaussian PDF centered at the redshift indicated by the linear fit and with a value of $\sigma$ given by the local horizontal dispersion (right panel of Figure~\ref{img:EW_PDF}).

Also in this case, we limit the range of validity to $\log(\mathrm{EW}_{\mathrm{min}})<\log(\mathrm{EW})<\log(\mathrm{EW}_{\mathrm{max}})$, where $\log(\mathrm{EW}_{\mathrm{min}})=$1.75 and $\log(\mathrm{EW}_{\mathrm{max}})=$2.75. It is possible to observe (Figure~\ref{img:EW_PDF}) that these thresholds allows us to approximately separate sources dominated by the \ha\ and \oii\ emission from those dominated by \oiii, helping to limit the contamination of the \oii\ sample from the other more numerous samples.

\subsubsection{Line flux ratio priors}
\label{SEC:lratios_prior}

Unless the $z$-PDF is particularly narrow, it is difficult to automatically classify an emission line using just the PDF itself. For example, while $\lambda_{\mathrm{H}\alpha}\sim\lambda_{\mathrm{[SII]}}$, it is very unlikely that \sii\ can be the unique single line observed in a spectrum, because \ha\ is commonly brighter than \sii. In general, detecting the H$\alpha$ emission line in a randomly selected spectrum is more common than detecting \oiii, and \oiii\ is more common than \oii\ or other lines such as the fainter H$\beta$, H$\gamma$, or \sii. Then, in order to obtain a proper classification, the expected flux ratios between different lines can be used as an additional source of information.

The average line flux ratios measured in the calibration ``\emph{gold}'' sample allows us to rescale the detection probabilities of the different species/transitions considered. For example, let us assume the case of a spectrum with only one line detected, and a similar value of the photo-$z$ PDF measured at $z_{\mathrm{H\alpha}}$ and $z_{\mathrm{[SII]}}$, with $z$-PDF$\sim$0 for all the other observable emission lines. Since the H$\alpha$ emission is  always stronger than that due to \sii, the line detected must be classified as H$\alpha$. Because the observed wavelength is known ($\lambda^{\mathrm{obs}}_{i}$), identifying the nature of the (strongest) emission line measured ($i$=H$\alpha$, in this example) automatically provides a measure of the spectroscopic redshift: $z=z_{\mathrm{H}\alpha}=(\lambda^{\mathrm{obs}}_{\mathrm{H}\alpha}/\lambda_{\mathrm{H}\alpha})-1$.

\begin{figure*}[!ht]
 \centering
 \includegraphics[width=7.4cm]{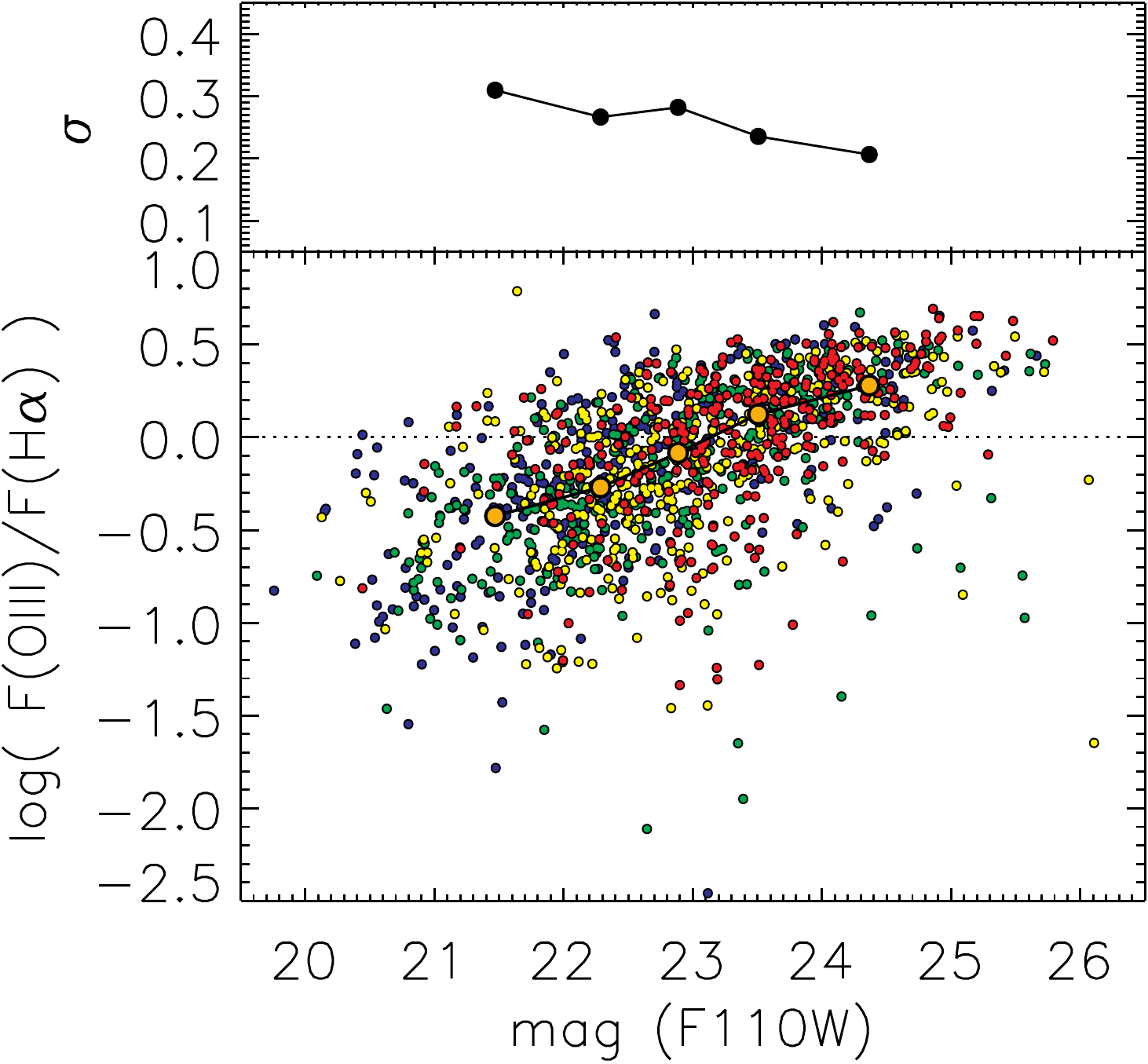}
 \includegraphics[width=7.4cm]{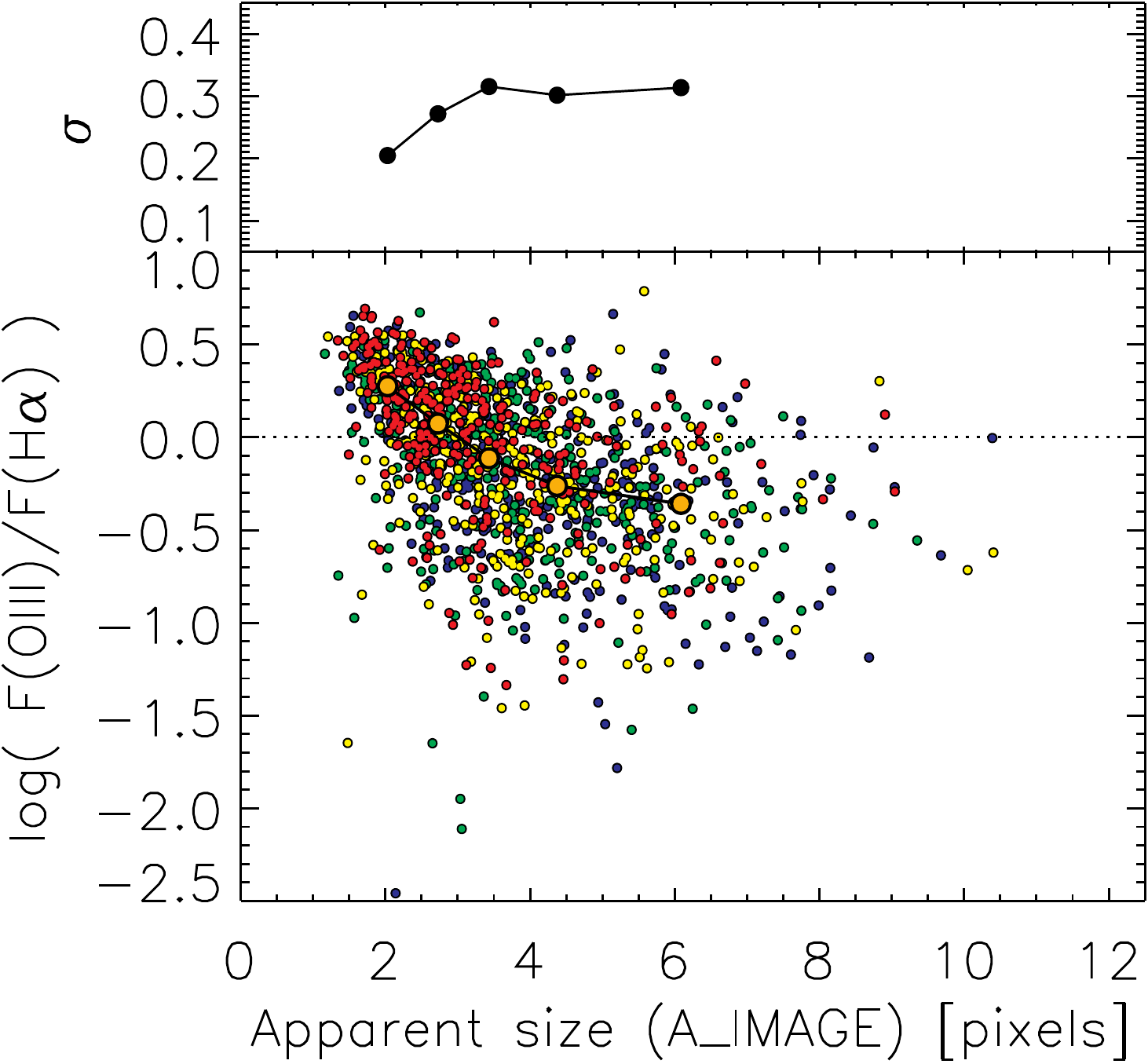}
 \includegraphics[width=7.4cm]{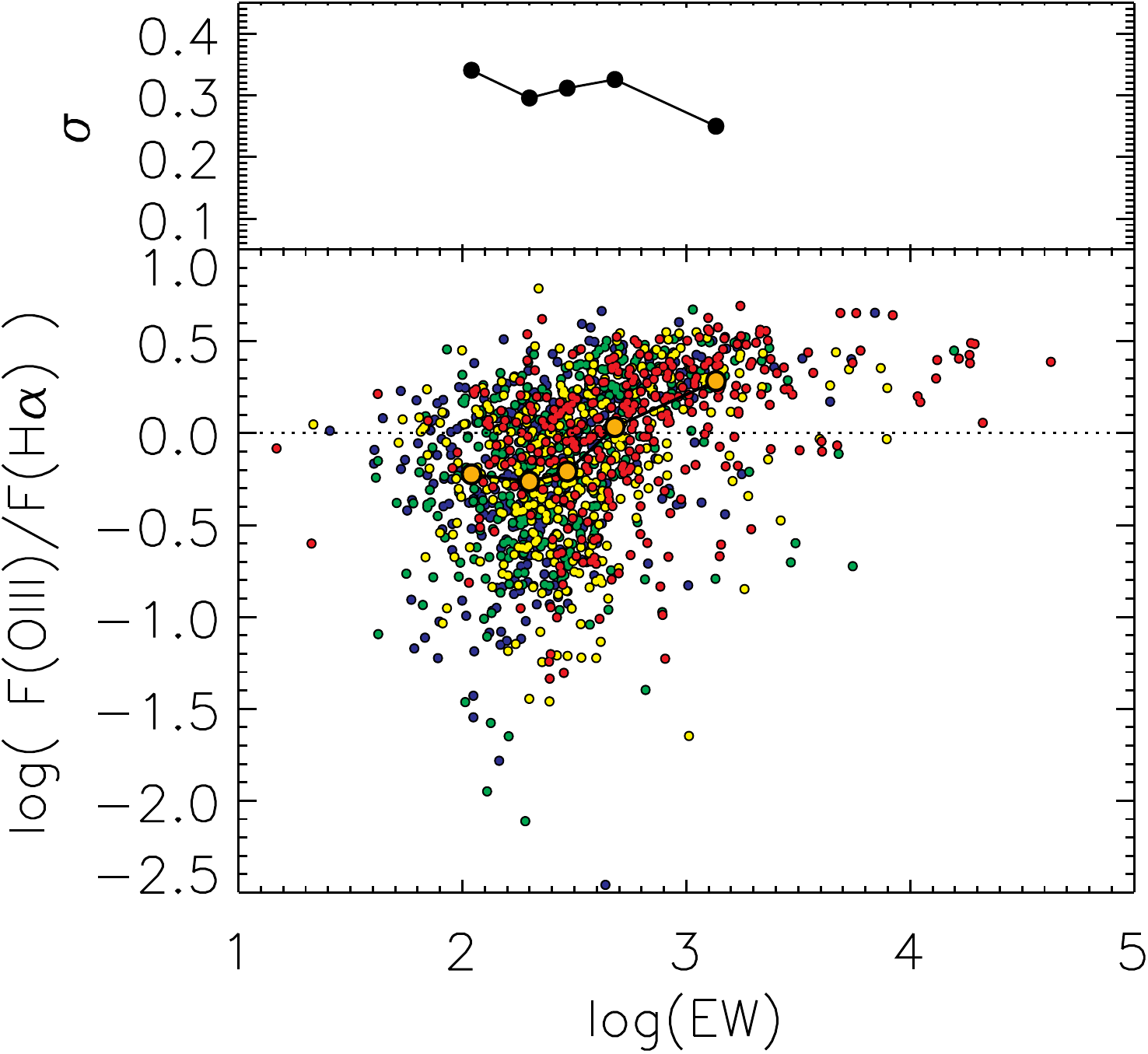}
 \includegraphics[width=7.4cm]{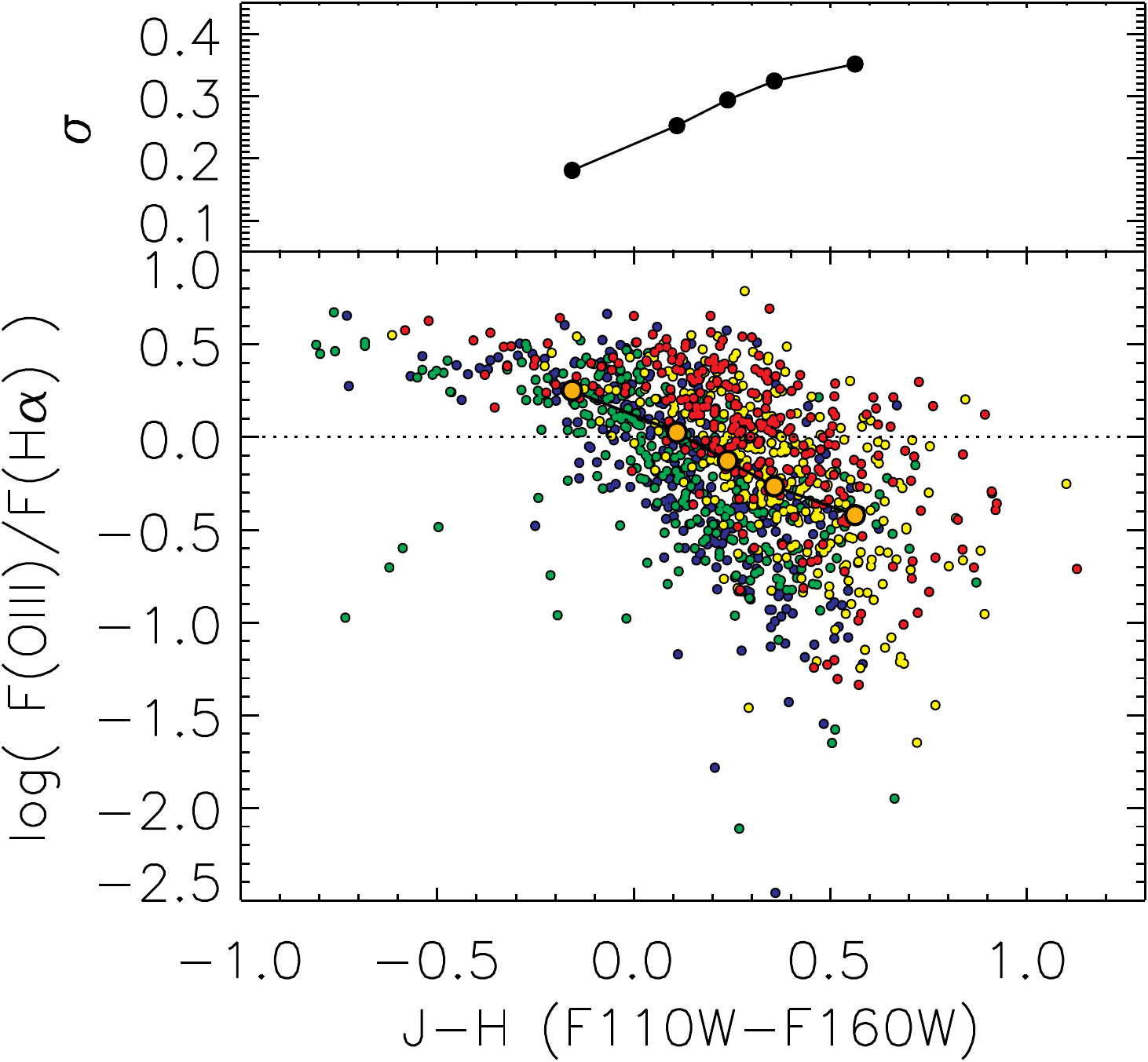}
 \caption{Ratio between H$\alpha$ and \oiii\ fluxes, measured as a function of apparent F110W magnitude (top left panel), size (top right), equivalent width (bottom left) and J-H (F110W-F160W) color index (bottom right), for the sources in the calibration ``\emph{gold}'' sample. A similar relation is computed using also the F110W-F140W color index, but it is not shown in these plots. For each relation, the dispersion ($\sigma$) is shown in the upper parts of each panel. The possible dependence of these relations on the redshift can not be directly taken into account, since the redshift is not known \emph{a priori}. However, using the observed wavelength prior (Section~\ref{SEC:empirical_corr}), it is possible to indirectly correct for such a possible effect. The relations are shown for 4 different bins of redshifts, from $z\sim$0.87 to $z\sim$1.42. In order of increasing redshift, blue ($<z>\sim$0.87), green ($<z>\sim$1.05), yellow ($<z>\sim$1.22), red ($<z>\sim$1.43). }
 \label{img:FLUX_RATIO_priors_A}
 \end{figure*}

\begin{figure*}[!ht]
 \centering
 \includegraphics[width=7.4cm]{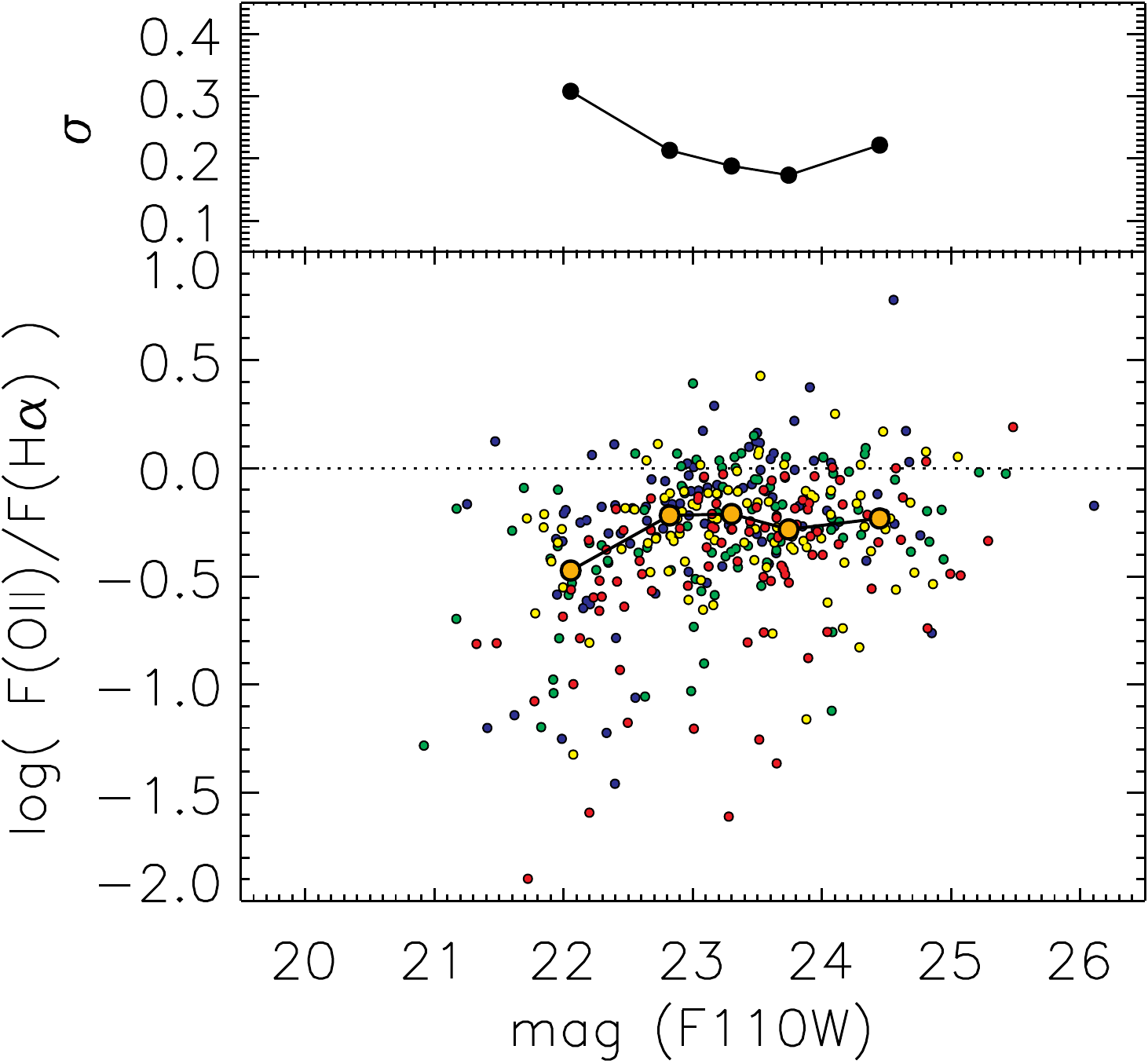}
 \includegraphics[width=7.4cm]{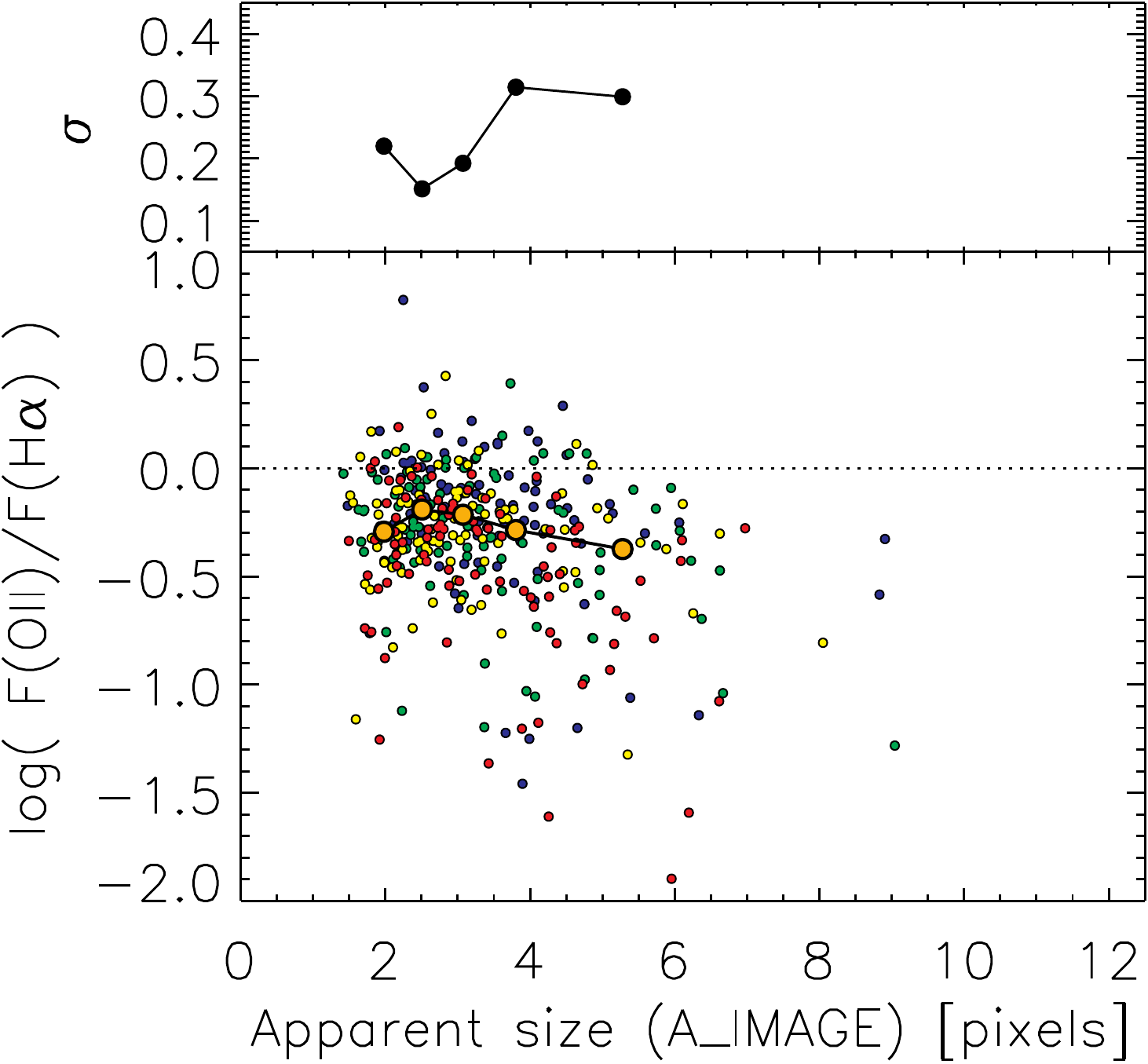}
 \includegraphics[width=7.4cm]{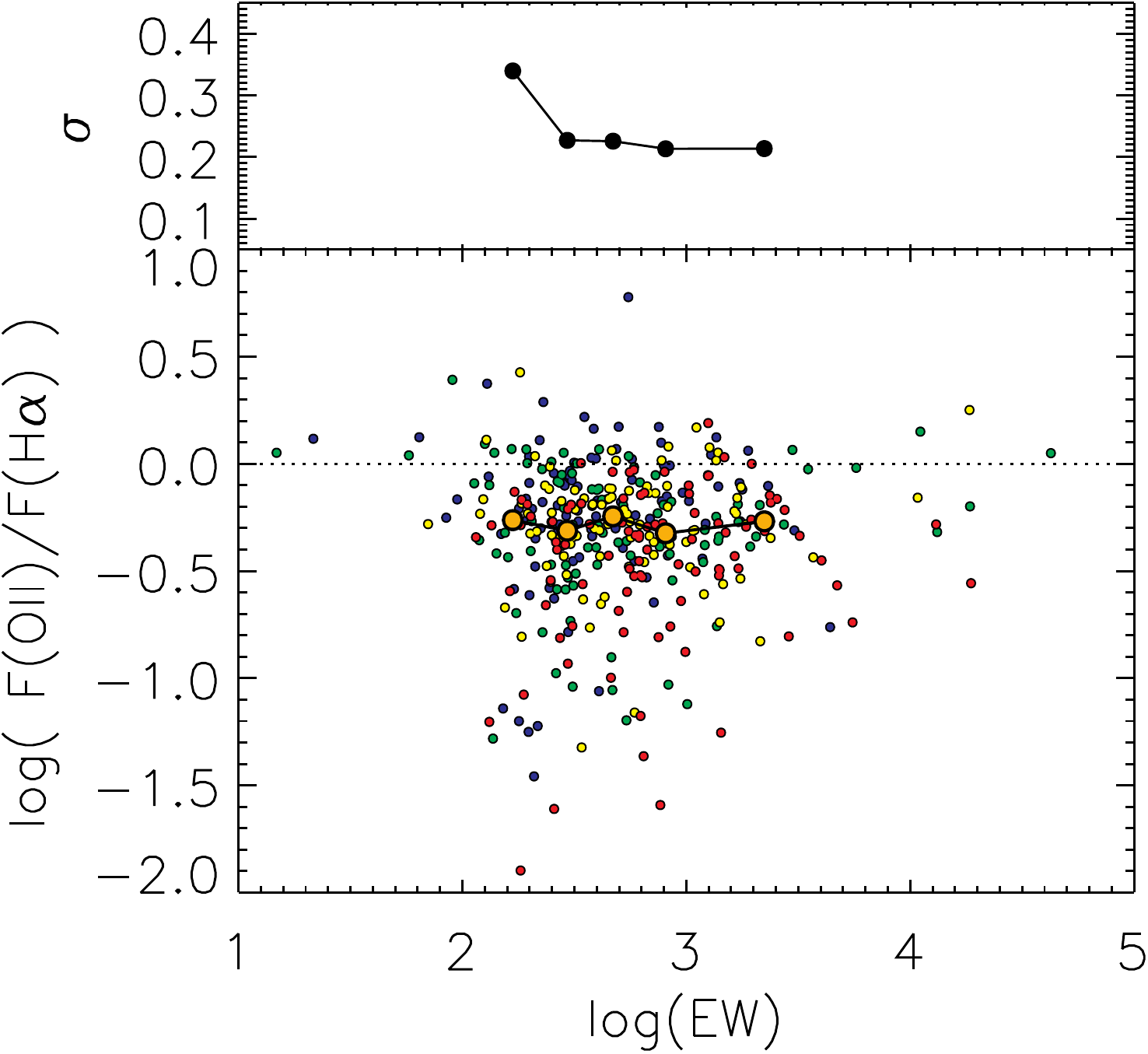}
 \includegraphics[width=7.4cm]{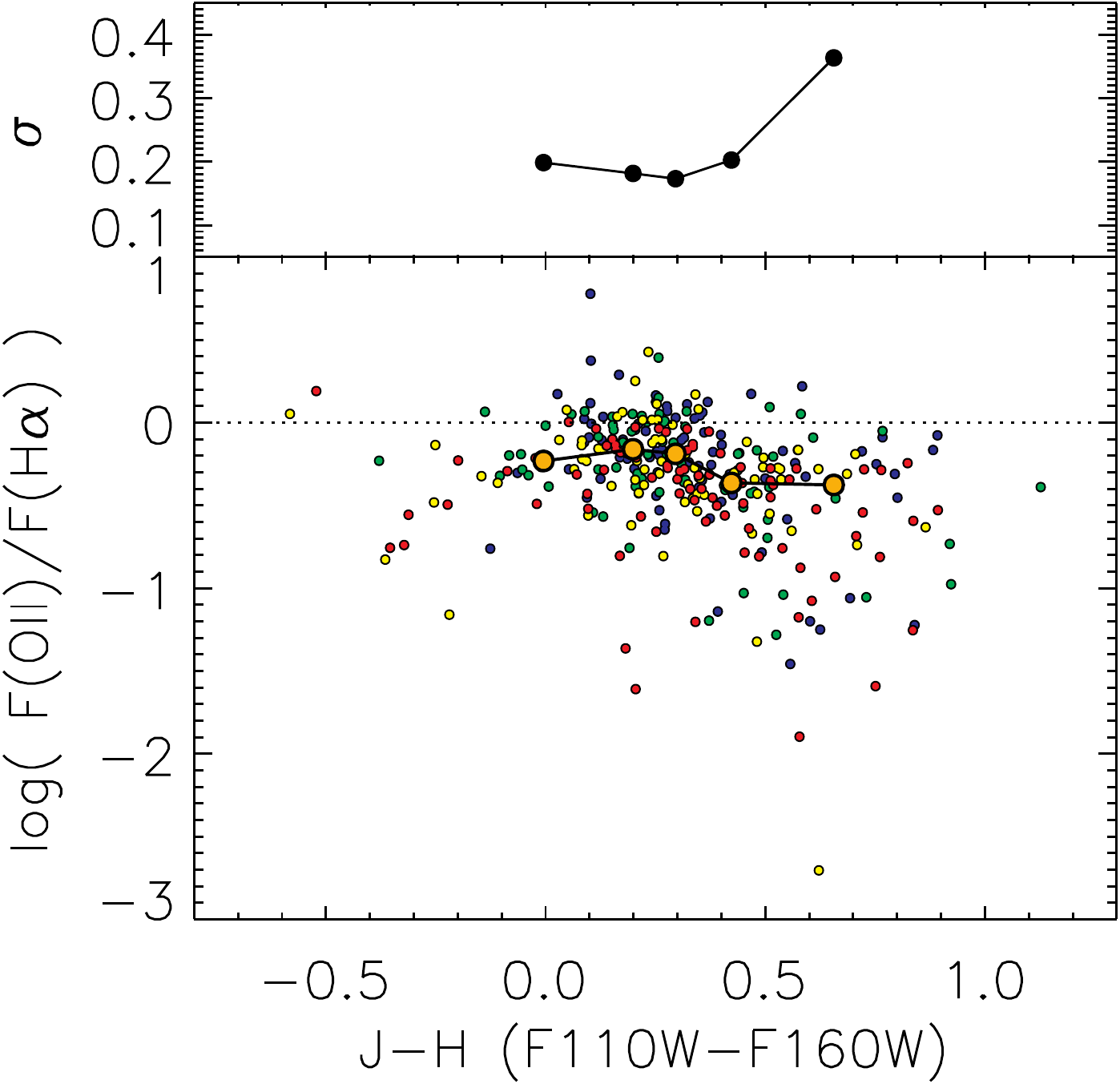}
 \caption{Ratio between H$\alpha$ and \oii\ fluxes, measured as a function of apparent F110W magnitude (top left panel), size (top right), equivalent width (bottom left) and J-H (F110W-F160W) color index (bottom right), for the sources in the calibration ``\emph{gold}'' sample. A similar relation is computed using also the F110W-F140W color index, but it is not shown here. For each relation, the dispersion ($\sigma$) is shown in the upper parts of each panel.  The possible dependence of these relations on the redshift can not be directly taken into account, since the redshift is not known \emph{a priori}. However, using the observed wavelength prior (Section~\ref{SEC:empirical_corr}), it is possible to indirectly correct for such a possible effect. The relations are shown for 4 different bins of redshifts, from $z\sim$1.31 to $z\sim$1.50. In order of increasing redshift, blue ($<z>\sim$1.32), green ($<z>\sim$1.37), yellow ($<z>\sim$1.43), red ($<z>\sim$1.50). }
 \label{img:FLUX_RATIO_priors_B}
 \end{figure*}

For each of the species/transitions \emph{i} considered, we computed the average observed flux ratio F$_{i}$/F$_{\mathrm{H}\alpha}$ as a function of J magnitude, apparent size, equivalent width, and J-H color. This flux ratio is calculated for sources in the calibration ``\emph{gold}'' sample and in the wavelength range of observability of the emission lines\footnote{We do not exploit the ratio between \oiii\ and \oii\ fluxes, computed in the calibration ``\emph{gold}'' sample, above $z\sim$1.50. This is due to a suspected bias in the original classification of strong \oii\ emitters above this redshift. When H$\alpha$ is redder than the upper $\lambda$ limit of the G141 grism, and if the \oiii\ and H$\beta$ lines are weaker than the \oii\ emissions, the latter tends to be wrongly misidentified as H$\alpha$, especially when the S/N ratio is low. }.  As shown in Figures~\ref{img:FLUX_RATIO_priors_A} and~\ref{img:FLUX_RATIO_priors_B}, the line flux ratios show some correlation with all the parameters considered.  Thus, if one or more of the parameters considered are available, it is possible to compute the expected relative strength of different lines.
This information can be directly translated into the relative probability (P$_{\mathrm{FR}}(i)$) that a given species/transition, $i$, corresponds to the strongest line measured in the spectrum. In particular, P$_{\mathrm{FR}}(i)$ corresponds to the weighted average of the various P$_{\mathrm{FR}}^{j}(i)$ obtained from each of the different indicators $j$ considered (J magnitude, apparent size, equivalent width and J-H color):
\begin{eqnarray}
\label{EQ:PFR1}
\mathrm{P}_{\mathrm{FR}}(i)=\ [\ \mathrm{P}_{\mathrm{FR}}^{\mathrm{J}}(i) W_{\mathrm{FR}}^{\mathrm{J}}(i) +\mathrm{P}_{\mathrm{FR}}^{\mathrm{size}}(i) W_{\mathrm{FR}}^{\mathrm{size}}(i) +\ \ \ \ \ \ \ \ \ \ \ \ \ \nonumber \\
+\mathrm{P}_{\mathrm{FR}}^{\mathrm{EW}}(i) W_{\mathrm{FR}}^{\mathrm{EW}}(i) +\mathrm{P}_{\mathrm{FR}}^{\mathrm{J-H}}(i) W_{\mathrm{FR}}^{\mathrm{J-H}}(i)\ ]\ /\sum_{j} W_{\mathrm{FR}}^{j}(i),\ \ \ \ \
\end{eqnarray} 
The weights $W_{\mathrm{FR}}^{j}(i)$ are the square inverse of the dispersion of the relations, computed as a function of the reference parameters (see upper panels of Figures~\ref{img:FLUX_RATIO_priors_A} and~\ref{img:FLUX_RATIO_priors_B}):
\begin{equation}
\label{EQ:WPFR1}
W_{\mathrm{FR}}^{j}(i)=\sigma_{j}^{-2}
\end{equation}

Given the observed wavelength of the strongest emission line $\lambda_{\mathrm{obs}}$, the spectroscopic redshift solution depends on the unknown nature of the species/transition \emph{i} responsible for such an emission as $z_{i}=(\lambda_{\mathrm{obs}}/\lambda_{i})-1$, where $\lambda_{i}$ is the rest frame wavelength  of the species/transition \emph{i}. 
Hence, 
the algorithm considers both the probability P$_{\mathrm{PDF}}(z_{i})$, associated with $z_i$ being the correct redshift solution, and P$_{\mathrm{FR}}(i)$, corresponding to the species/transition \emph{i} being the responsible for the strongest emission:
\begin{equation}
\mathrm{P}(z_{i})=\mathrm{P}_{\mathrm{FR}}(i)\times\mathrm{P}_{\mathrm{PDF}}(z_{i}).
\end{equation}
 
The previous equation does not provide the complete description of the algorithm, for which we refer to Section~\ref{sect:op_descr}. In fact, the complete computation includes the \emph{a posteriori} optimization (Section~\ref{SEC:empirical_corr}) and the limitations described in Section~\ref{SEC:f_ratio_limits}


\subsubsection{Observed wavelength prior}
\label{SEC:empirical_corr}

The calibration of all the methods described is performed by using the high quality data of the calibration "\emph{gold}" sample. However, since these data correspond to real observations, the phase space used to classify the emission lines can not be homogeneously calibrated.
Additionally, we use a large but simplified (quasi--linear) set of functions describing the correlations between the observational parameters and the outputs. While this approach prevents us from \emph{overfitting} the calibration data set (because the output does not depend on a discretized configuration of the inputs, but on their weighted average), the same method results in an uneven precision along the dimensions of the parameter space.

In particular, possible biases could be introduced at different $\lambda_{\mathrm{obs}}$ (i.e. at different redshifts). 
%
%
For example, 
while the average flux ratios are calibrated as a function of magnitude, color, size, and EW, no redshift evolution is taken into account for these relations, since there is not \emph{a priori} knowledge of the redshift itself. 
Similar biases are visible in the combined photo-$z$ PDF, showing systematic over-- and under--estimations of the photometric redshift at different values of the spectroscopic reference.


Some of these redshift dependent biases can be corrected by using the wavelength of the observed lines as a prior. Combining all the methods described in the previous sections, the algorithm computes probability ratios between the different species (the probability associated with the most probable species is assumed equal to 1.0). We can exploit the wavelength position of the detected lines for \emph{a posteriori} fine-tuning of these probability ratios. This approach allows for the optimization of the algorithm.

In the WISP survey, the strongest emission lines measured are mostly either H$\alpha$ or \oiii\ (1293 and 949 objects, respectively, in the test "\emph{gold}" sample). For an additional small fraction of sources, \oii\ is the most prominent line (24 objects in the test "\emph{gold}" sample). 
For this reason, to improve the overall accuracy of the algorithm, the most effective approach consists in the fine-tuning of the ratio between the probabilities computed for H$\alpha$ and \oiii. Secondarily, another correction can be applied to the ratio between the \oiii\ and \oii\ probabilities to further improve the performances of the algorithm.

\begin{figure*}[!ht]
 \centering
 \includegraphics[width=8.5cm]{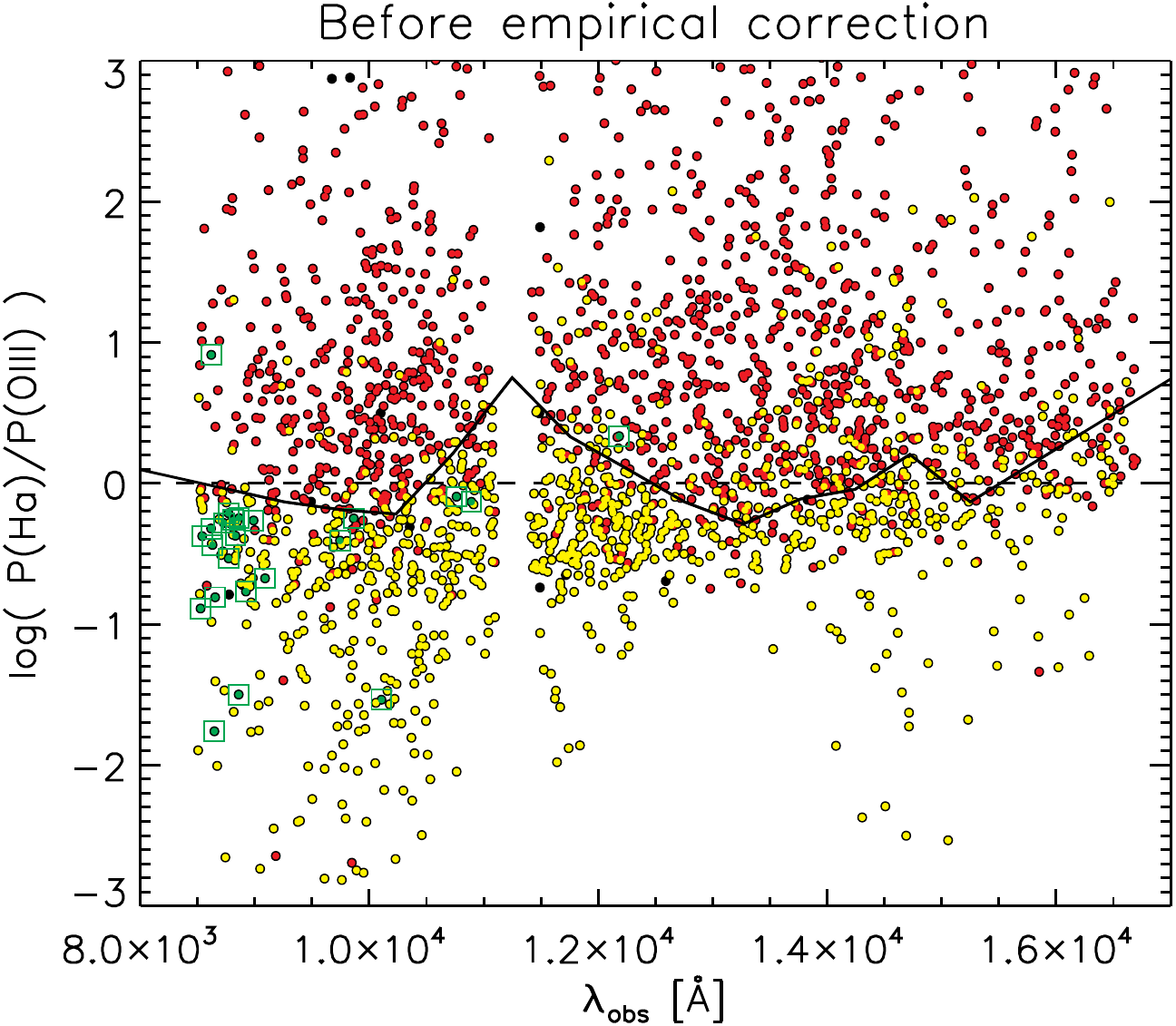}
 \includegraphics[width=8.5cm]{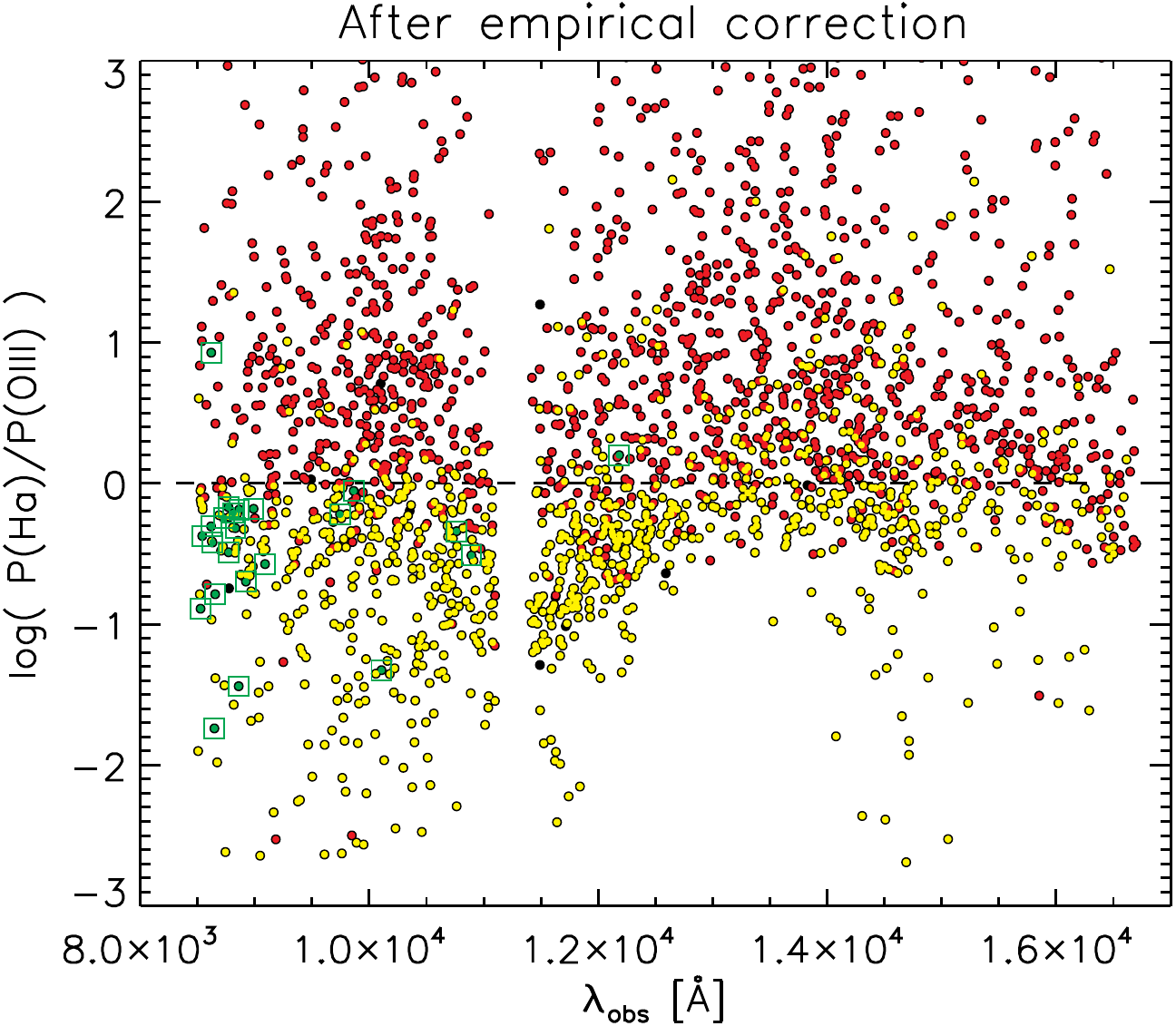}
 \caption{{\bf Left panel:} Measured P(H$\alpha$)/P(\oiii) for the sources in the test "\emph{gold}" sample, before applying the $\lambda_{\mathrm{obs}}$-based fine-tuning correction (P$^{\mathrm{best}}_{\mathrm{H}\alpha/\mathrm{[OIII]}}(\lambda_{\mathrm{obs}})$, black curve, Table~\ref{tbl:Empirical_corr_Ha_OIII}). The identification of the strongest emission line, detected at $\lambda_{\mathrm{obs}}$, is represented with yellow circles for \oiii, red circles for H$\alpha$ and green circles surrounded by squares for \oii. {\bf Right panel:} values of P(H$\alpha$)/P(\oiii) after the application of the P$^{\mathrm{best}}_{\mathrm{H}\alpha/\mathrm{[OIII]}}(\lambda_{\mathrm{obs}})$ correction.  }
 \label{img:PHA_POIII_NOCORR}
 \end{figure*}

In the left panel of Figure~\ref{img:PHA_POIII_NOCORR}, we show the original H$\alpha$ over \oiii\ probability ratio obtained for all the sources of the test "\emph{gold}" sample, before the fine-tuning correction is applied: P$_{\mathrm{H}\alpha/\mathrm{[OIII]}}$=P(H$\alpha$)/P(\oiii). It is possible to see that the majority of the emission lines are already correctly identified (P$_{\mathrm{H}\alpha/\mathrm{[OIII]}}$$<$0 for \oiii\ and P$_{\mathrm{H}\alpha/\mathrm{[OIII]}}$$>$0 for H$\alpha$). However, these results can be improved by deriving, for each bin of observed wavelength, the value of P$_{\mathrm{H}\alpha/\mathrm{[OIII]}}$ that better separates the two emission lines: P$^{\mathrm{best}}_{\mathrm{H}\alpha/\mathrm{[OIII]}}(\lambda_{\mathrm{obs}})$ (black line in the left panel of Figure~\ref{img:PHA_POIII_NOCORR}). In Table~\ref{tbl:Empirical_corr_Ha_OIII}, we report the tabulated values of P$^{\mathrm{best}}_{\mathrm{H}\alpha/\mathrm{[OIII]}}(\lambda_{\mathrm{obs}})$ that we computed in equally spaced wavelength bins ($\Delta\lambda=$500\AA) ranging from 8500 to 16500 \AA\ (the left column in Table~\ref{tbl:Empirical_corr_Ha_OIII} represents the mean $\lambda$ of the spectral lines in each bin). The right panel of Figure~\ref{img:PHA_POIII_NOCORR}, shows the data distribution after the application of the  fine-tuning correction P$^{\mathrm{best}}_{\mathrm{H}\alpha/\mathrm{[OIII]}}(\lambda_{\mathrm{obs}})$.

While we do not modify the value of P$_{\mathrm{H}\alpha}$, the correction factor P$^{\mathrm{best}}_{\mathrm{H}\alpha/\mathrm{[OIII]}}(\lambda_{\mathrm{obs}})$ is applied to both P$_{\mathrm{[OIII]}}$ and P$_{\mathrm{OII}}$. This choice allows us to keep unaltered the probability ratio between \oiii\ and \oii\ that is independently corrected.

\begin{deluxetable}{lr}
\tabletypesize{\footnotesize}

\tablecolumns{2}
\tablewidth{0pc}
\tablecaption{Fine-tuning of the P(H$\alpha$)/P(\oiii) ratio}
\tablehead{\colhead{$<\lambda_{\mathrm{obs}} [\emph{\AA}]>$} & \colhead{$\log$P$^{\mathrm{best}}_{\mathrm{H}\alpha/\mathrm{[OIII]}}(\lambda_{\mathrm{obs}})$} }
\startdata
\label{tbl:Empirical_corr_Ha_OIII}
8763.10 & -0.042 \\ 
9259.92 & -0.132   \\ 
9771.42 & -0.189   \\ 
10233.7 & -0.218   \\ 
10730.2 & 0.211    \\ 
11248.4 & 0.750    \\ 
11755.0 & 0.331    \\ 
12234.3 & 0.111    \\ 
12755.0 & -0.132   \\ 
13267.0 & -0.289   \\ 
13767.4 & -0.111   \\ 
14242.6 & -0.043  \\ 
14721.9 & 0.205    \\ 
15250.4 & -0.132   \\ 
15743.8 & 0.117    \\ 
16240.9 & 0.364    \\ 
\enddata
\end{deluxetable}

\begin{figure*}[!ht]
 \centering
 \includegraphics[width=8.5cm]{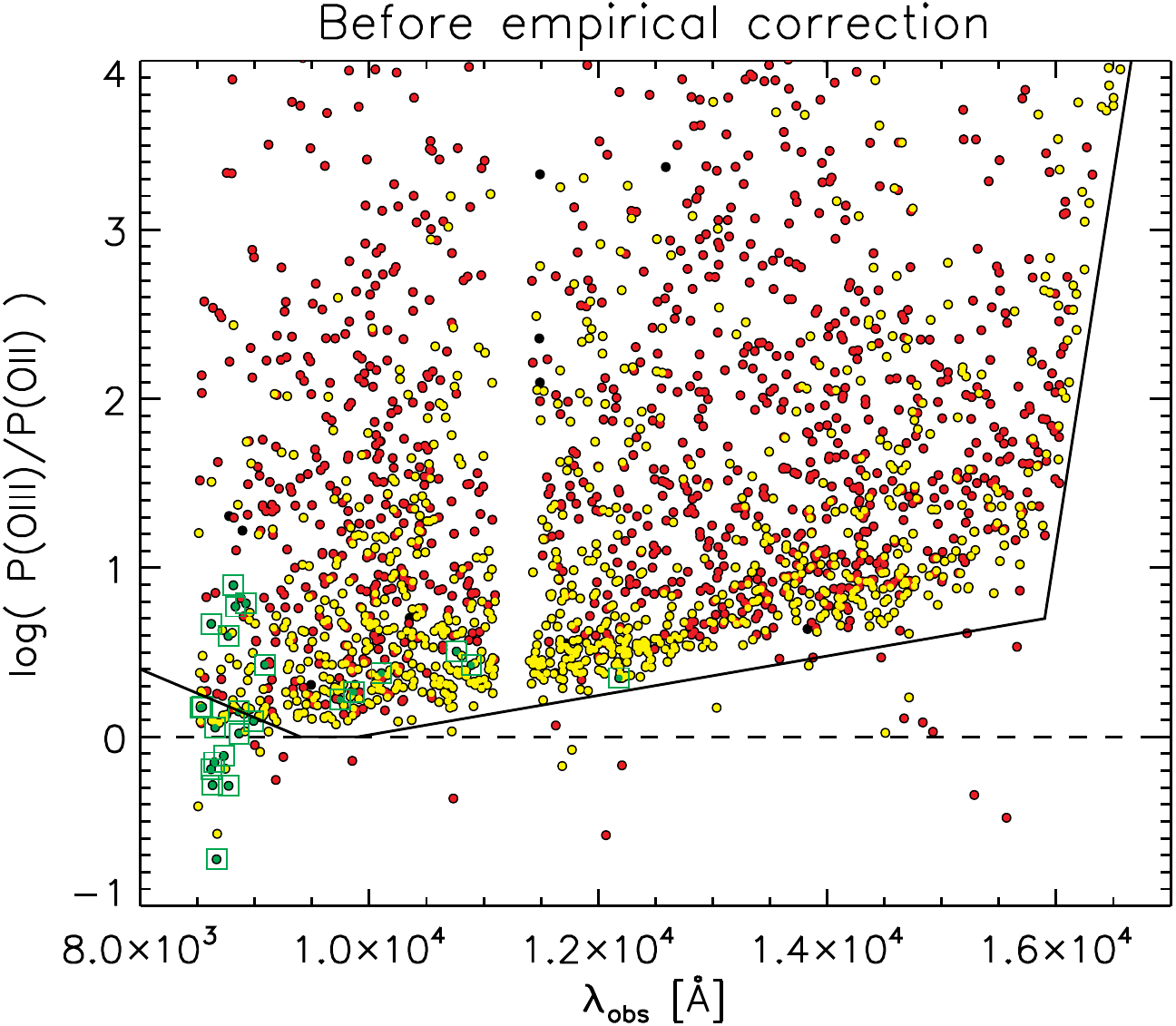}
 \includegraphics[width=8.5cm]{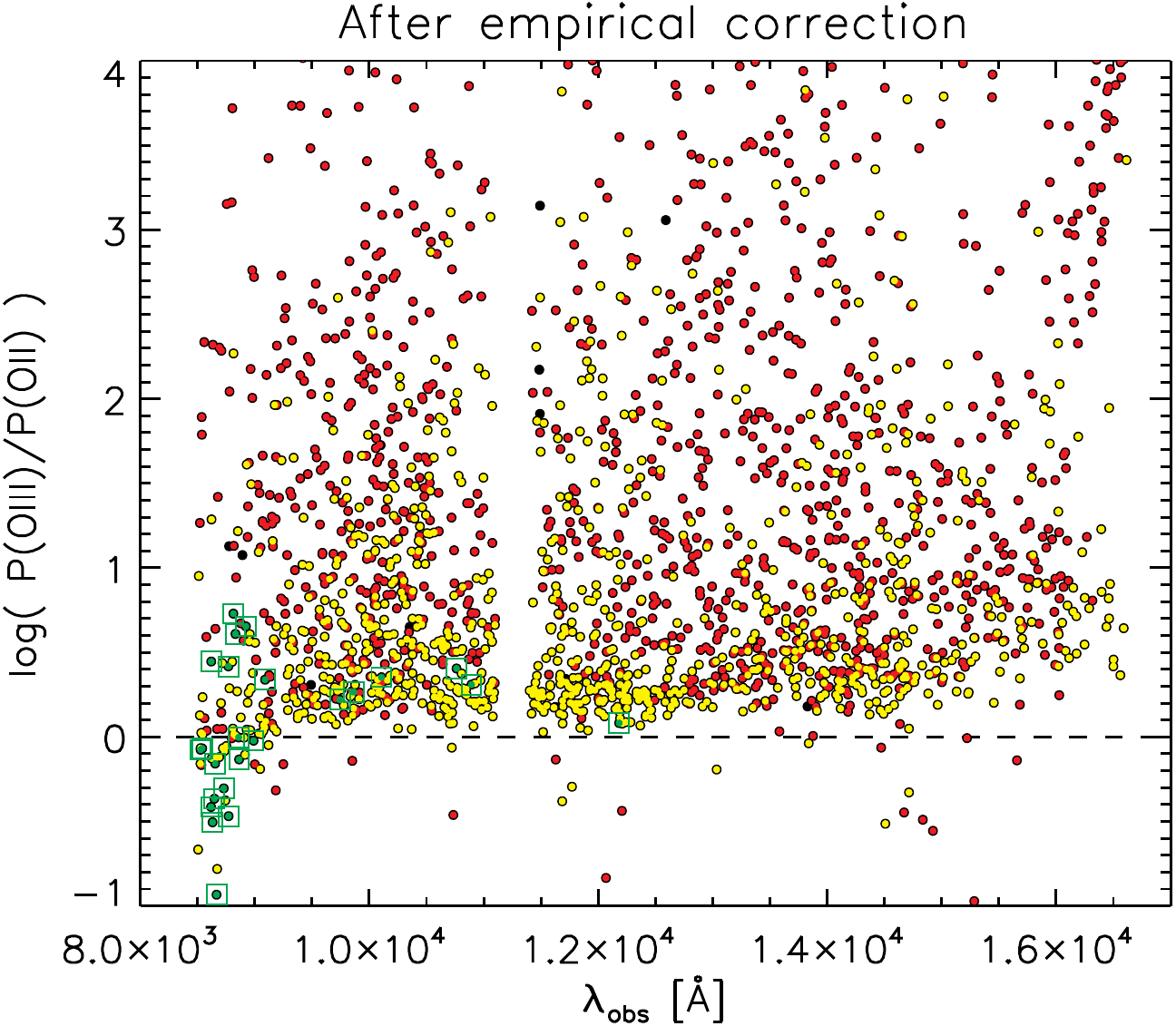}
 \caption{{\bf Left panel:} Measured P(\oiii)/P(\oii) ratio for the sources in the test "\emph{gold}" sample, before applying the $\lambda_{\mathrm{obs}}$-based fine-tuning correction (P$^{\mathrm{best}}_{\mathrm{[OIII]}/\mathrm{OII}}(\lambda_{\mathrm{obs}})$, black curve, Table~\ref{tbl:Empirical_corr_OIII_OII}). The identification of the strongest emission line, detected at $\lambda_{\mathrm{obs}}$, is represented with yellow circles for \oiii, red circles for H$\alpha$ and green circles surrounded by squares for \oii. {\bf Right panel:} values of P(\oiii)/P(\oii) after the application of the P$^{\mathrm{best}}_{\mathrm{[OIII]}/\mathrm{OII}}(\lambda_{\mathrm{obs}})$ correction.  }
 \label{img:POIII_POII_NOCORR}
 \end{figure*}

Similarly to what was done for the H$\alpha$ and \oiii\ pair, we applied a correction to the \oii\ over \oiii\ probability ratio. In the left panel of Figure~\ref{img:POIII_POII_NOCORR}, we show the original \oiii\ over \oii\ probability ratio obtained for all the sources of the test "\emph{gold}" sample, before the fine-tuning correction is applied: P$_{\mathrm{[OIII]}/\mathrm{[OII]}}$=P(\oiii)/P(\oii). In this case, given the lack of sources with prominent \oii\ emission above $\lambda_{\mathrm{obs}}\sim$9000\AA, the probability ratio can not be precisely fine--tuned. However, above $\lambda_{\mathrm{obs}}\sim$11500\AA, a systematic deviation of this ratio from P$_{\mathrm{[OIII]}/\mathrm{[OII]}}$=0 is immediately evident and easy to correct. In Table~\ref{tbl:Empirical_corr_OIII_OII}, we report the values of P$_{\mathrm{[OIII]}/\mathrm{[OII]}}$ that we applied to P(\oii) to fine tune the P(\oiii)/P(\oii) ratio.

\begin{deluxetable}{lr}
\tabletypesize{\footnotesize}

\tablecolumns{2}
\tablewidth{0pc}
\tablecaption{Fine-tuning of the P(\oiii)/P(\oii) ratio}
\tablehead{\colhead{$<\lambda_{\mathrm{obs}} [\emph{\AA}]>$} & \colhead{$\log$P$^{\mathrm{best}}_{\mathrm{[OIII]}/\mathrm{OII}}(\lambda_{\mathrm{obs}})$} }
\startdata
\label{tbl:Empirical_corr_OIII_OII}
8000  & 0.4 \\
9400  & 0.0 \\
9900  & 0.0 \\
15900 & 0.7 \\
17000 & 5.5 \\
\enddata
\end{deluxetable}

\subsubsection{Limitations}
\label{SEC:f_ratio_limits}
The wavelength range covered by WISP is limited.
For this reason, the expected flux ratios can not always be used to identify the strongest \emph{measured} line. 
For example, even if the flux ratios indicate that \ha\ is the \emph{expected} strongest line, \ha\ may fall outside of the observable wavelength range. The brightest \emph{measured} line may be due to other species/transitions, (e.g. \oiii\ or \oii). 


 
In particular, H$\alpha$ falls outside the grism at $z>1.6$, corresponding to $\lambda_{\mathrm{obs}}$(\ha)$>$17000\AA\ and $\lambda_{\mathrm{obs}}$(\oiii)$>$12900\AA. 
Therefore, when 1) the observed wavelength of the strongest (single) observed emission line is $\lambda_{\mathrm{obs}}>$12900\AA, and 2) the expected F$_{\mathrm{H}\alpha}$/F$_{\mathrm{[OIII]}}$ flux ratio is $>1$, there is a non--negligible chance that H$\alpha$ is indeed stronger than \oiii\ (and \oii), but outside the range of observability.
Given these circumstances, when the conditions 1) and 2) are verified at the same time, we set 
P$_{\mathrm{FR}}(\mathrm{H}\alpha)$/P$_{\mathrm{FR}}(\mathrm{[OIII]})= 1$, 
while keeping the values computed for P$_{\mathrm{FR}}(\mathrm{[OIII]})$ and P$_{\mathrm{FR}}(\mathrm{[OII]})$. This is equivalent to 
not considering the expected flux ratio between \ha\ and \oiii).

Similarly to the pair \ha\ - \oiii, the same problem is expected to affect the  \oiii\ - \oii\ pair when \oiii\ falls outside the range of observability. However, this mistake is only theoretically possible, since the depth of the WISP survey make the observability of a $z\gtrsim$2.5 source unlikely. For this reason, we do not apply any correction to the expected P$_{\mathrm{FR}}(\mathrm{[OIII]})$/P$_{\mathrm{FR}}(\mathrm{[OII]})$ ratio. 

On the opposite side of the wavelength range covered by the grisms, the same problem can arise if \oiii\ is the \emph{intrinsic} strongest line but it is located below $\lambda\sim 8000$\AA, corresponding to $\lambda_{\mathrm{obs}}$(\ha)$<$10536\AA. While the flux ratios could correctly indicate \oiii\ as the expected strongest line, the strongest \emph{measured} line that the algorithm must indicate is \ha. We estimated the effect of this possibility by excluding the flux ratios when $\lambda<10536$\AA\ and F$_{\mathrm{H}\alpha}$/F$_{\mathrm{[OIII]}}<1$. The results that we obtain confirm that the overall effect is negligible on both the test sample (accuracy$\sim$81.7\%) and the \emph{``single line''} sample (for which the recovered distribution in $z$ does not change significantly). For this reason we do consider the flux ratios indicators also when $\lambda<10536$\AA.

We emphasize on the fact that the probability P$_{\mathrm{FR}}$, obtained from the expected flux ratios and corrected as described above, is not used alone. 
In fact, the outputs of the \emph{classification} block are combined with the outputs of the \emph{regression} and \emph{optimization} blocks, as we will better describe in detail in Section~\ref{sect:op_descr}.

The test "\emph{gold}" sample includes 632 sources for which the limitations described above apply, out of a total of 2283 galaxies. For this subsample, after applying the correction described, we measure an accuracy (i.e. fraction of spectral lines correctly recovered by the algorithm) of 86.0, that is even higher than the average accuracy obtained testing the complete "\emph{gold}" sample (82.6\%, see Section~\ref{SECT:PRECISION}). 

Another limitation concerns the availability of the observations needed to compute the J-H color index. We can compute J-H only when one combination of F110W-F140W or F110W-F160W is available. In all the other cases we rely only on the flux ratio priors based on the other indicators.

Finally, when the J magnitude is not available, we consider the H band (F140W or F160W) for the magnitude indicator. In this case, the H magnitude is recalibrated to J as shown in Figures~\ref{img:MAG_corr}.

\subsection{Operation description}
\label{sect:op_descr}

Given the strongest line measured in a spectrum (sometimes the only detected line), the main output of the algorithm is the relative probability, for each of the  species/transition considered, to be the responsible for such an emission.

In Section~\ref{sec:ALGORITHM_INPUTS}, we described the modules of the algorithm, their organization in separate blocks and their independent calibration. 
In this Section we detail on how the different modules are combined with each other, and how the algorithm gets the final probabilities from the outputs of different blocks.



\begin{figure}[!ht]
 \centering
 \includegraphics[width=8.5cm]{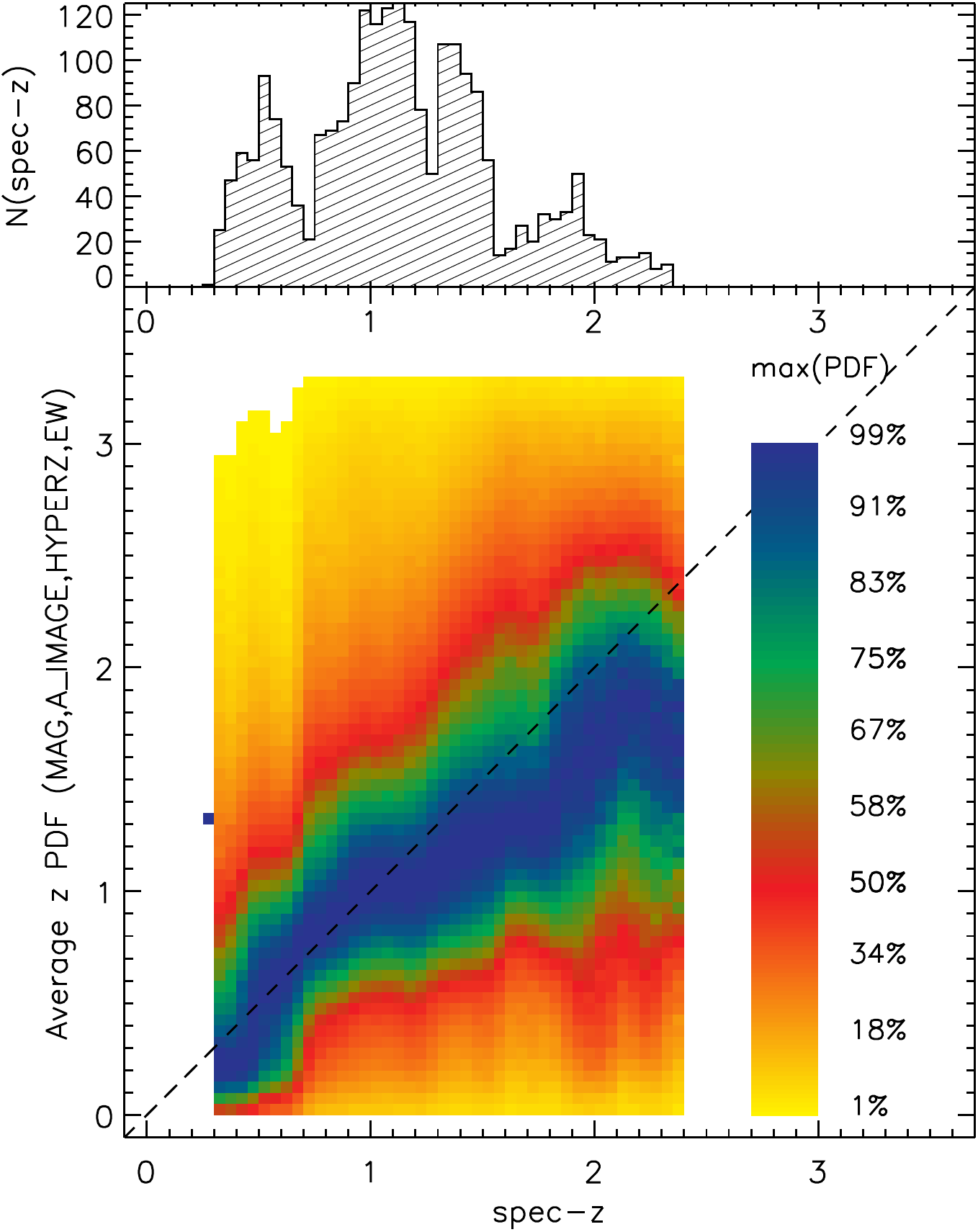}
 \caption{Average \pdfn\ resulting from the combination of the \pdfn s obtained using all the redshift indicators described in Sections~\ref{SEC:photoz},~\ref{SEC:mag_prior},~\ref{SEC:size_prior},~\ref{SEC:EW_PDF} (SED fitting, apparent J magnitude, size, and equivalent width respectively), and singularly shown in Figure~\ref{img:All_PDFs}. The histogram in the upper panel represents the distribution of the spectroscopic sources used in the experiment ("\emph{gold}" sample).}
 \label{img:COMB_PDFs}
 \end{figure}
\begin{figure*}[!ht]
 \centering
 \includegraphics[width=16.5cm]{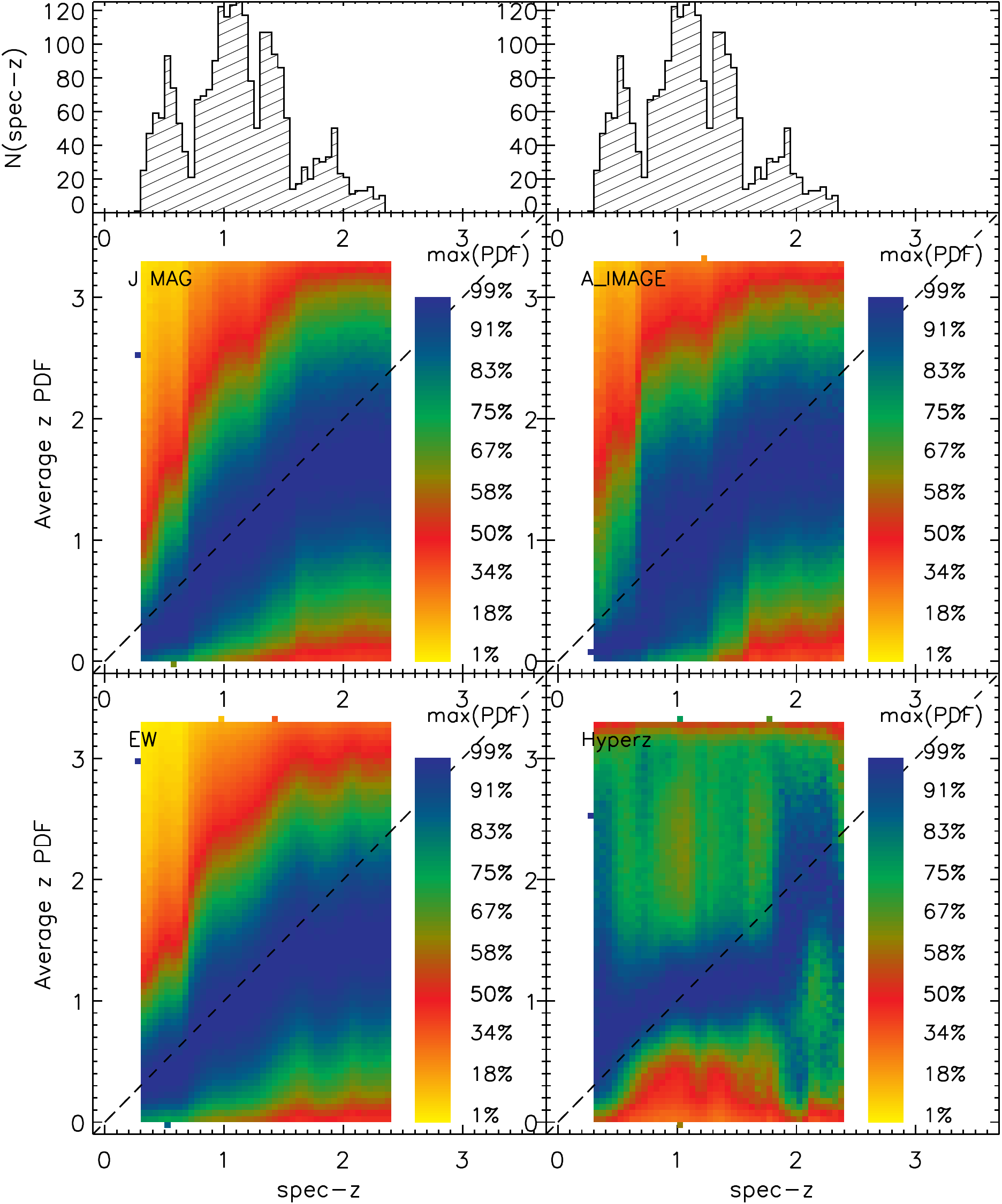}
 \caption{Average \pdfn\ obtained using different methods, as a function of the actual spectroscopic redshift of the sources in the "\emph{gold}" sample. The \pdfn s are computed in bins of spectroscopic redshift (x-axes) and represented through a color scale. The duplicated histograms in the upper panels represent the distribution of the spectroscopic sources used in the experiment. {\bf Top left:} average \pdfn\ obtained using the J magnitude (F110W band) as a redshift indicator (Section~\ref{SEC:mag_prior}). {\bf Top right:} average \pdfn\ obtained using the A\_IMAGE parameter (apparent size in the F110W band) (Section~\ref{SEC:size_prior}). {\bf Bottom left:} average \pdfn\ obtained using the equivalent width (Section~\ref{SEC:EW_PDF}) {\bf Bottom right:} average \pdfn\ obtained using the \emph{hyperz} SED fitting method (Section~\ref{SEC:photoz}). While the low precision SED-fitting technique encounters problems between z$\sim$1 and z$\sim$2, the other redshift indicators are better correlated with the redshift in the same range.}
 \label{img:All_PDFs}
 \end{figure*}

The output of the first block is a combination of redshift probability distribution functions ($z$-\pdfn), i.e. the (peak--normalized) probability associated with each value of $z$, regardless of the line ID.
The final \pdfn\ is obtained by combining the \pdfn s independently derived from SED-fitting, apparent magnitude, apparent size, and equivalent width (see Sections~\ref{SEC:photoz}, \ref{SEC:mag_prior}, \ref{SEC:size_prior}, \ref{SEC:EW_PDF}, and Figures~\ref{img:COMB_PDFs} and \ref{img:All_PDFs}). As such, this function does not provide any information to identify the expected strongest emission line. 

The second block statistically associates a set of flux ratios with a given combination of apparent magnitude, size, equivalent width and J-H color index (see Section~\ref{SEC:lratios_prior} and Figures~\ref{img:FLUX_RATIO_priors_A} and \ref{img:FLUX_RATIO_priors_B}). This second block does not provide information on the expected redshift, but it predicts which among the considered species/transitions is more likely the responsible for the strongest emission in the spectrum considered

The third block performs the fine-tuning of the relative probabilities computed for the different species/transitions (see Section~\ref{SEC:empirical_corr}, Tables~\ref{tbl:Empirical_corr_Ha_OIII} and \ref{tbl:Empirical_corr_OIII_OII}, and Figures~\ref{img:PHA_POIII_NOCORR} and \ref{img:POIII_POII_NOCORR}). This function depends only on the observed wavelength of the strongest emission line ($\lambda_{\mathrm{obs}}$).

Schematically, for each of the species/transition \emph{i}, the three functions are combined as follows:
\begin{equation}
\label{EQ:THREE_FUNCT_COMBO}
\mathrm{P}(i)  \equiv \mathrm{P}(z_{i})=\mathrm{P}_{\mathrm{PDF}}(z_{i})\times \mathrm{P}_{\mathrm{FR}}(i)\times\mathrm{P}_{\mathrm{corr}}(i,\lambda_{\mathrm{obs}}),
\end{equation}
where $\mathrm{P}(i)$ is directly related to the probability that the species/transition \emph{i} is responsible for the strongest emission line in a spectrum, or equivalently, to the  probability  that $z_{i}$ ($\equiv \frac{\lambda_{\mathrm{obs}}}{\lambda_{i}}-1$) is the correct spectroscopic redshift. 
$\mathrm{P}_{\mathrm{PDF}}(z_{i})$ is the value of the $z$-\pdfn\ in $z_{i}$, while $\mathrm{P}_{\mathrm{FR}}(i)$ is the probability obtained from the flux ratio priors, and $\mathrm{P}_{\mathrm{corr}}(i,\lambda_{\mathrm{obs}})$ represents the $\lambda_{\mathrm{obs}}$ dependent fine-tuning correction.
The $z$-\pdfn\ is the combination of the SED fitting output with the priors obtained from the apparent J magnitude, size and equivalent width: 
\begin{equation}
\label{eqn:PDF_combo}
\mathrm{P}_{\mathrm{PDF}}(z_{i})=\mathrm{P}^{\mathrm{sed\_fit}}_{\mathrm{PDF}}(z_{i})\mathrm{P}^{\mathrm{J}}_{\mathrm{PDF}}(z_{i})\mathrm{P}^{\mathrm{size}}_{\mathrm{PDF}}(z_{i})\mathrm{P}^{\mathrm{EW}}_{\mathrm{PDF}}(z_{i}).
\end{equation}
$\mathrm{P}_{\mathrm{FR}}(i)$, derived from the measured flux ratios, is the weighted average of the values inferred from apparent magnitude, size, equivalent width and J-H color, as described by Equations~\ref{EQ:PFR1} and~\ref{EQ:WPFR1}.
For each source, the values of P(\emph{i}) are normalized to the maximum value of P(\emph{i}) obtained among the different species/transitions \emph{i}, so that if, e.g.,  the transition \emph{n} is the most probable one responsible for the strongest emission, with the transition \emph{m} having only half of the probability with respect to \emph{n}, then the algorithm indicates P(\emph{n})=1 and P(\emph{m})=0.5 respectively (i.e., the sum over $i$ of the factors P(\emph{i}) is larger than 1.0 \footnote{It is always possible to renormalize the single values of P$_{i}$ to their sum, obtaining actual values of probability, since all the terms are known. However, we emphasize that such a convenction would be misleading: the algorithm outputs relative probabilities among the options taken into account, but it does not consider \emph{all} the possibilities, such as additional transitions (e.g. Ly$\alpha$) or possible spurious detections. }). 

The output probability ratios are not directly and unequivocally related to the expected flux ratio between two transitions. This can be understood if we consider the hypothetical example of a source characterized by an extremely narrow and precise photo-$z$ PDF. In this case, the probability associated with the emission line identified as the strongest would be very high, independent of the flux expected from the other transitions.

\begin{figure}[!ht]
 \centering
 \includegraphics[width=8.5cm]{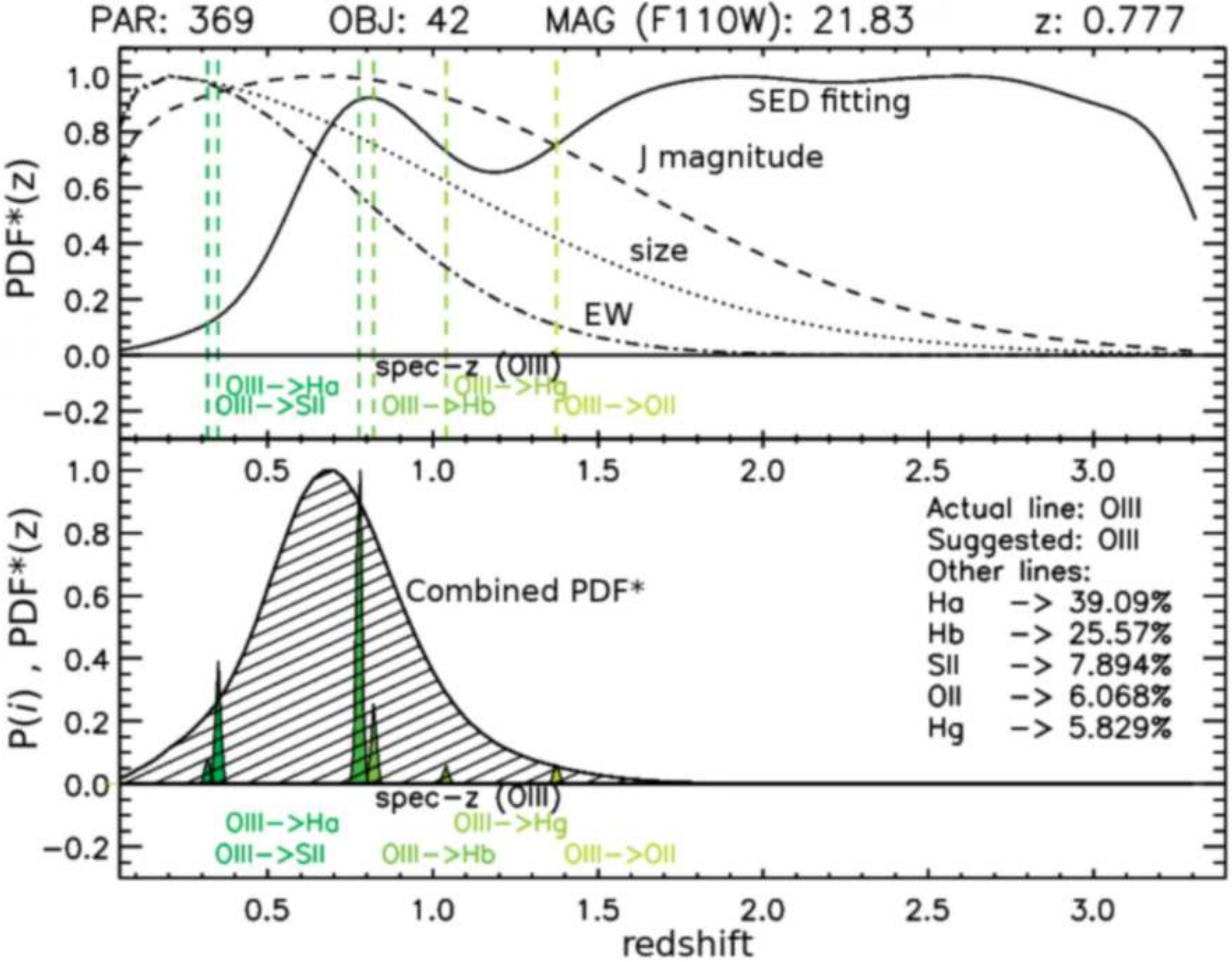}
 \caption{Example output of the algorithm. The {\bf upper panel} shows the four $z$-\pdfn s obtained from SED fitting ($\mathrm{P}^{\mathrm{SED\_FIT}}_{\mathrm{PDF}}(z)$, black continuous line), apparent J magnitude ($\mathrm{P}^{\mathrm{J}}_{\mathrm{PDF}}(z)$, black dashed line), apparent size ($\mathrm{P}^{\mathrm{size}}_{\mathrm{PDF}}(z)$, black dotted line), and equivalent width ($\mathrm{P}^{\mathrm{EW}}_{\mathrm{PDF}}(z)$, black dotted-dashed line ).
The redshift solution originally indicated by the WISP reviewers is indicated by a ``spec-$z$'' label, with the corresponding species/transition reported in parentheses (in this example, \oiii). All the possible redshift solutions $z_{i}$ are also shown using vertical dashed lines labelled as: ``original solution -$>$ other possible solution'' (example: \oiii\ -$>$ H$\alpha$).
In the {\bf lower panel}, the three $z$-\pdfn s are combined in a unique \pdfn\ ($\mathrm{P}_{\mathrm{PDF}}(z)$, continuous black line and shaded area), as described by Equation~\ref{eqn:PDF_combo}. The possible solutions are represented using Gaussian profiles ($\sigma=\Delta z$) with amplitude equivalent to the normalized value of P($i$). For each of the species/transition, the corresponding normalized value of relative probability (Equation~\ref{EQ:THREE_FUNCT_COMBO}) is reported in the legend (the probability is expressed as a percent fraction of the maximum value of P$(i)$ obtained, corresponding to 100\%). }
 \label{img:PDF_example}
 \end{figure}

Figure~\ref{img:PDF_example} shows the typical output of the algorithm. While in the upper panel of this Figure, the $z$-\pdfn s are independently shown, in the lower panel it is possible to see their combination (see Equation~\ref{eqn:PDF_combo}). The relative probabilities, associated with the different possible \emph{i} solutions, are represented using Gaussian functions (where $\sigma$ corresponds to the redshift uncertainty $\Delta z$ in the original WISP catalog). Their peak values are proportional to the corresponding values of P(\emph{i})

\section{Accuracy and uncertainties}
\label{SECT:PRECISION}
We computed the accuracy of the algorithm by running it on the test "\emph{gold}" sample and comparing the output predictions with the original WISP classifications. The latter  sample includes the identifications performed by WISP reviewers when they agreed on the redshift. The coverage by both grisms is required, and at least two emission lines must be detected. This selection guarantees that the uncertainties estimated can be almost completely attributed to the algorithm itself and not to the original WISP classification. In any case, if a small fraction of wrong identifications were attributable to the original WISP classification, we would be underestimating the actual accuracy of the algorithm (unless the amount of misidentification is big enough to sensibly modify the calibration of the algorithm itself). 
We limited the wavelength range of both the calibration and the test phases to two intervals: $8500$ \AA\ $<\lambda_{\mathrm{obs}}<11000$ \AA\ and $11400$ \AA\ $<\lambda_{\mathrm{obs}}<16700$ \AA. This limitation mitigates the effects of the low grism transmission,  including  inaccurate flux-ratio estimates and  possible  false detections.

For the test "\emph{gold}" sample, we  know the identity of the strongest emission line  in each spectrum. In the test run, we treat these sources as if only one line, the strongest, was measured (i.e., the algorithm is blind to any information concerning the other emission lines).

It is important to note that the sample that we use to estimate the accuracy of the algorithm is made by \emph{real} emission lines. This may not be the case in different samples, in particular when only one spectral feature (a possible false detection) is detected. This question is discussed in more detail in Section~\ref{SECT:FLcontam}.

The test "\emph{gold}" sample is divided into subgroups for each of which a different species/transitions is responsible for the strongest emission measured. 
The sources showing emission lines generated by the H$\alpha$ transition represent the most numerous subgroup (N$_{H\alpha}=$1293), followed by the  \oiii\ group (N$_{\mathrm{[OIII]}}$=949), \oii\ (N$_{\mathrm{OII}}$=24) and other species/transitions (N$_{\mathrm{other}}$=16 in total for H$\beta$, H$\gamma$, \sii)\footnote{One spectrum showing a strong \siii-9532 emission is excluded from the test}. 

In our test, the algorithm is run over the complete sample. 
For each spectrum, the species/transitions that the algorithm indicates as responsible for the most prominent line is the one for which the algorithm itself found the highest value of probability.

Table~\ref{tbl:EM_LINE_LIST} summarizes the inputs and the results of our test on real lines. For each of the subgroups, we measured the recovery fraction, or completeness (the fraction of emission lines correctly identified by the algorithm) and the contamination (fraction of wrong identifications in the recovered sample). All the species/transitions summarized in Table~\ref{tbl:EM_LINE_LIST} are considered in our tests. However, the algorithm is not meant to properly work on the few rare objects characterized by peculiar emissions. Since the \sii, H$\beta$ and H$\gamma$ fluxes are almost always weaker than those measured for H$\alpha$, \oiii\ and \oii, only very few sources characterized by a prominent emission from \sii, H$\beta$ and H$\gamma$ can be used in the calibration process (i.e. our algorithm is not expected to work for these species)\footnote{None of the 16 sources characterized by \sii, H$\beta$ or H$\gamma$ stronger than the other transitions in the test "\emph{gold}" sample was correctly identified by the algorithm. Only one source was wrongly classified as \sii.}

For the most numerous samples, H$\alpha$ and \oiii, the completeness is high (88.3\% and 77.0\% respectively), and the contamination is contained (15.2\% and 17.9\% respectively). The \oii\ sample still shows an acceptable completeness (50\%) but a very high contamination (73.3\%). This high level of contamination can be explained by the fact that even a very small fraction of wrong identifications in large samples (such as H$\alpha$ and \oiii) can cause a strong contamination in a numerically smaller sample (e.g. \oii). 
For the other species/transitions considered, we do not expect the algorithm to produce any reliable result, for at least two reason. First, the very limited amount of such emitters in the original WISP calibration "\emph{gold}" sample does not allow us for an opportune calibration of the algorithm itself on these species/transitions. Second, the strong contamination from numerically larger samples is expected to compromise the reliability of the recovered samples (similarly to what happens for the \oii\ transition).

 \begin{deluxetable*}{cclllllll}
 \tabletypesize{\footnotesize}
 \tablecolumns{9}
 \tablewidth{0pc}
 \tablecaption{Accuracy test on the species/transitions considered by the algorithm, ordered by importance ($N_{\mathrm{TGS}}$)\tablenotemark{[a]}.}

\tablehead{ species/ & $\lambda_{\mathrm{obs}}$ & N$_{\mathrm{TGS}}$\tablenotemark{[b]} & N$_{\mathrm{I}}$\tablenotemark{[c]} & N$_{\mathrm{CI}}$\tablenotemark{[d]} & N$_{\mathrm{WI}}$\tablenotemark{[e]} & Completeness  & Contamination & Accuracy (Purity) \\
        transitions  & [\AA] & & & & & (N$_{\mathrm{CI}}$/N$_{\mathrm{TGS}}$) & (N$_{\mathrm{WI}}$/N$_{\mathrm{I}}$) & (N$_{\mathrm{CI}}$/N$_{\mathrm{I}}$) }
 \startdata
 \label{tbl:EM_LINE_LIST}
 \bf{H$\alpha$}  & \bf{6564.5}          & \bf{1293} & \bf{1346} & \bf{1142} & \bf{204} & \bf{88.3\%}  &  \bf{15.2\%} & \bf{84.8\%} \\
 \bf{\oiii} & \bf{4960.3 - 5008.2} & \bf{949}  & \bf{890}  & \bf{731}  & \bf{159} & \bf{77.0\%}  &  \bf{17.9\%} & \bf{82.1\%} \\
 \bf{\oii}\tablenotemark{[f]} & \bf{3727.1 - 3729.9}  & \bf{24}   & \bf{45}   & \bf{12}   & \bf{33 } & \bf{50.0\%}  &  \bf{73.3\%} & \bf{26.7\%} \\
 \sii              & 6718.3 - 6732.7    & 7         & 1         & 0 (0\%)  & 1   (100\%)  & 0\%          &  100\% & 0\% \\
 H$\beta$          & 4861.4             & 5         & 0         & 0 (0\%)  & 0   ( - )    & 0\%          &  - & 0\% \\
 H$\gamma$         & 4340.5             & 4         & 0         & 0 (0\%)  & 0   ( - )    & 0\%          &  - & 0\% \\
 \enddata
\tablenotetext{a}{ All the estimates reported refer to samples made of \emph{real} lines.}
\tablenotetext{b}{ Number of sources, in the test "\emph{gold}" sample (TGS), showing this transition as the most prominent emission line.)}
\tablenotetext{c}{ Number of sources, in the test "\emph{gold}" sample, for which the algorithm associates to the possibility of this transition to be the responsible for the most prominent emission line measured, the highest relative probability (Identified, I).}
\tablenotetext{d}{ Number of sources, in the test "\emph{gold}" sample, for which the strongest emission due to this transition is correctly identified (CI).}
\tablenotetext{e}{ Number of sources, in the test "\emph{gold}" sample, for which the strongest emission is mistaken for this transition (Wrongly Identified, WI).}
\tablenotetext{f}{ Due to the small amount of data included in the calibration and test sample, the accuracy estimated should be considered an upper limit to the actual value.}
 \end{deluxetable*}

 The overall accuracy of the algorithm can be measured as the fraction of spectra for which the species/transition responsible for the strongest emission is correctly predicted by the algorithm. Considering all the sources in our test we found this overall accuracy to be 82.6\%. This value demonstrates how the algorithm can strongly improve the accuracy obtained by considering the SED fitting strategy alone ($\sim$50\%, if we consider only \ha, \oiii, and \oii), and the accuracy that we would obtain by always classifying every strongest line as \ha\ ($\sim$57\%), as in the WISP default classification option. 
We can compute the accuracy also for each single sample (in this case, accuracy=purity=1--contamination). As for the completeness and contamination, we report these estimates in Table~\ref{tbl:EM_LINE_LIST}. 

\begin{figure*}[!ht]
 \centering
 \includegraphics[width=8.5cm]{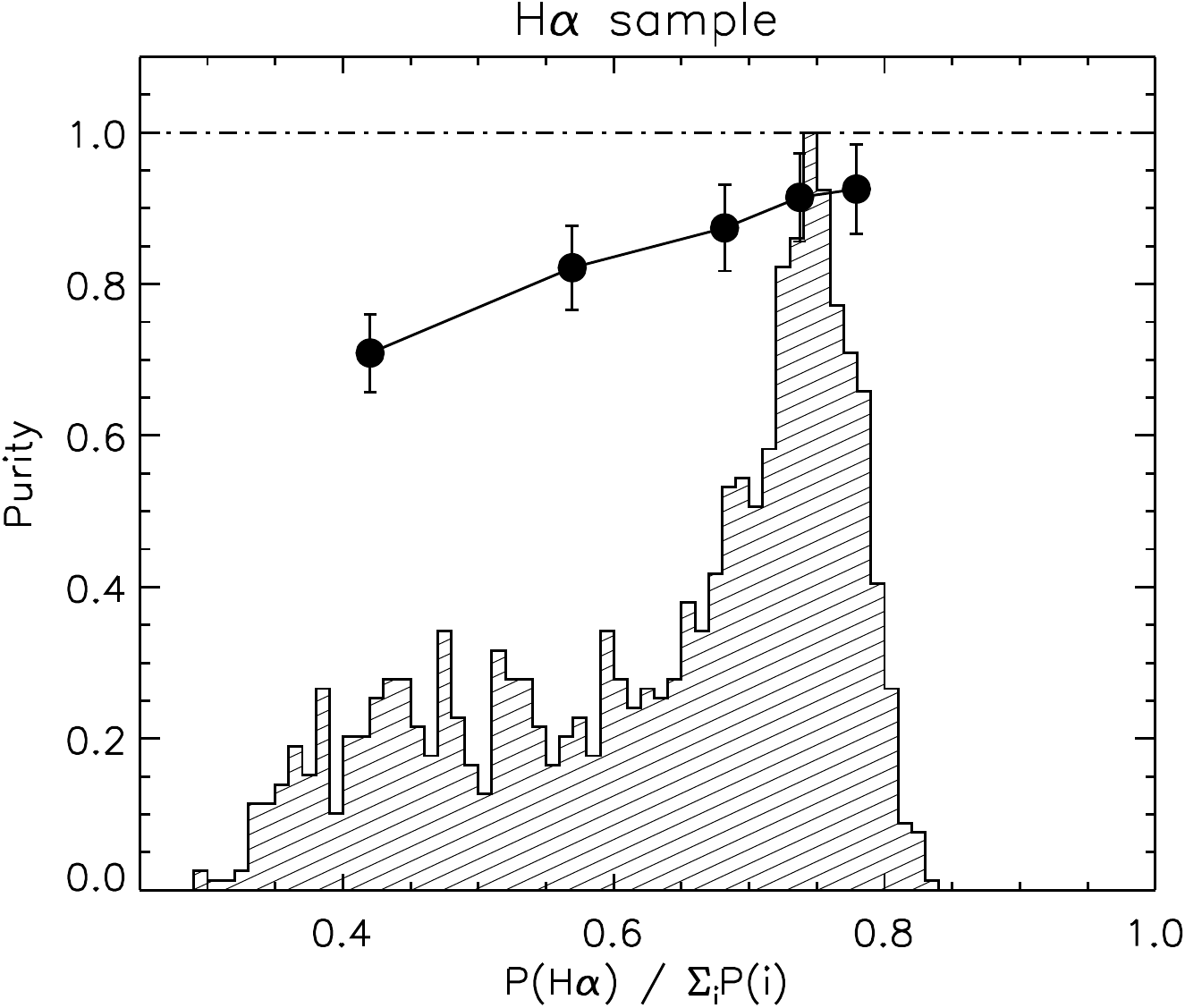}
 \includegraphics[width=8.5cm]{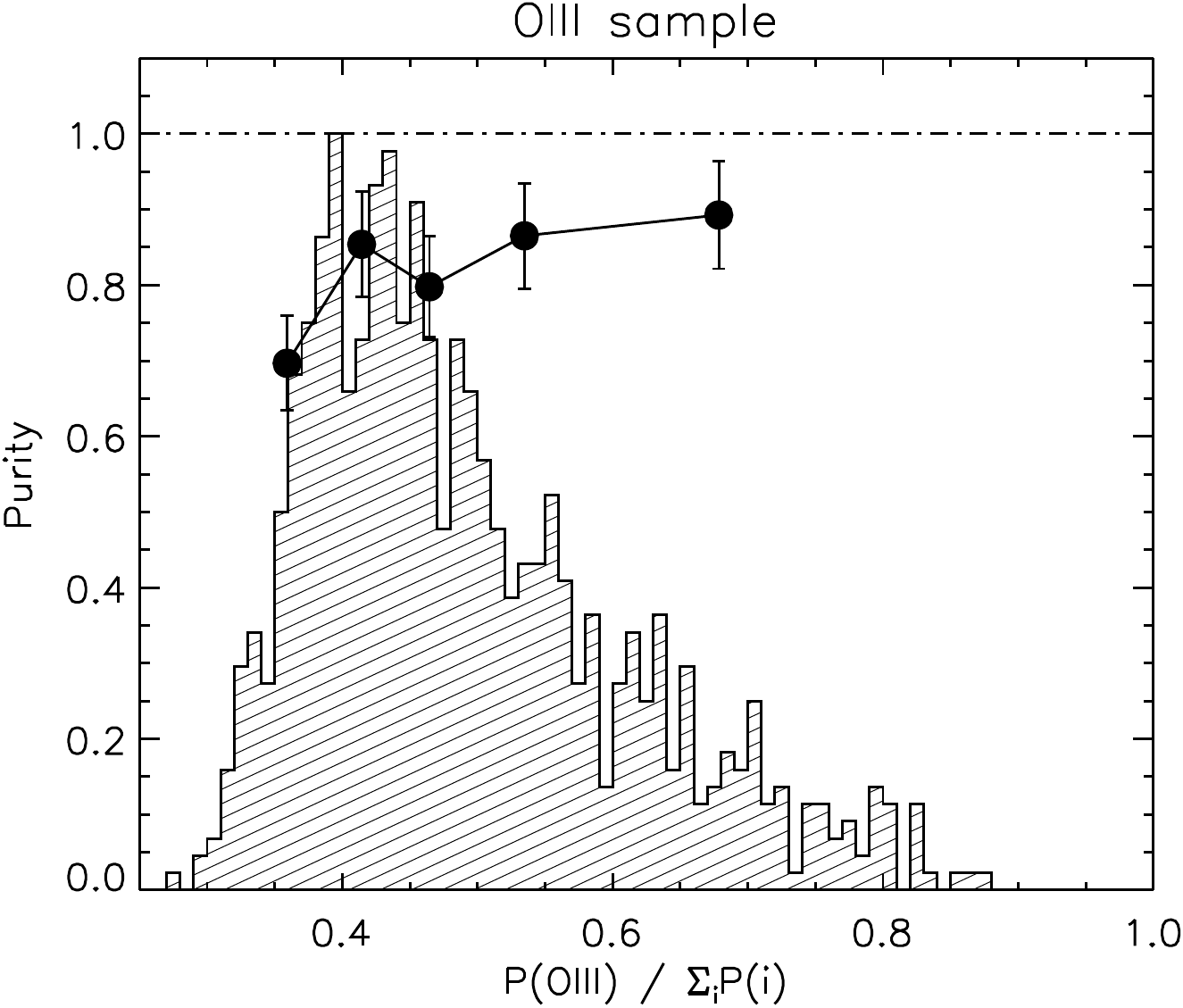}
 \caption{Purity of the recovered H$\alpha$ and \oiii\ samples (i.e. fraction of emission lines correctly identified as due to H$\alpha$ and \oiii\ by the algorithm) as a function of the probability values P(H$\alpha$) ({\bf left panel}) and P(\oiii)({\bf right panel}), normalized to the sum of the probabilities $\sum_{i}$P(\emph{i}). The black filled circles represent the average purity obtained in equally populated bins of probability, each of them containing 20\% of the data. Poissonian uncertainties are shown. The background histograms represent the normalized probability distribution of the overall sample (\ha +\oiii). }
 \label{img:PROB_VS_PRECISION}
 \end{figure*}

The probability indicators can be used to select samples with different levels of purity. This can be achieved by selecting different probability thresholds. Figure~\ref{img:PROB_VS_PRECISION} shows how increasing values of P(H$\alpha$) and P(\oiii), normalized to the sum of the probabilities $\sum_{i}$P(\emph{i}), correspond to samples characterized by increasing levels of purity (accuracy). Given the limited amount of data, we cannot show a similar Figure for the \oii\ sample. For this case, we refer the reader to the average value reported in the last column of Table~\ref{tbl:EM_LINE_LIST}.

\subsection{The not independent calibration and test samples}
\label{SECT:NOT_INDIP_SAMPLES}
\begin{figure}[!ht]
 \centering
 \includegraphics[width=8.7cm]{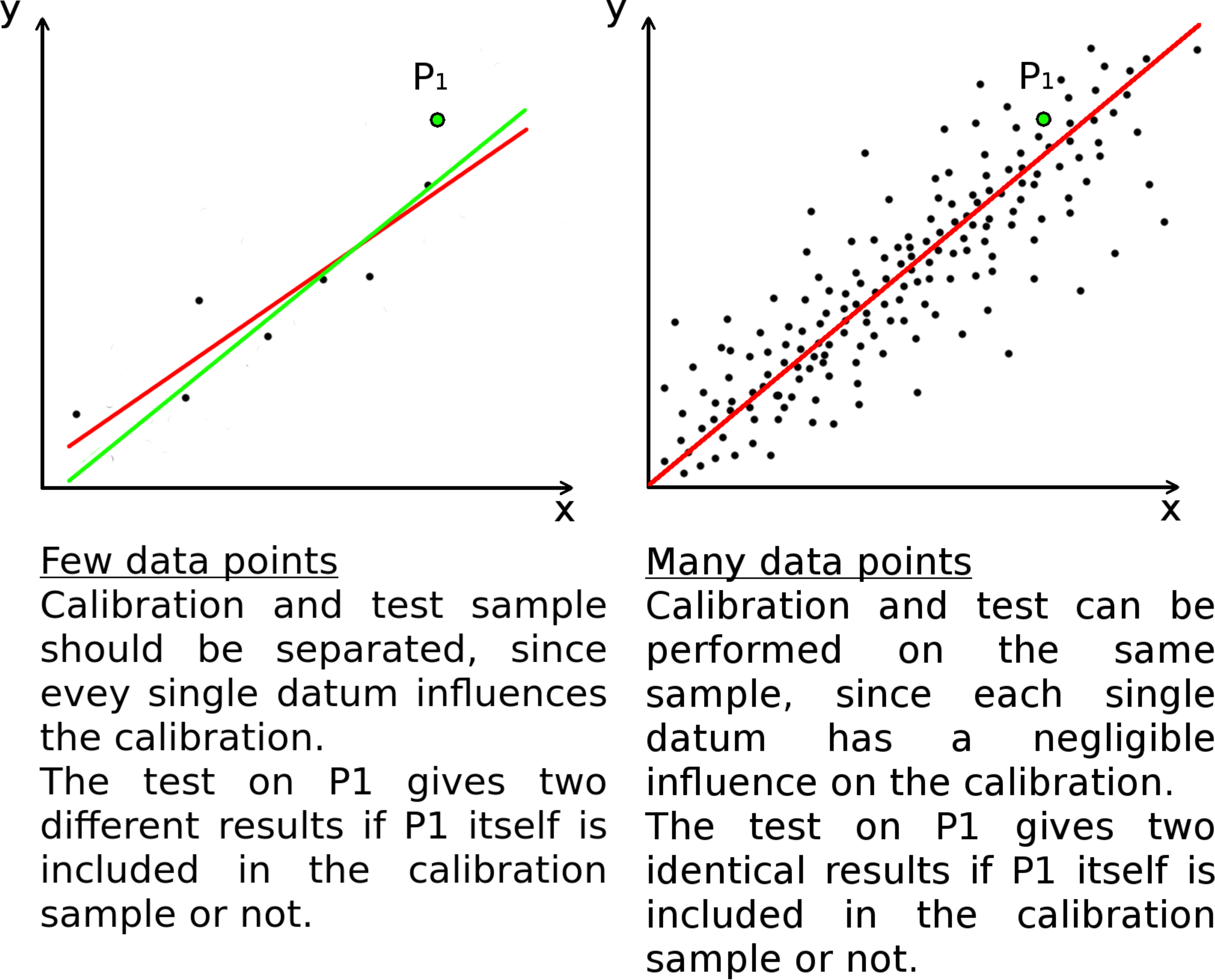}
 \caption{The best approach to validate a relation is to test it on a data sample that is independent from the sample used to compute the relation itself. This is due to the fact that every datum tends to be closer to the relation, if the relation was calibrated using that datum ({\bf left panel}), or if the function considered is particularly complex (overfitting problem). However, when the function is simple (such as the linear relation in these plots) and the amount of data is sufficiently large, every datum considered singularly, has a negligible effect when calibrating the relation ({\bf right panel}). In this context, when testing the relation on a datum, there is no difference if that specific datum was included in the calibration data-set or not.}
 \label{img:test_cal_samples}
 \end{figure}

As specified in Section~\ref{SEC:cal_test_samples}, the \emph{calibration} and the \emph{test} samples are not independent of each other.
In particular, all the data in the calibration sample are also included in the test sample. This decision is justified by the argument that follows. The algorithm is characterized by a modular structure, where each module represents a simple relation (usually linear) between different observational quantities (e.g. apparent size versus redshift, or magnitude versus line flux ratios). These relations are computed using a large amount of data so that each single datum produces a negligible effect on the calibration of a specific module. In other words, removing a datum from the calibration sample would not change the behavior and outputs of the algorithm, even when testing the software on the removed datum. This condition is visually represented in Figure~\ref{img:test_cal_samples}. The test that we perform would be unreliable (representing an upper limit to the actual accuracy) only if few data were available, or if we were considering particularly complex relations prone to overfitting issues.

On the other hand, the accuracy can be assessed by dividing the sample into two independent subsets and calibrating the relations using only one of the two subsamples. This approach guarantees that the accuracy (estimated on the other independent subsample) is not biased by possible overfitting or by the use of too complex functions. However, given the smaller amount of data used for the calibration itself (half of the total), the relations can not be as accurate as they are when considering the complete sample. In other words, the accuracy estimated using the method just described represents an underestimate of the actual accuracy that can be achieved by calibrating the algorithm using the entire data sample. 
For the reasons explained, we consider the accuracy computed 
using the full "\emph{gold}" sample a more reliable estimation of the actual accuracy of our algorithm. 

To convince the skeptical reader, we performed two validation tests by dividing the sample as described above (see Tables~\ref{tbl:EM_LINE_LIST_T1} and \ref{tbl:EM_LINE_LIST_T2} in Appendix~\ref{SECT:AppendixA}). In these tests, we obtained an overall accuracy of  80.6\% and 80.4. These values are lower, but consistent, with the 82.6\% accuracy reached using the same sample for both calibrating and testing the algorithm. The lower accuracy obtained in these two tests does not indicate that when the sample is not divided the algorithm is overfitting (i.e. that we are overestimating the accuracy). Vice versa, the small difference is mostly due to the less precise calibration achievable exploiting only half of the available data.

The exception to the arguments reported above is represented by the sample of strong \oii\ emitters, for which the small amount of data available, especially at $z>1.5$, limits our ability to calibrate the algorithm (in particular, regarding the empirical correction described in Section~\ref{SEC:empirical_corr}) and to measure the associated uncertainty. For this reason, the precision reported for the \oii\ sample in Table~\ref{tbl:EM_LINE_LIST} should be considered an approximate upper limit to the actual value. However, we point out that given the small amount of data, the same uncertainty associated with the precision estimate does affect also any other estimate obtained dividing the complete sample in two independent subsamples (this is confirmed by comparing the precision obtained for the \oii\ sample in Tables~\ref{tbl:EM_LINE_LIST}, \ref{tbl:EM_LINE_LIST_T1} and \ref{tbl:EM_LINE_LIST_T2},  ).

\section{Single lines}
\label{SEC:NEW_CLASSIF}

\subsection{Characterization of the single--line sample}
\label{SECT:SL_Charact}

After calibrating the algorithm using the secure identifications of the "\emph{gold}" sample, 
we run the software on the sample of spectra showing single unidentified emission lines.  To this purpose, similarly to what is required for the ``\emph{gold}'' sample, we consider only 345 \emph{``single line''} sources covered by observations in both the G102 and the G141 grisms at the same time.

It is important to note that for many spectra, after the original identification of the main emission line, the flux of additional lines could be automatically measured, even if a proper detection of these lines (S/N$>2\sigma$) was not available. In many cases, some of these \emph{undetected} but visible (and \emph{measurable}) lines helped the WISP reviewers during the by-eye classification process. However, considering these lines as secure identifications would be misleading (e.g. a bump in the noise can easily be mistaken for \hb\ or \sii). Therefore, these spectra are not included either in the \emph{gold}, or in the \emph{single line} sample.

\begin{figure*}[!ht]
 \centering
 \includegraphics[width=8.5cm]{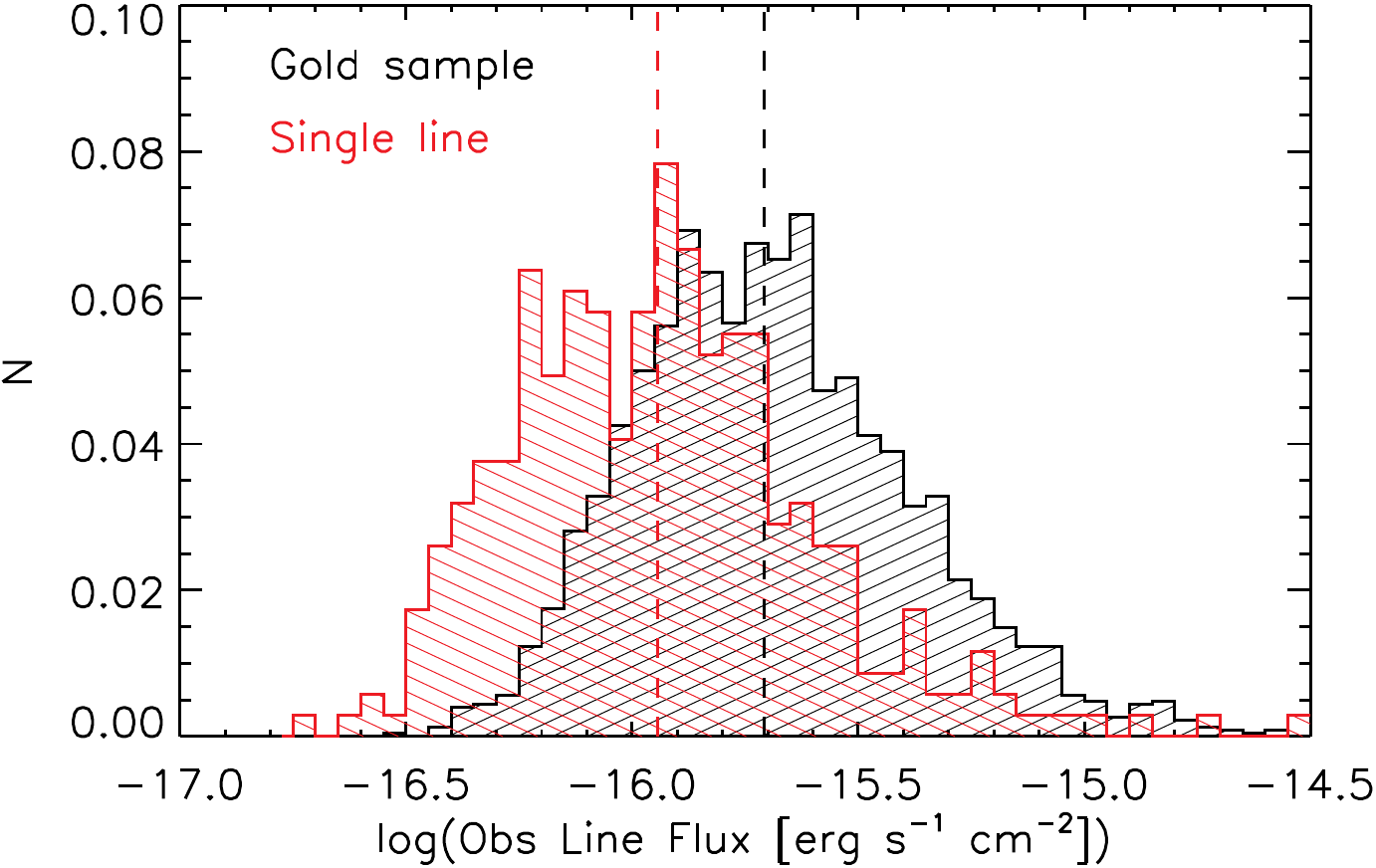}
 \includegraphics[width=8.5cm]{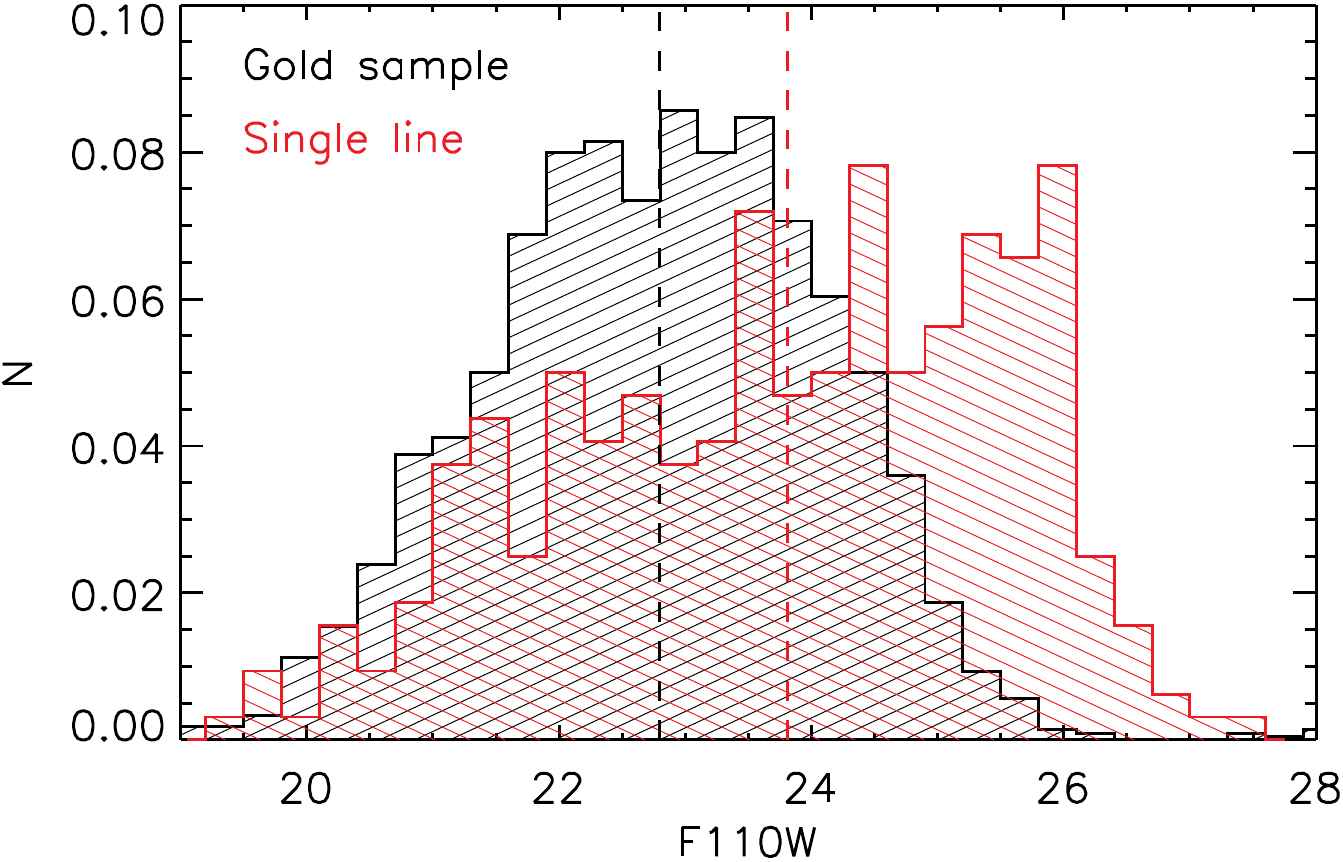}
 \includegraphics[width=8.5cm]{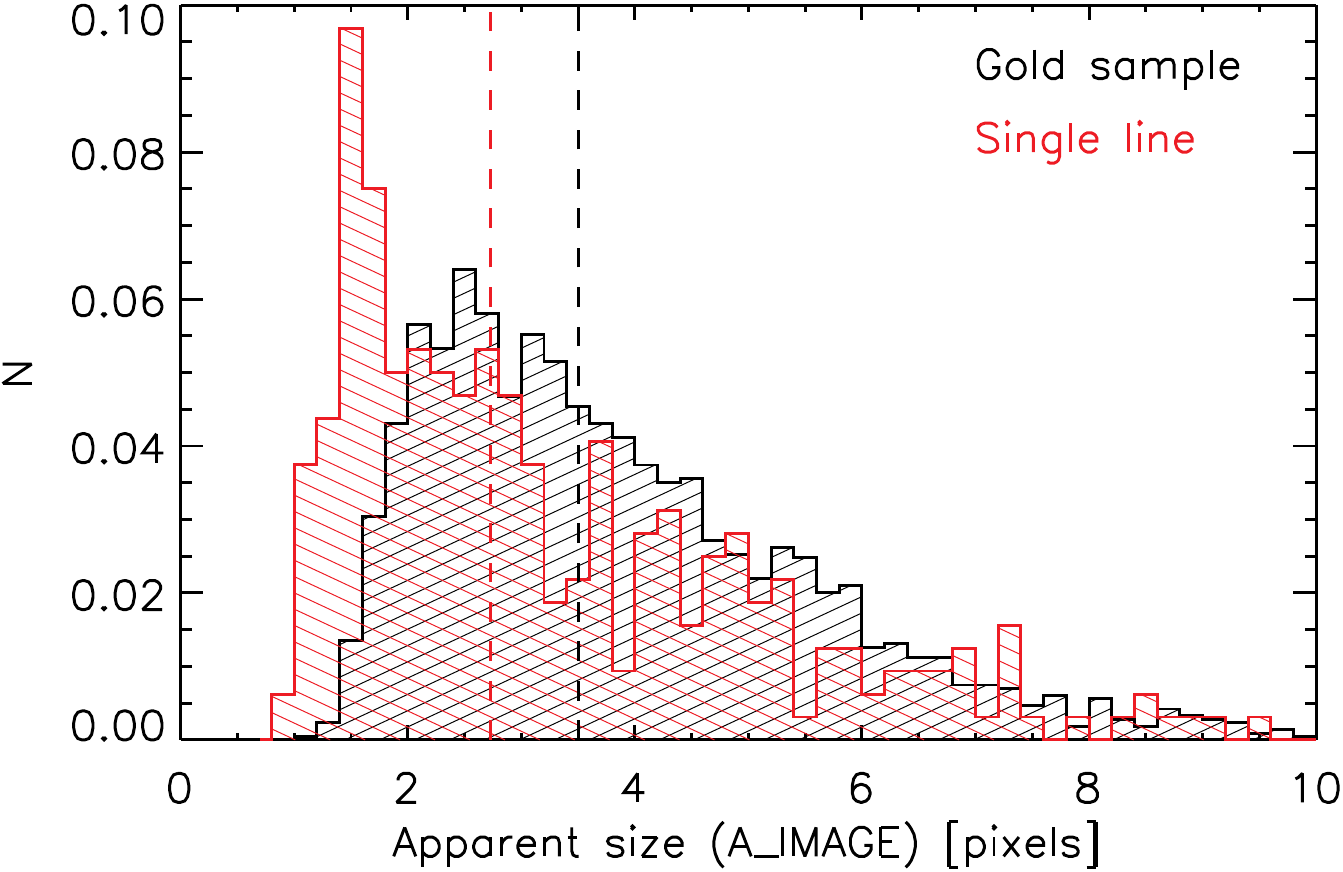}
 \includegraphics[width=8.5cm]{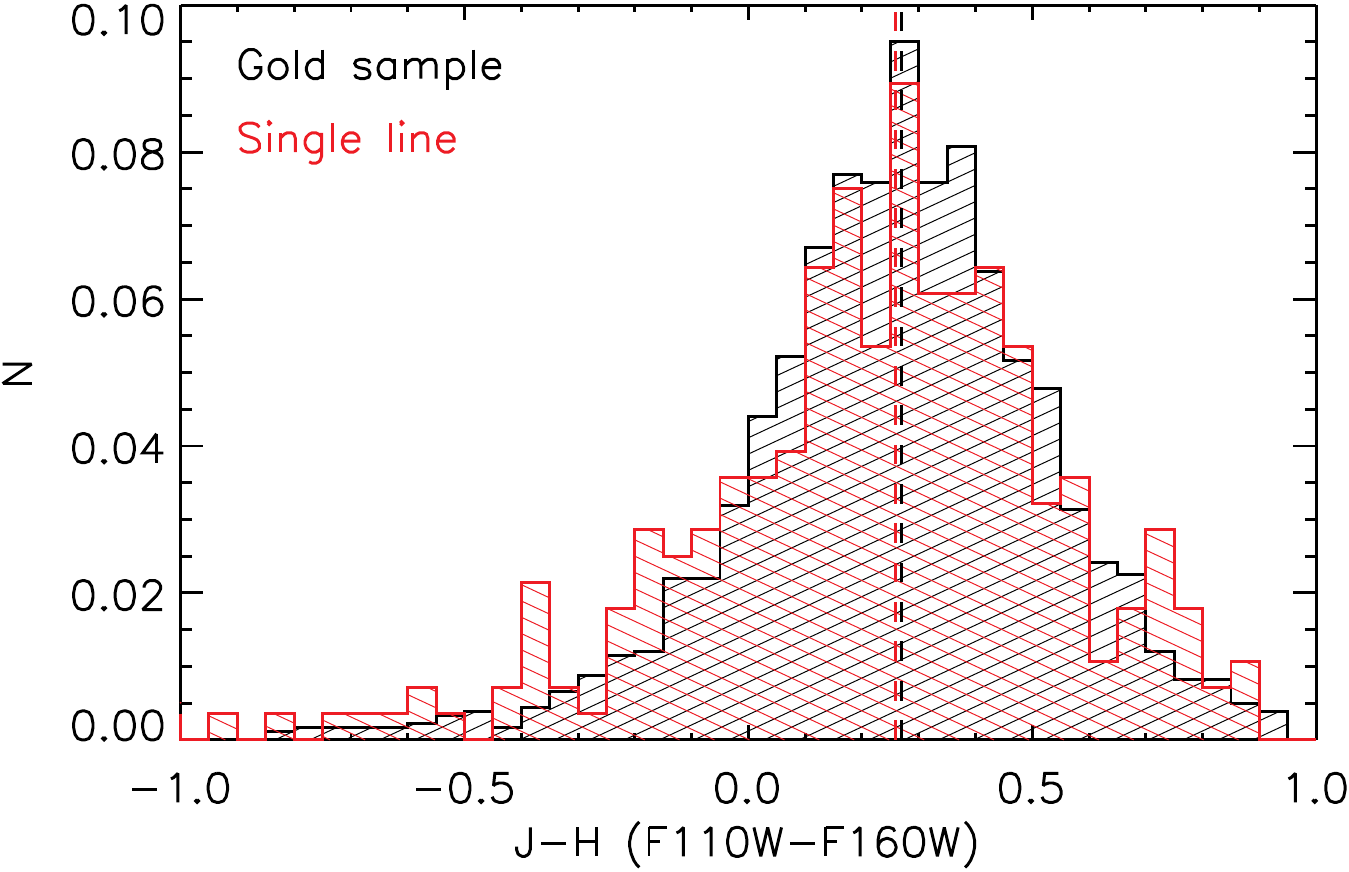}
 \caption{Normalized distribution of the flux of the strongest line (upper left panel), J magnitude (upper right panel), apparent size (bottom left), and J-H color index (bottom right). The distributions are shown for the calibration sample (in black), characterized by two or more lines detected, and for the sample of spectra with only one line detected (in red). The vertical dashed lines represent the median values for the two samples. 
The similar J-H color distribution indicates that there are not strong redshift differences between the two samples. Instead, the differences observed in the distributions of line flux, total magnitude, and apparent size suggest an average difference in intrinsic size, stellar mass, or both.
}
 \label{img:OBS_PROP_COMPAR}
 \end{figure*}

In order to test the homogeneity of the two samples, we compare their main properties. Figure~\ref{img:OBS_PROP_COMPAR} shows the distributions of flux of the strongest line, apparent magnitude, apparent size, and J-H color index. 
While the J-H color distribution is similar for the two samples, the distributions of apparent magnitude, size and line flux are not identical. On average, the strongest line of a spectrum is weaker (in absolute terms) when only one line is detected. 
This behavior is not unexpected, as weaker emission lines will not be detected in a spectrum where the strongest line is already at the level of the S/N detection threshold.
On the other hand, because the ratios between the fluxes due to different transitions are not completely arbitrary (see plots in Figures~\ref{img:FLUX_RATIO_priors_A} and~\ref{img:FLUX_RATIO_priors_B}), when the strongest line is characterized by a particularly high value of S/N, the weaker lines are also more likely to be detected above the S/N threshold.

\begin{figure}[!ht]
 \centering
 \includegraphics[width=8.3cm]{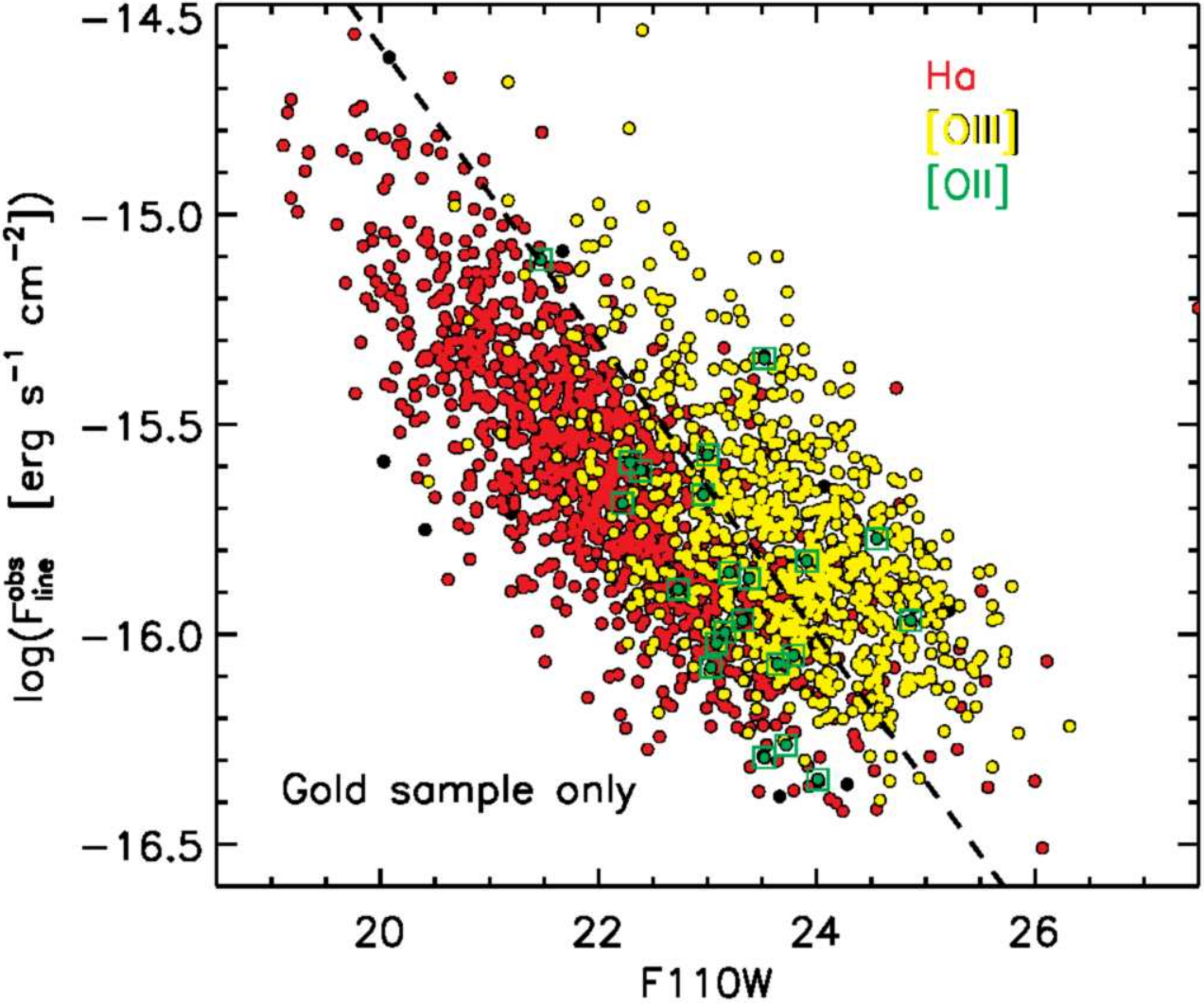}
 \includegraphics[width=8.3cm]{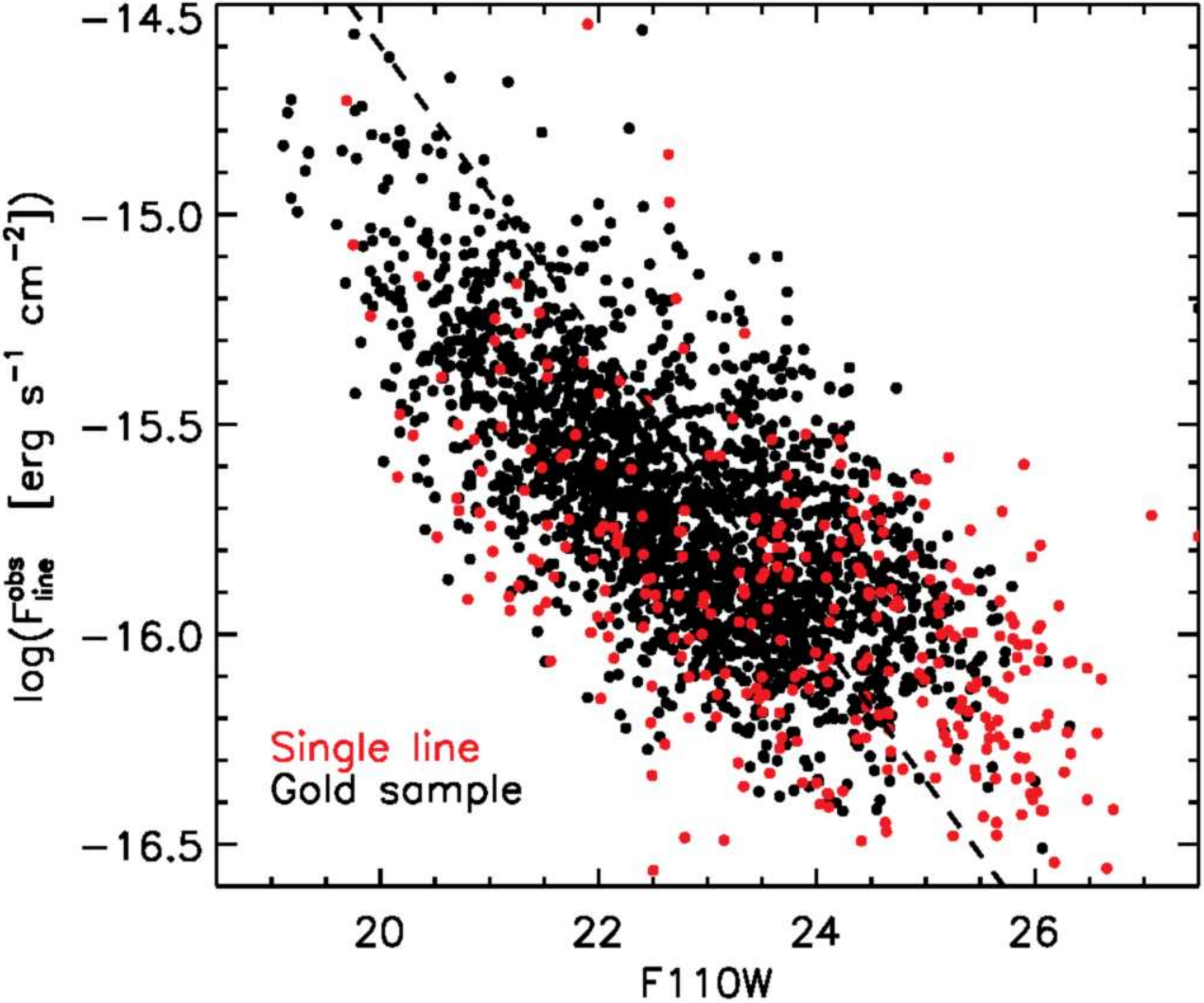}
 \caption{The broad anticorrelation existing between the flux of the strongest spectral line detected and the observed J magnitude.
In the {\bf upper panel}, we consider the ``\emph{gold}'' sample, using red filled circles when the brightest line is \ha, yellow when it is \oiii, and green circles enclosed in squares when it is \oii. Two different normalizations are clearly visible for \ha\ and \oiii. For \oii\ lines, the behavior looks more controversial. In the {\bf bottom panel}, the relation is shown for the ``\emph{gold}'' (black filled circles) and for the "\emph{single--line}" (red filled circles) samples. We can approximately separate \ha\ from \oiii\ dominated sources, by empirically dividing the plots in two regions (black dashed lines). Almost all of the "\emph{single--line}" sources at high magnitudes (J$\sim$24.5 corresponds to the  average magnitude of the single line sample at $0.8<z<1.2$) are located on the \oiii\ side of the plot, along the natural continuation of the relation valid for sources dominated by the \oiii\ emission itself. This evidence independently confirms the prevalence of \oiii\ among the "\emph{single--line}" sources at 0.8$<z<$1.2 (compare with the stacked spectrum in Figure~\ref{img:stack_OIII_spect}).}
 \label{img:LineF_Vs_Jmag}
 \end{figure}

Similarly to the flux of the strongest line, the apparent magnitude is usually higher when only one line is detected. Again, this is due to the fact that, given the broad correlation between observed line flux and continuum (see Figure~\ref{img:LineF_Vs_Jmag}), fainter sources are also characterized by weaker emission lines. 
Hence, in these cases, all emission lines are more likely to be undetected.
The opposite argument applies to the strongest emission lines, which are more commonly associated with brighter J magnitudes.

Because the similar J-H color distribution (bottom right panel of Figure~\ref{img:OBS_PROP_COMPAR}) is consistent with a homogeneous redshift distribution, the different apparent size distribution (bottom left panel of Figure~\ref{img:OBS_PROP_COMPAR}) indicates that these effects are probably due to a different average \emph{intrinsic} size of the two samples.


 \subsection{Classification of the single lines}
\label{SECT:SL_Classif}
\begin{figure*}[!ht]
 \centering
 \includegraphics[width=8.5cm]{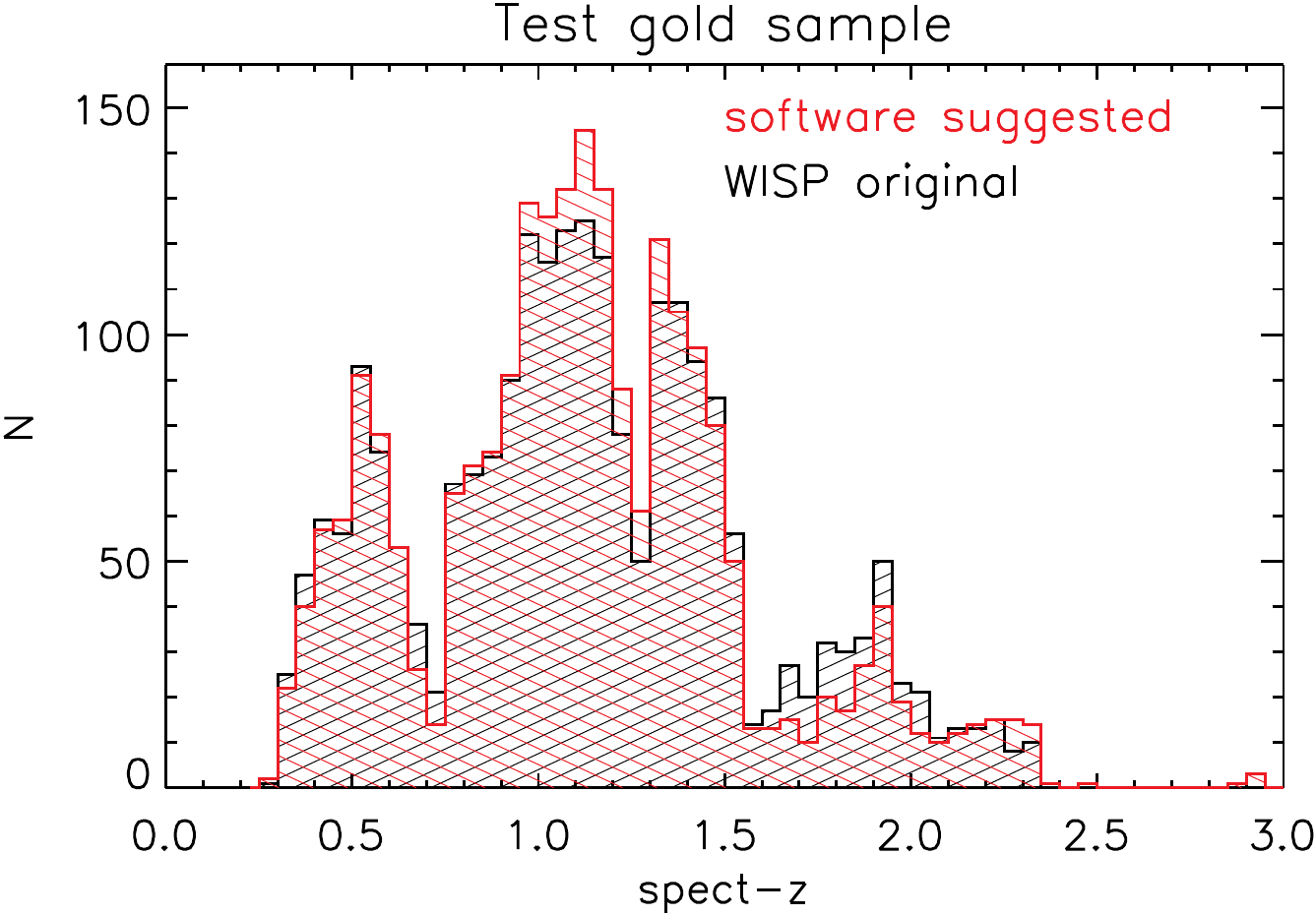}
 \includegraphics[width=8.5cm]{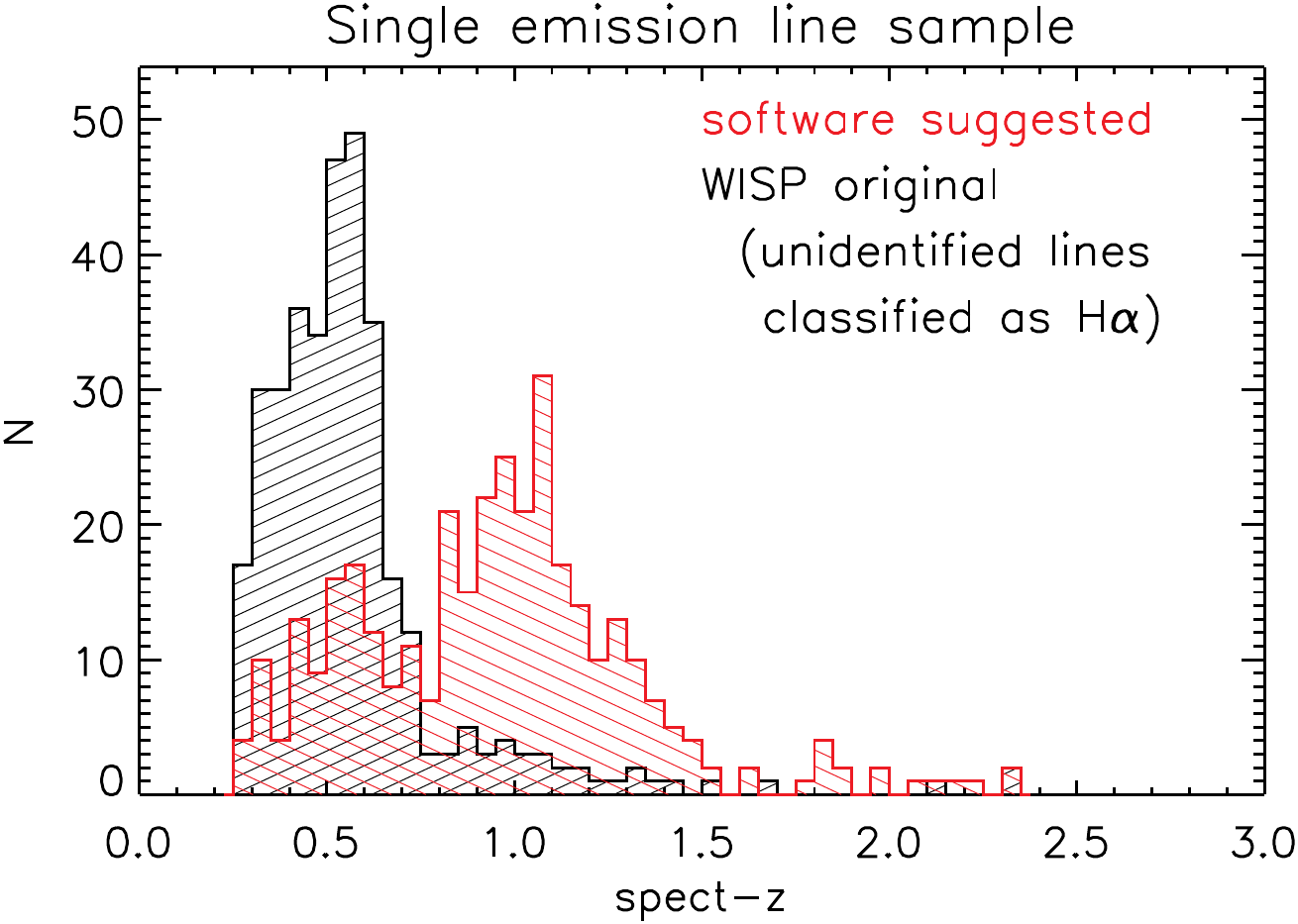}
 \caption{Redshift distribution of the sources resulting from the WISP original (human) and WISP default classification (black shaded histograms in the left and right panels, respectively), compared with the distributions resulting from the automatic classification performed by our algorithm (red shaded histograms). The {\bf left panel} shows the original and the recovered redshift distributions of the sources in the test "\emph{gold}" sample. The two distributions agree with each other, indicating that the algorithm does not introduce evident redshift dependent biases. In the {\bf right panel}, we consider the sample of sources characterized by only one spectral line detected. In this case, a default identification (\ha) is compared with the automatic classification of our algorithm. The strong differences between the two distributions highlight the limits of using a default option when the emission line is not identified. In fact, while the distribution recovered by the algorithm is consistent with that observed for the ``\emph{gold}'' sample, the default option is clearly strongly biased.}
 \label{img:z_Histograms}
 \end{figure*}

As suggested by the similar J-H color distribution (bottom right panel of Figure~\ref{img:OBS_PROP_COMPAR}), the automated classification performed by the algorithm confirms that the ``\emph{single--line}'' and the ``\emph{gold}'' samples are characterized by similar redshift distributions. This can be observed by comparing the red histogram in the right panel of Figure~\ref{img:z_Histograms} with the black histogram in the left panel of the same Figure.


When multiple lines are detected, we can safely consider the human classifications as a reference. Then, the similar distributions in the left panel of Figure~\ref{img:z_Histograms} represent an additional reliability test of the automatic classification, as it indicates that no significant biases are introduced by the algorithm.

On the other hand, when only one single line is detected, the software used for the by-eye classification performed by the WISP reviewers is set to initially indicate \ha\ as a default choice. The human reviewer can modify (or confirm) this default option by considering the presence of additional spectral lines (not detected above 2$\sigma$) or features (e.g. the asymmetric shape of the unresolved \oiii\ doublet). This default option is justified by the fact that indeed \ha\ is the actual strongest emission for more than half of the WISP "\emph{gold}" sample ($\sim$56.7\%). However, if we were considering the default \ha\ option as always valid when additional spectral features were not present (black histogram in the right panel), the redshift distribution of the entire ``\emph{single--line}'' sample would present strong differences with the distribution of the sources in the ``\emph{gold}'' sample (black shaded histograms in the left panel). In particular, the bulk of the sources located between $z\sim$0.7 and $z\sim$1.5 in the test "\emph{gold}" sample would almost completely disappear in the ``\emph{single--line}'' sample. Because we can safely accept the original classification of the "\emph{gold}" sample, it follows that assuming the unidentified single lines is always generated by \ha\ emission, results in a relevant bias.

An independent confirmation of this bias can be found by observing the plots of Figure~\ref{img:LineF_Vs_Jmag} (flux of the brightest detected line versus apparent magnitude). In these plots, \ha\ and \oiii\ lines are located, with some overlap, in two different regions of the bidimensional space. From the bottom panel of the same figure we can observe that, below J$\sim$24.5, the ``\emph{gold}'' and the ``\emph{single--line}'' samples are similarly distributed, indicating that the fraction of \ha\ and \oiii\ lines must be comparable for these two samples. The smaller fraction of ``\emph{single--line}'' sources above J$\sim$24.5 is located in a region of the plot where \oiii\ lines dominate. These observations confirm that the ``\emph{single--lines}'' can not be mostly \ha, as assumed in the original WISP default classification. Instead, they are more probably evenly distributed between \ha\ and \oiii, with an excess of \oiii\ above J$\sim$24.5.

Summarizing, the distributions of the J-H color index, apparent size, magnitude, and line flux (Figures~\ref{img:OBS_PROP_COMPAR} and~\ref{img:LineF_Vs_Jmag}) further supported by the similar redshift distributions of ``\emph{gold}'' and ``\emph{single--line}'' samples (as recovered by the algorithm) indicate that the two samples are characterized by galaxies with intrinsic differences in size (and stellar mass), but with similar redshift distributions. Additionally, we point out that the human classification of single lines, driven by the use of a ``default option'' (such as \ha), biases the sample.

 \subsection{Contamination from false detections}
 \label{SECT:FLcontam}

 Our comparison between the \emph{gold} and the \emph{single line} samples (Section~\ref{SECT:SL_Charact}), and the classification performed by the algorithm (Section~\ref{SECT:SL_Classif}) do not take into account the contamination due to false detections. The identification of multiple lines guarantees that the \emph{gold} sample is not affected (or marginally affected) by this problem. On the contrary, given the detection of only one line, for the \emph{single line} sample there is no such a guarantee.

   Most of the contaminants were eliminated from the \emph{single line} sample during the original by-eye classification. False detections can be easily identified when the line detected in one of the two grisms is too bright, if compared with the overall emission of the source in the corresponding photometric band (J or H). In other cases, they can be identified because the shape (or size) of the detected spectral feature is completely at odds with the shape (size) of the emitting source. Most of the times, contaminants are also decentered with respect to the midline of the bidimensional spectrum.

   However, in some cases, false detections can not be easily identified as described above. This happens when both the spectrum and the contaminant are faint and point like. In these cases, it is impossible to compare the apparent shapes of the spectral line with the shape of the source in the image. Additionally, determining the precise position of the source and corresponding midline of the bidimensional spectrum is not as easy as with brighter sources.

   Sometimes, false detections can be due to the stochastic nature of noise, but most of the time the contamination is due to the overimposition of a zero-order spectrum of a different source located just outside the field. Because most of the sources in any (randomly located) field are faint and point like, then observing faint, rather than bright, contaminated spectra is much more likely. Consequently, contaminants are usually characterized by high magnitudes, faint line fluxes, and small apparent sizes. Given the depth of the WISP survey, these characteristics are typical of spectra dominated by the \oiii\ emission (see Figure~\ref{img:FLUX_RATIO_priors_A}). Therefore, we expect the algorithm to identify almost all of the false detections as \oiii, whereas they are classified as \ha\ in the WISP default classification.

   The current version of the algorithm is not designed to identify false detections. However, in Section~\ref{SEC:single_lines_08z12}, we estimate an upper limit to the fraction of contaminants expected for the \emph{single line} sample in the redshift range $0.8<z<1.2$ ($\sim$30-35\%).

   Finally, we remind readers that the future Euclid and WFIRST surveys are going to be significantly less contaminated by false detections. This will be achieved by dispersing multiple spectra of the same fields along different directions.
 
\subsection{Single lines at $0.8<z<1.2$}
\label{SEC:single_lines_08z12}

We focus our analysis on a subsample of "\emph{single--line}" galaxies that our algorithm locates in the redshift range $0.8<z<1.2$. Also in this case, we consider only sources covered by observations in both the G102 and the G141 grisms.

As illustrated in Figure~\ref{img:z_Histograms}, the redshift bin considered would be almost empty, if all these sources were classified as \ha\ when no distinguishable spectral features were visible (default option).
On the contrary, the same bin includes the peak of the redshift distribution of the ``\emph{single--line}'' sample when we consider the automatic line identification performed by the algorithm. In other words, most of the lines in this range are classified as \oiii\ by our software, whereas the WISP default classification indicates them as \ha.

This redshift range is particularly relevant also because, given the wavelength coverage of grisms G102 and G141, \hb, \oiii, and \ha\  are contemporarily located in the visibility range of the WISP survey. We exclude sources located above $z=1.2$ because, at these higher redshifts (and up to $z\sim 1.5$), the \oiii\ and \hb\ lines fall into the noisier spectral region located between the passband of the two grisms. Additionally, limiting the redshift bin to $z<1.2$, we can obtain a more homogeneous sample, in terms of redshift distribution. 

\subsubsection{Stacking}
\label{SECT:Stack}
\begin{figure*}[!ht]
 \centering
 \includegraphics[width=17.0cm]{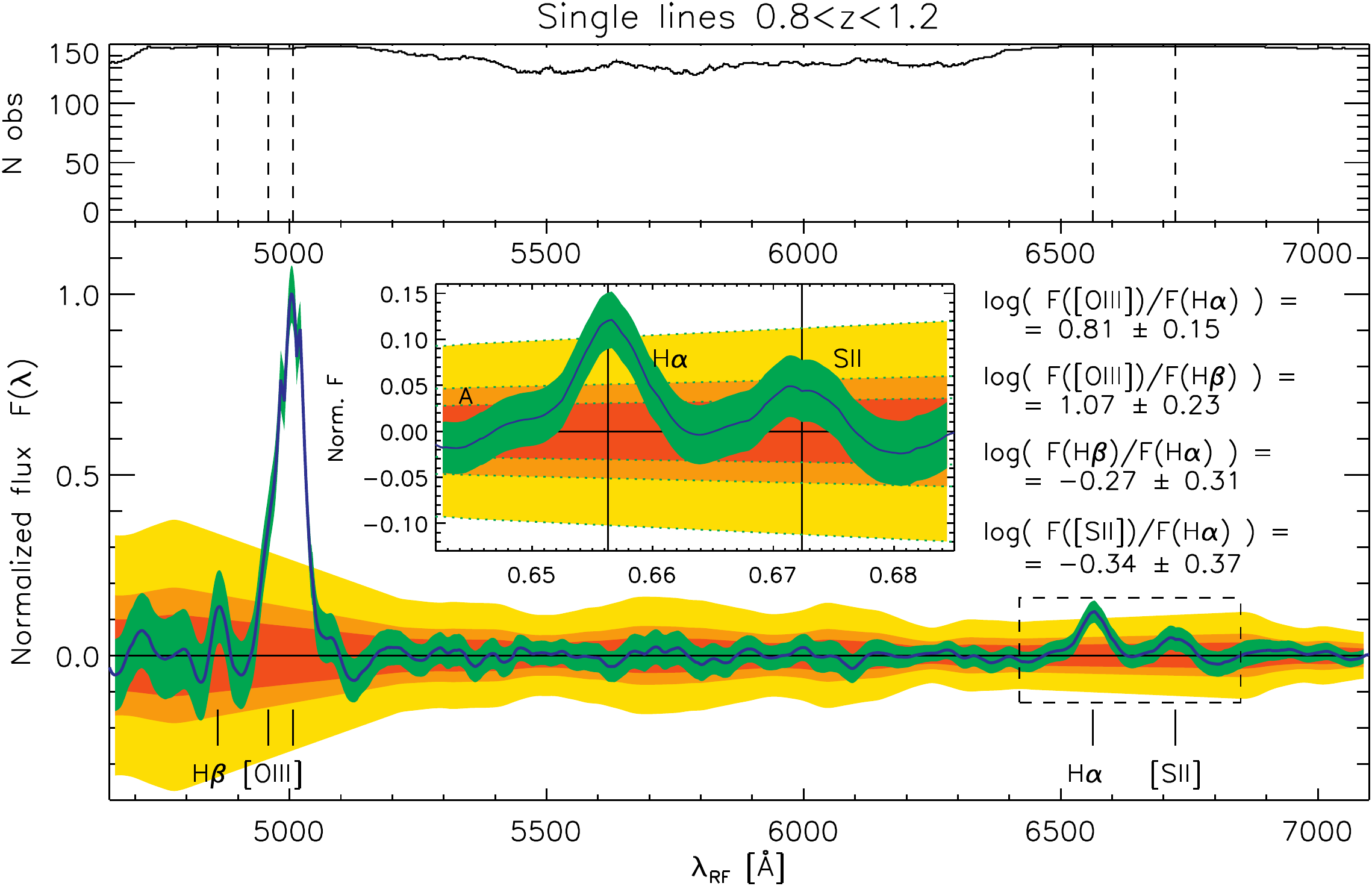}
 \caption{Stacked spectra of ``\emph{single--line}'' sources located in the redshift range $0.8<z<1.2$. The spectra are stacked consistently with the solution suggested by the algorithm. 
The position of the main spectral emissions is indicated by black vertical lines (in order of increasing wavelength: H$\beta$, \oiii, H$\alpha$, \sii). The {\bf upper panel} shows the number of spectra considered for the stacking, at each wavelength. In the {\bf lower panel}, we show the stacked spectrum (blue line), resulting from the weighted average of the normalized single spectra, after rejecting the values above and below $1\sigma$ from the median value (i.e. above the 84th percentile and below the 16th percentile). Single spectra are normalized by the flux of the detected emission line. In order to obtain the average spectrum F$(\lambda)$, at each wavelength every spectrum is weighted using the square inverse of the local noise (1/$\sigma(\lambda)^{2}$). 
In the resulting spectrum, the peak of the strongest emission line (\oiii) is normalized to 1.0. The red, orange, and yellow shaded areas represent the 3$\sigma$, 5$\sigma$, and 10$\sigma$ thresholds, respectively, where $\sigma(\lambda)$ is locally computed from the final stacked spectrum (the positions of the main lines are masked in this process). The green shaded area represents the local value of $\pm 3\sigma(\lambda)$ added to F$(\lambda)$. In the inset, we magnify the spectral region surrounding the \ha\ emission. Some raw line ratios  (i.e. not corrected for the biases explained in Section~\ref{SECT:Biases}), together with their associated $3\sigma$ uncertainties,  are indicated in the plot. These ratios are computed considering the Gaussian fits to the detected lines. }
 \label{img:stack_OIII_spect}
 \end{figure*}

We stacked the ``\emph{single--line}'' spectra detected in this redshift range using the method described in Appendix~\ref{SECT:AppendixB}. Before the stacking, we removed from the sample about 14\% of spectra characterized by heavy contamination or other evident problems. The redshift selection and the exclusion of bad spectra reduces the original ``\emph{single--line}'' sample to 148 valid spectra that we stacked according to the line identification performed by the algorithm. The result of this process is shown in Figure~\ref{img:stack_OIII_spect}.

 While none of the single spectra is characterized by \hb\ or \sii\ as the strongest (unique) emission line, both these two lines are detected well above the local $3\sigma$ threshold, in the stacked spectrum.  The detection of these lines would not be surprising even if the algorithm's classification was failing in the majority of the cases (provided that it was correct in a significant fraction of the cases). However, the detection of these lines at such high intensities (compared with \oiii) indicates that the classification performed by the algorithm is mostly correct.

 For example, from the \emph{gold} sample we expect log(\oiii/\hb)$\sim$0.8 at J$\sim$24.5 (corresponding to the typical magnitude of the stacked sample). From the stacking, we measure log(\oiii/\hb)$\sim 1.07$. However, \hb\ can contribute to the value measured in the stacked spectrum only when the brightest emission lines in the single spectra are correctly classified. Therefore, we must conclude that the algorithm is correctly classifying lines in \emph{at least} $\sim$54\% of the cases. In reality, due to the presence of some biases affecting this ratio (see Section~\ref{SECT:Biases}), the actual accuracy is higher.

\begin{figure*}[!ht]
 \centering
 \includegraphics[width=17.0cm]{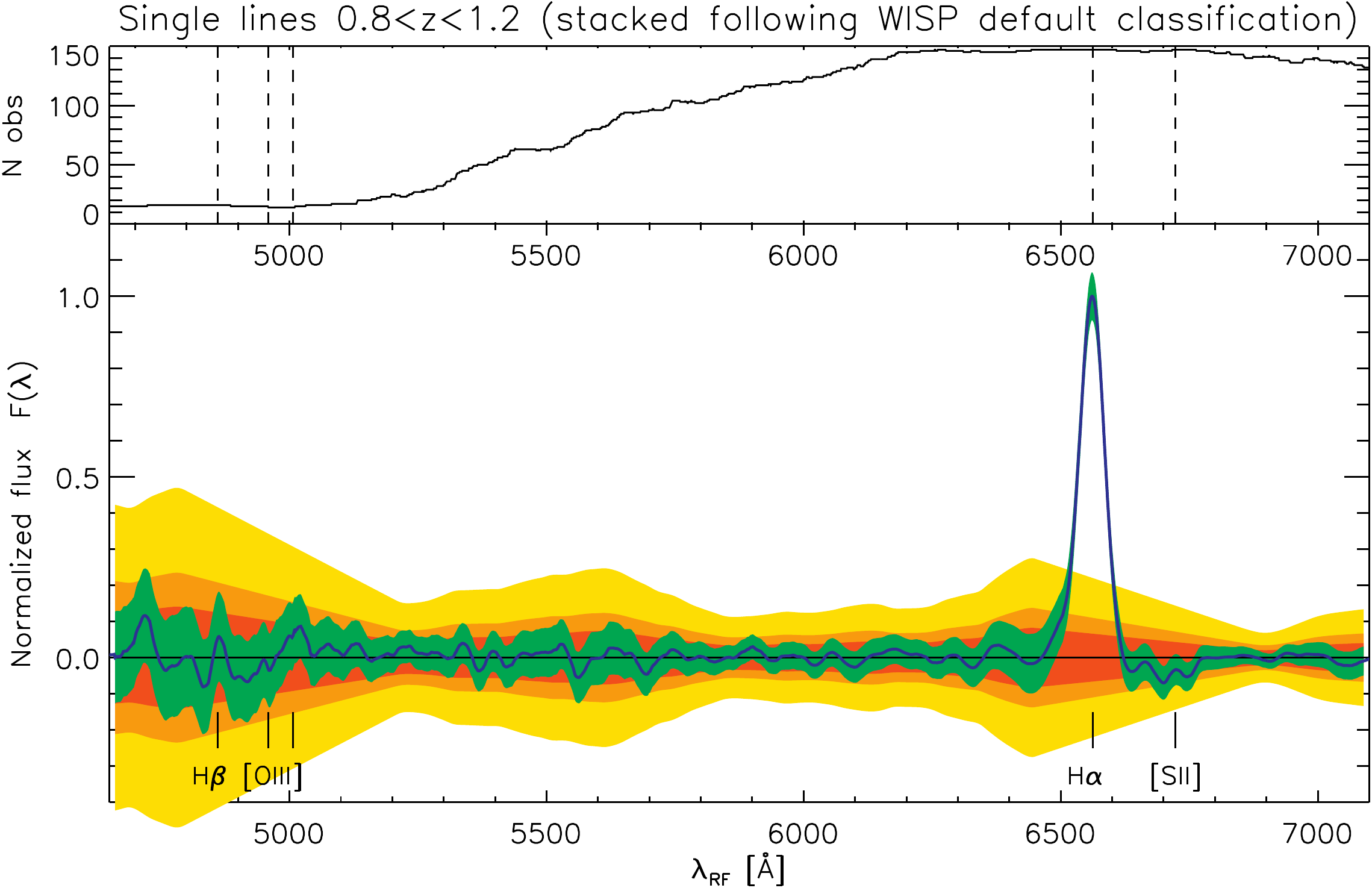}
 \caption{Spectrum obtained by stacking the same ``\emph{single--line}'' sources used for Figure~\ref{img:stack_OIII_spect}. In this case, the WISP default classification is assumed. The same fine recentering strategy is used (see Appendix~\ref{SECT:AppendixB}). In this case, \sii\ is not even visible, whereas the same line is detected above 3$\sigma$ if we follow the classification performed by our algorithm. Instead, following the WISP default classification, \hb\ and \oiii\ would fall outside the wavelength range covered by the grism most of the times, so that a comparison between the two classifications is not possible. This comparison indicates that our algorithm performs better than the WISP default classification. For a complete description of the features visible in these plots, the reader can refer to Figure~\ref{img:stack_OIII_spect}. }
 \label{img:alter_stack_spect}
 \end{figure*}

 In order to test the alternative possibility (i.e., most of the lines due to \ha\ instead of \oiii), we stacked the same spectra as if the original WISP classification were correct. In this case, \oiii\ and \hb\ would be located at $\lambda<$8000\AA\ most of the times, and for this reason they can not be reliably considered in this test. On the contrary, the \sii\ line would always be located inside the wavelength range covered by the two grisms. The stacked spectrum shown in Figure~\ref{img:alter_stack_spect}, shows that \sii\ is not even \emph{visible} (much less \emph{detected}). We conclude that the WISP default classification must be wrong in much more than about half of the cases.

 Some relevant flux ratios, which we measure from the stacked spectrum, are listed in Table~\ref{tbl:line_ratios} (last column). We can compare these values with the flux ratios expected at similar magnitudes for the \emph{gold} sample (second column). It is immediately evident that some flux ratios (especially \oiii/\ha\ and \oiii/\hb) are particularly high if compared with the expected values. The reasons for such discrepancies are explained in Section~\ref{SECT:Biases}.

\begin{deluxetable*}{l|ccccc|c}
\tabletypesize{\footnotesize}
\tablecolumns{2}
\tablewidth{0pc}
\tablecaption{Some flux ratios for the main emission lines ($0.8<z<1.2$)}
\tablehead{\multicolumn{1}{c|}{}  & \multicolumn{5}{c|}{Expected\tablenotemark{[a]} at J=24.5}  & \colhead{Measured\tablenotemark{[b]}} \\
\multicolumn{1}{c|}{Line Ratio}  & \colhead{Unbiased (Gold Sample)\tablenotemark{[c]}} & \colhead{With bias 1\tablenotemark{[d]}} & \colhead{With biases 1 and 2\tablenotemark{[e]}} & \colhead{+30\% f.d.\tablenotemark{[f]}} & \multicolumn{1}{c|}{+35\% f.d.\tablenotemark{[f]}} & \colhead{(stacking)} }
\startdata
\label{tbl:line_ratios}
$\log($\oiii$/$\ha$)$ & 0.30  & 0.56$\pm$0.02  & 0.59$\pm$0.02  & 0.79$\pm$0.03 & 0.85$\pm$0.03 &  0.81$\pm$0.15  \\ 
$\log($\oiii$/$\hb$)$ & 0.79  & 0.77$\pm$0.04  & 0.84$\pm$0.04  & 0.98$\pm$0.06 & 1.06$\pm$0.18 &  1.07$\pm$0.23  \\ 
$\log($\hb$/$\ha$)$   & -0.49 & -0.21$\pm$0.03 & -0.25$\pm$0.04 & -0.22$\pm$0.07 & -0.23$\pm$0.09 &  -0.27$\pm$0.31 \\ 
$\log($\sii$/$\ha$)$  & -0.79 & -0.49$\pm$0.06 & -0.52$\pm$0.08 & -0.57$\pm$0.15 & -0.75$\pm$0.23 &  -0.34$\pm$0.37 \\
\enddata
\tablenotetext{a}{The associated errors correspond to the dispersion of the values obtained in the different simulations.}
\tablenotetext{b}{The associated errors correspond to the actual uncertainties measured from the stacking.}
\tablenotetext{c}{The unbiased flux ratios expected for the stacked \emph{single line} sample are measured from the \emph{gold} sample, at apparent J$\sim$24.5. These values represent the input of our simulation.}
\tablenotetext{d}{The spectra in the \emph{single line} sample are specifically selected for not showing any additional line besides the brightest one. This particular selection biases the \emph{single line} sample (bias 1). At J$\sim$24.5, in the redshift range $0.8<z<1.2$, most of the \emph{single lines} are identified as \oiii. Therefore, only spectra with a particularly damped \ha\ can be included in this sample. }
\tablenotetext{e}{The accuracy of the algorithm (Table~\ref{tbl:EM_LINE_LIST}) affects the recovered line ratios (bias 2). When a single line is mistaken for another, the other lines in the same spectrum can not contribute to the stacked spectrum.}
\tablenotetext{f}{False detections (f.d.) modify the line ratios in the stacked spectrum (bias 3). When a detected line is not due to a real emission, there are no other lines in the same spectrum that can contribute to the stacked spectrum. Almost all the f.d. are classified as \oiii\ by the algorithm. In this table we report the results of the simulations for two different levels of contamination from f.d. (30\% and 35\%). Biases 1 and 2 are taken into account.}
\end{deluxetable*}

\subsubsection{Biases}
\label{SECT:Biases}

 The line ratios that we measure from the stacked spectrum are affected by three major biases that we estimated by running a series of simulations. In Table~\ref{tbl:line_ratios} we compare the expected flux ratios, which we input into the simulations, with the outputs and with the ratios measured from the stacked spectrum.

 The first bias affecting the line ratios (bias 1) is due to the particularly restrictive sample selection. In fact, the \emph{single line} sample is made by spectra characterized by only one line detected above 2$\sigma$, and with no other measurable lines visible (not even below the same 2$\sigma$ threshold). In fact, when additional lines are not detected but visible (and measurable), the spectra are classified as \emph{multiple lines} (although they are not included in the \emph{gold} sample).

 Due to the typically high magnitude of the stacked \emph{single line} sources (J$\sim$24.5), most of these spectra are characterized by \oiii\ stronger than \ha, as shown in Figure~\ref{img:FLUX_RATIO_priors_A}.
Hence, the sample used for the stacking is mostly made by sources characterized by a particularly suppressed \ha\ emission. In fact, if \ha\ were not suppressed, then it would simply be measurable, causing the source to be classified as ``multiple--line'', and the spectrum to be excluded from the stacking.

 In order to measure this selection bias, for each of the 148 stacked \emph{single line} spectra, we simulated 21 spectra characterized by a similar noise distribution along $\lambda$. Then, we added simulated emission lines to these pure-noise spectra. We place all the simulated sources at $z=1.0$ (i.e. the average redshift of the stacked \emph{single line} sample). For each spectrum, we set the flux of the strongest emission line (\oiii) so that S/N$\sim$7.67, corresponding to the typical S/N characterizing the \emph{single line} spectra. The width $\Delta\lambda(\lambda)$ of the simulated lines is also set in accordance with the median size of the \emph{single lines}. The flux of all the other lines is set in accordance with the expected line ratios (second column in Table~\ref{tbl:line_ratios}), that we measure from the \emph{gold} sample, at similar magnitudes. Then, from the simulated sample (3108 spectra in total), we selected all of the spectra characterized by S/N(\ha)$<2$, corresponding to the WISP detection threshold\footnote{This criterion should provide a conservative estimate of the actual bias, as some of these lines are visible and measurable (although not detected). Similar spectra would not be included in the real \emph{single line} sample.}.

 We repeated the simulation described 100 times. On average, in each of the 100 simulations, 196$\pm$6 spectra (6.3\%) were selected using the criterion described. For an approximate comparison, the 323 real \emph{single lines}, located in the combined wavelength range 8500\AA$<\lambda<$11100\AA\ and 11400\AA$<\lambda<$16700\AA, represent 6.1\% of the overall sample of spectra with strongest line detected in the same range (5327 sources). Finally, we stacked the selected simulated spectra by replicating the procedure used for the real ones (we did not apply the recentering procedure because all simulated spectra were already placed at $z$=1.0). In Table~\ref{tbl:line_ratios} (third column), we report the average flux ratios measured from the 100 simulated stacked spectra.

 The flux ratios are affected by an additional bias (bias 2), due to the intrinsic accuracy of the algorithm (see Table~\ref{tbl:EM_LINE_LIST}). Every time the algorithm mistakes \ha\ for \oiii\ (or vice versa), this modifies the flux ratios in the stacked spectrum. In fact, while the strongest emission is incorrectly attributed to \oiii\ (or \ha), there is no contribution to the stacked spectrum from all the other lines.

 In order to estimate this bias, we run 100 simulations similar to those described above (for bias1). In this case, we simulated the incorrect identifications by removing all the emission lines from some of the selected spectra. Then, we added only one simulated line (\ha\ or \oiii), which we set to S/N=7.67. The amount of false identifications is set considering the expected purity of the \ha\ and \oiii\ samples (last column in Table~\ref{tbl:EM_LINE_LIST}), and in accordance with the fraction of actual \oiii\ and \ha\ identifications in the (real) stacked \emph{single line} sample. The combined effects of selection bias and accuracy bias are reported in the fourth column of Table~\ref{tbl:line_ratios}.

 The third bias is due to the presence of false identifications (see Section~\ref{SECT:FLcontam}). While the algorithm is not trained to identify false spectral lines, we can simulate their effect on the recovered flux ratios. Similar to bias 2, false identifications are characterized by the absence of additional lines aside from the strongest one. The value of S/N for the false identifications is assumed similar to that characterizing the rest of the \emph{single lines} (false identifications with peculiar S/N were easily identified and eliminated during the WISP original by-eye classification). For the reasons explained in Section~\ref{SECT:FLcontam}, we expect almost all the false identifications in this sample to be classified as \oiii\ by the algorithm. 

 In Table~\ref{tbl:line_ratios} we report the flux ratios expected from the stacked spectrum by considering the combined effects of bias 1, bias 2 and 30\% or 35\% contamination from false detections. These values are obtained by running 50 additional simulations for each of the two levels of contamination. The \oiii/\ha\ ratio is more consistent with a 30\% contamination, while \oiii/\hb\ agrees better with a 35\% contamination. Due to the higher uncertainty associated with the other flux ratios, these indicators are consistent with both possibilities.


 As previously explained, bias 1 could be underestimated due to the conservative criteria adopted in the simulations. Additionally, the accuracy estimated on the \emph{gold} sample could degrade when classifying \emph{single lines}. In both these two cases, the fraction of false detections that we estimate would represent  an upper limit to the actual level of contamination.


\section{Discussion}
\label{SECT:Discussion}
\subsection{Analogies with the future Euclid and WFIRST surveys}
The primary purpose of the algorithm that we described in this paper is the automatic classification of the strongest emission lines detected in WISP spectra. However, the modular structure of the algorithm is specifically designed to allow for an easy addition, replacement, and removal of each single module (and block). Hence, the algorithm can be easily adapted to different surveys.
Given the similarities between WISP and the Euclid and WFIRST surveys, the experiment that we performed represents an important pilot study in the context of these future missions, in particular for what concerns the maximization of their scientific return. 
Euclid and WFIRST will probe the nature of dark energy by carrying out spectroscopic surveys over an unprecedented sky area  \citep{2011arXiv1110.3193L,2012SPIE.8442E..0TL,2012arXiv1208.4012G,2015arXiv150303757S}, using the H$\alpha$ and \oiii\ emission lines as tracers of the large-scale structure over a wide range of redshifts (0.5$\lesssim z \lesssim$2.9). In this context, the correct identification of such lines is going to be crucial.

\begin{figure*}[!ht]
 \centering
 \includegraphics[width=18.0cm]{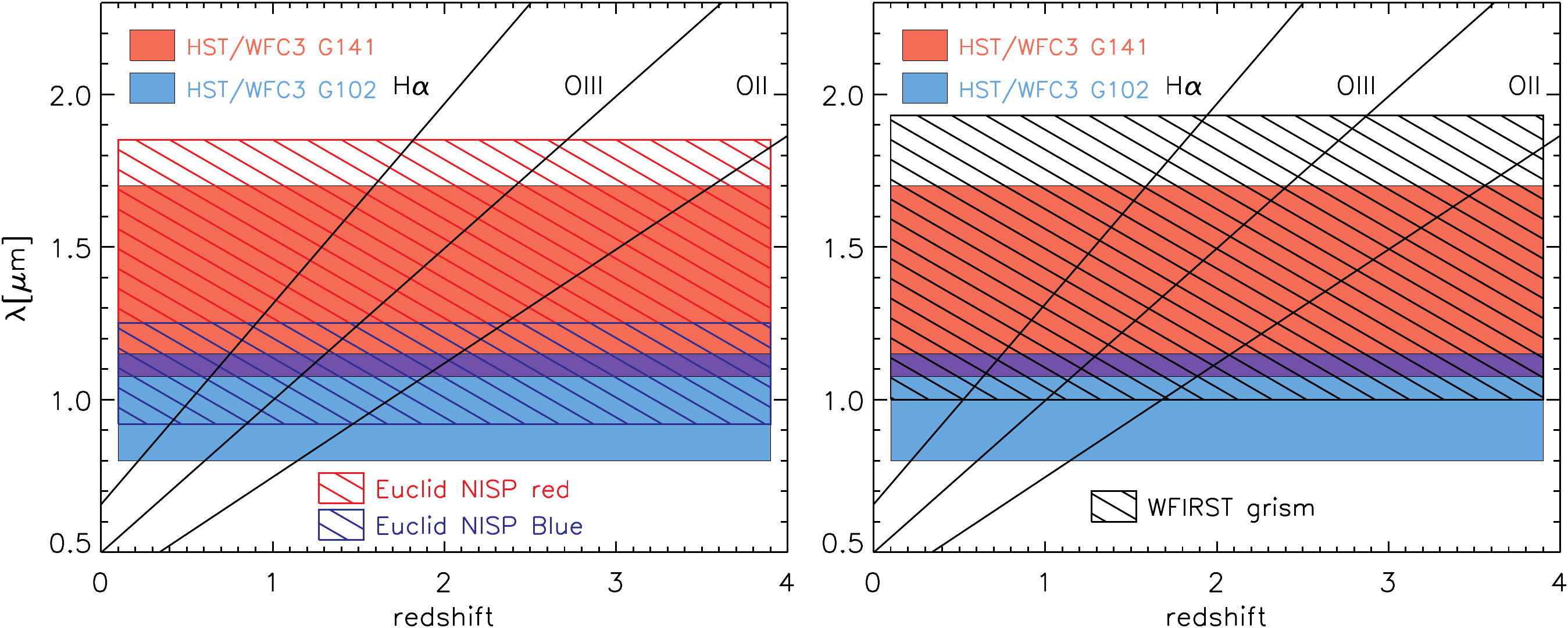}
 \caption{Comparison between the spectral coverages of WISP (HST/WFC3), Euclid NISP, and WFIRST. The wavelength positions of the main spectral lines considered in this paper (H$\alpha$, \oiii\ and \oii) are represented as a function of the redshift, using thick black lines. }
 \label{img:WISP_EUCLID_WFIRST_GRISM}
 \end{figure*}

The WISP survey represents one of the most important proxies for the future Euclid and WFIRST missions. The similar spectral range covered by the grisms of the HST/WFC3, Euclid/NISP, and WFIRST is shown in Table~\ref{tbl:WISP_EUCLID_WFIRST_GRISM} and Figure~\ref{img:WISP_EUCLID_WFIRST_GRISM}.
The presence of two grisms with similar bandwidth (blue and red) makes Euclid/NISP even more similar to HST/WFC3. 

\begin{deluxetable*}{l|cc|cc|c}
\tabletypesize{\footnotesize}
\tablecolumns{6}
\tablewidth{0pc}
\tablecaption{HST-WISP, Euclid, and WFIRST grism main parameters}
\tablehead{ \colhead{} & \multicolumn{2}{c}{WISP (HST/WFC3)} & \multicolumn{2}{c}{Euclid}&  WFIRST  \\
\colhead{} & \colhead{G102} & \colhead{G141} & \colhead{Blue} & \colhead{Red} & \colhead{} }
\startdata
\label{tbl:WISP_EUCLID_WFIRST_GRISM}
$\lambda$ range [nm]    & 800-1150         &  1075-1700        & 920-1250 & 1250-1850     &  1000-1930  \\
$\lambda/\Delta\lambda$ & 210 (at 1000 nm) & 130 (at 1400 nm) &  \multicolumn{2}{c|}{380} &  435-865 \\
\enddata
\end{deluxetable*}

Nevertheless, Euclid and WFIRST will probably benefit from a larger amount of ancillary data than WISP. Compared with our analysis, the photometric redshifts ($z$-PDF) obtained from the SED-fitting approach considered singularly will allow for a more accurate identification of spectral lines, with a smaller fraction of catastrophic failures \citep[$\sim$5-10\% catastrophic failures, $\sigma_{z}/(1+z)\leq$0.03-0.05;][]{2011arXiv1110.3193L}.
 However, as described in Section~\ref{SEC:photoz}, our algorithm does integrate the SED fitting approach with additional and independent sources of information. Consequently, whatever the actual accuracy obtained using all of the ancillary data will be, we expect a reduced fraction of catastrophic failures when applying an algorithm similar to the one described in this paper, if compared with the SED-fitting strategy used alone. 

 Our algorithm is not trained to identify false detections (see Section~\ref{SECT:FLcontam}). In WISP, we estimate these contaminants to represent $\lesssim$35\% of the \emph{single line} sample. Most of this contamination is due to zero-order spectra corresponding to sources located just outside the WISP fields. Both Euclid and WFIRST will strongly limit such type of contamination by dispersing spectra along different directions.
 
 One of the two primary goals of Euclid is to study the accelerated expansion of the universe through the analysis of barionic acoustic oscillations (BAO) \footnote{Euclid is optimized also to study the dark matter distribution in the universe through weak gravitational lensing (WL).}. To this purpose, the spectroscopic redshift precision required on single sources is $\sigma_{\Delta z}\leq 0.001(1+z)$. Instead, the constraint on the fraction of catastrophic failures is quite large: $f_{c}<20\%$ \citep{2011arXiv1110.3193L}. However, the uncertainty associated with $f_{c}$ ($\sigma_{\mathrm{f}_{c}}$) must be known to 1\%. A more extensive discussion on the effects of line misidentification, for both Euclid and WFIRST, can be found in \cite{2019ApJ...879...15A}. With our analysis, we demonstrate that the goal of limiting the fraction of catastrophic failures ($f_{c}$) below 20\% can be achieved, even considering the poor photometric coverage characterizing the WISP survey. On the other hand, in order to estimate $f_{c}$ with the precision required, a larger calibration sample would be needed.

 Besides WISP, there are currently two other spectroscopic surveys that represent important proxies for the Euclid and WFIRST missions.
The 3D-HST \citep[P.I. P. van Dokkum:][]{2012ApJS..200...13B,2014ApJS..214...24S,2016ApJS..225...27M} and the AGHAST surveys \citep[P.I. B. Weiner:][]{2009hst..prop11600W} obtained spectroscopic observations, through the WFC3/G141 grism, of about 150 pointings located in the CANDELS \citep{2011ApJS..197...35G,2011ApJS..197...36K} fields. 

On the one hand, the pointings of the 3D-HST and AGHAST surveys are covered, with some exceptions, only by the G141 grism and by the H band filter (F140W), whereas WISP uses both the G102 and G141 grisms and both the J and H filters (F110W and F140W/F160W). On the other hand, WISP is a pure-parallel survey, meaning that the pointings are usually randomly located in the sky, while 3D-HST and AGHAST can exploit the richness of ancillary data already available in these major cosmological fields. 

For a comparison, the Euclid surveys will use observations in the  Y , J, and H bands of the NISP instrument plus the photometric coverage of VIS, between $\lambda\sim$5500\AA\ and $\lambda\sim$9000\AA. Additional ground based observations will be obtained in the g, r, and i bands from ground based telescopes. Therefore, the SED of the sources covered by the future Euclid surveys will not be as continuously sampled as those in the CANDELS fields. 

Another point favoring WISP as a proxy for the WFIRST and Euclid missions is the total sky area covered, which is more than three times wider that that explored by 3D-HST and AGHAST combined (at the moment, the available WISP data covers a total area of $\sim$1520 arcmin$^{2}$). It is worth noting, in any case, that given the analogies among these three surveys, they can be combined to obtain even more accurate forecasts, as it has recently been done in Bagley et al. (2019, submitted).

Our tests (see Table~\ref{tbl:EM_LINE_LIST} for a summary) show that, using the current strategy, it is possible to obtain acceptable results concerning the identification of the H$\alpha$ and \oiii\ emission lines in WISP. Most of the algorithm's capability is based on the assumption that the strongest emission line in a galaxy spectrum (among the lines considered) falls in the wavelength range covered by the two grisms combined. For the WISP dataset, this assumption represents a good approximation of the real situation. However, for a minor part of the WISP data, this assumption is not correct. In these cases, we adopt the less precise strategy described in Section~\ref{SEC:f_ratio_limits}. We highlight the fact that the acceptability of this important assumption depends on the wavelength range covered by the combined grisms. Hence, for the Euclid case, the precision achieved using the methods described could be undermined by the use of only one of the two grisms (as it is planned for the Euclid wide survey).

The accuracy of the algorithm is low when we consider the \oii\ sample. This outcome is mostly due to the impossibility of precisely calibrating the algorithm by using the very small number of sources in the "\emph{gold}" sample (N$^{\mathrm{[OII]}}_{\mathrm{TGS}}$=24) characterized by an \oii\ emission stronger than that of H$\alpha$ and \oiii. Besides the inaccurate calibration, given the uneven distribution of sources in the \ha, \oii, and \oiii\ samples, a small fraction of misidentified \ha\ and \oiii\ lines can significantly contaminate the much smaller \oii\ sample.
As a consequence, in order to use similar algorithms in the context of the future Euclid and WFIRST missions, an initial phase of careful characterization of the rarest objects should be considered.

\subsection{Possible improvements and future perspectives}
The algorithm we discussed can be improved as follows. 
On the one hand, additional information could be considered by the algorithm. For example, we could  
include the spectral continuum (for the brightest sources) and the rich sets of ancillary photometric and spectroscopic data that are (and that are going to be) available in the fields that will be covered by Euclid and WFIRST. The study of the $\chi^{2}$ obtained by fitting different spectral models (at different $z$) could in principle be used as an additional source of information (we propose an alternative to this method in Appendix~\ref{SECT:AppendixB}). On the other hand, the algorithm could also be improved by taking advantage of the symbiotic use of different strategies, such as supervised and unsupervised machine-learning. 

Finally, the wide area covered by Euclid and WFIRST will allow us to better characterize the rarest objects (such as those in the \oii\ sample described above) from the very initial phases of the surveys, making it possible to limit the contamination affecting these samples of sources. An additional important possibility, which was not explored in this paper, is the ability to quantify their probability, for the single lines identified, of being real emissions or spurious detections.

Even without any further improvement to the actual accuracy ($\sim$82.6\%), determining the nature of most of the single emission lines is crucial for many reasons.
For example, it makes it possible to determine the accurate redshifts of the faintest sources (i.e. the most difficult to explore), or to improve the definition of the faint end of the H$\alpha$ and \oiii\ luminosity functions, eliminating particularly pernicious biases on some selected data samples.

In this paper, we run our software on a sample of spectra where only one spectral line is detected. 
However, this algorithm can also be useful in the more general case, when multiple lines are (barely) visible, or even detected, but there is no consistency among the classification of different reviewers. 

\section{Summary and conclusions}
\label{SECT:Conclusion}

We presented an algorithm that can be used to identify the strongest emission lines detected in near-IR grism spectra. The algorithm exploits ancillary information that is usually not considered when determining the spectroscopic redshift. For the algorithm described in this work, we considered low-precision photometric redshifts from SED fitting, priors based on the apparent magnitude, size, equivalent width, color indices, and expected line flux ratios. 
The classification is optimized by including empirical \emph{a posteriori} corrections based on the wavelength position of the detections.

Such an approach is very useful to identify a spectral line when it is the only emission detected in a spectrum, especially when the precision of the photometric redshifts from SED fitting is low. 
However, the same approach can also be helpful in the general case, 
when the identification of multiple emission lines is difficult for other reasons (such as contamination, low S/N, etc.). 


The WISP survey represents one of the most relevant proxies for the future Euclid and WFIRST missions, especially in terms of wavelength range covered, depth reached, and availability of near-IR photometric bands. In this context, our work is intended as a pilot study toward the maximization of the scientific return of these future surveys.

The approach that we adopted allows us to identify \oiii\ and H$\alpha$ emission lines in WISP with good accuracy ($\sim$82.6\% of correct identifications on real lines). The ancillary data that will be available for Euclid and WFIRST, together with their observing strategies, should significantly improve this result.

As we illustrated, more or less pure samples (corresponding to less or more complete samples, respectively) can be obtained by using specific probability indicators outputted by the algorithm. 
However, the accuracy of our algorithm is still low for the rarest objects (\oii). This problem is mostly due to the limited number of sources that can be used to calibrate the algorithm. In this sense, the large datasets available from the future Euclid and WFIRST surveys should sensibly improve the performance of similar algorithms.

We also expect further improvements of the accuracy from the exploitation of additional observational quantities and independent techniques that we do not consider in this pilot study, such as the combined use of different machine-learning approaches, including unsupervised machine-learning strategies, the analysis of the spectral continuum, the comparison of alternative models fitted to the spectra ($\chi^{2}$ analysis), or the use of the \emph{``contrast factor''} that we describe in Appendix~\ref{SECT:AppendixB}.


 \acknowledgements
 I.B. thanks Alvio Renzini, Alberto Franceschini, Paolo Cassata, and Andrea Grazian for the useful discussion about the thematics discussed in the paper.
 C.S. acknowledges financial support from NASA, through the STScI program number HST-AR-14311.001-A. STScI is operated by the Association of Universities for Research in Astronomy, Incorporated, under NASA contract NAS5-26555.
 M.B. acknowledges support from INAF under PRIN SKA/CTA FORECaST and from the Ministero degli Affari Esteri della Cooperazione Internazionale - Direzione Generale per la Promozione del Sistema Paese Progetto di Grande Rilevanza ZA18GR02.
 M.R. acknowledges financial support from NASA, through a grant from the Space Telescope Science Institute (STScI), program number HST-GO-14178.026-A.

\appendix

 \twocolumngrid

\section{Alternative accuracy tests.}
\label{SECT:AppendixA}

Similarly to Table~\ref{tbl:EM_LINE_LIST}, where the accuracy is computed using the same sample (with few differences) to both calibrate and test the algorithm, in Tables~\ref{tbl:EM_LINE_LIST_T1} and \ref{tbl:EM_LINE_LIST_T2} we report the accuracy measured by dividing the "\emph{gold}" sample in two independent subsamples. In the first test (Table~\ref{tbl:EM_LINE_LIST_T1}), we use one of the two subsamples to calibrate the algorithm and the other to assess the accuracy. For the second test (Table~\ref{tbl:EM_LINE_LIST_T2}), we invert the two subsamples. In both cases, all of the relations, including the \emph{a posteriori} empirical correction (Section~\ref{SEC:empirical_corr}), are recomputed considering only the calibration data set. The overall accuracy that we obtain considering all of the species/transitions is 80.6\% and 80.4\% for the first and the second test, respectively. In Tables~\ref{tbl:EM_LINE_LIST_T1} and \ref{tbl:EM_LINE_LIST_T2} we report only the accuracy obtained for the most relevant transitions (\ha, \oiii, and \oii). As explained in Section~\ref{SECT:NOT_INDIP_SAMPLES}, we consider the accuracy obtained with these two tests to be a lower limit to the actual accuracy of the algorithm. For more reliable estimates, we refer the reader to Table~\ref{tbl:EM_LINE_LIST}.

 \begin{deluxetable*}{cclllllll}
 \tabletypesize{\footnotesize}
 \tablecolumns{9}
 \tablewidth{0pc}
 \tablecaption{Alternative accuracy test 1\tablenotemark{[a]}}
\tablehead{ species/ & $\lambda_{\mathrm{obs}}$ & N$_{\mathrm{TGS}}$ & N$_{\mathrm{I}}$ & N$_{\mathrm{CI}}$ & N$_{\mathrm{WI}}$& Completeness  & Contamination & Accuracy (Purity) \\
        transitions  & [\AA] & & & & & (N$_{\mathrm{CI}}$/N$_{\mathrm{TGS}}$) & (N$_{\mathrm{WI}}$/N$_{\mathrm{I}}$) & (N$_{\mathrm{CI}}$/N$_{\mathrm{I}}$) }
 \startdata
 \label{tbl:EM_LINE_LIST_T1}
 H$\alpha$ & 6564.5          & 639 & 643 & 543 & 100   & 85.0\%  &  15.6\% & 84.4\% \\
 \oiii     & 4960.3 - 5008.2 & 461 & 453 & 354 & 99    & 76.8\%  &  21.9\% & 78.1\% \\
 \oii      & 3727.1 - 3729.9 & 11  & 19  & 3   & 16    & 27.3\%  &  84.2\% & 15.8\% \\
 \enddata
\tablenotetext{a}{ This test (to be compared with Table~\ref{tbl:EM_LINE_LIST}) is performed by dividing the "\emph{gold}" sample into two independent subsets, using one to calibrate the algorithm, and the other to test it.}
 \end{deluxetable*}

 \begin{deluxetable*}{cclllllll}
 \tabletypesize{\footnotesize}
 \tablecolumns{9}
 \tablewidth{0pc}
 \tablecaption{Alternative accuracy test 2\tablenotemark{[a]}}
\tablehead{ species/ & $\lambda_{\mathrm{obs}}$ & N$_{\mathrm{TGS}}$ & N$_{\mathrm{I}}$ & N$_{\mathrm{CI}}$& N$_{\mathrm{WI}}$ & Completeness  & Contamination & Accuracy (Purity)  \\
        transitions  & [\AA] & & & & & (N$_{\mathrm{CI}}$/N$_{\mathrm{TGS}}$) & (N$_{\mathrm{WI}}$/N$_{\mathrm{I}}$) & (N$_{\mathrm{CI}}$/N$_{\mathrm{I}}$) }
 \startdata
 \label{tbl:EM_LINE_LIST_T2}
 H$\alpha$ & 6564.5          & 654 & 679 & 565 & 114   & 86.4\%  &  16.8\% & 83.2\% \\
 \oiii     & 4960.3 - 5008.2 & 488 & 453 & 366 & 87    & 75.0\%  &  19.2\% & 80.8\% \\
 \oii      & 3727.1 - 3729.9 & 13  & 20   & 6  & 14    & 46.2\%  &  70\%   & 30.0\% \\
 \enddata
\tablenotetext{a}{ This test (to be compared with Table~\ref{tbl:EM_LINE_LIST}) is performed by dividing the "\emph{gold}" sample into two independent subsets, using one to calibrate the algorithm, and the other to test it. In this second test, the calibration and test samples are inverses of each other, with respect to the first test (Table~\ref{tbl:EM_LINE_LIST_T1})}
 \end{deluxetable*}


\section{Stacking technique and automatic spectra recentering}
\label{SECT:AppendixB}

In Section~\ref{SEC:single_lines_08z12}, we describe the result obtained by stacking the spectra of the ``\emph{single--line}'' sources in the range $0.8<z<1.2$. Here, we describe how the original spectra are treated and then combined to obtain the final result described. 

\begin{figure*}[!ht]
 \centering
 \includegraphics[width=17.5cm]{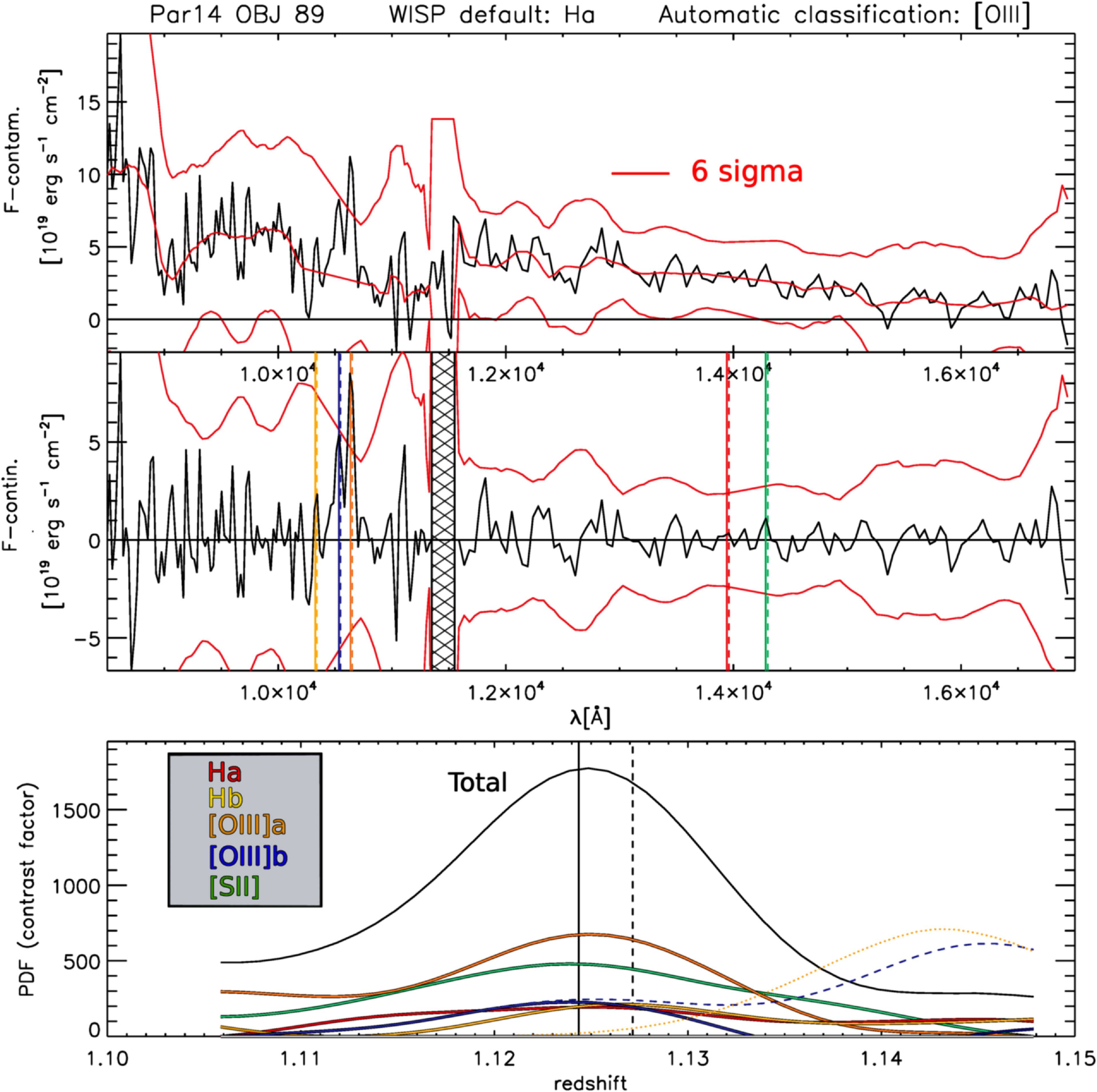}
 \caption{Example of ``\emph{single--line}'' spectrum used to compute the stacked spectrum described in Section~\ref{SEC:single_lines_08z12} and shown in Figure~\ref{img:stack_OIII_spect}. The contamination-subtracted spectrum is shown in the  {\bf top panel}. We considered locally computed $\pm 6\sigma$ deviations (upper and lower red curves) from the median (central red curve) to exclude unrealistic values, unless when the flux excess was located in the close proximity to expected spectral features. The continuum-subtracted spectrum is shown in the {\bf middle panel}. The position of the only line detected (in this example, it is the only line exceeding our 6$\sigma$ threshold) is initially obtained through a simple Gaussian fit. Following the identification performed by the algorithm, the automatic recentering of the spectrum is obtained using a ``contrast factor'' $FC$ ({\bf bottom panel}), which we compute for each of the lines expected in the range of visibility (see text and Figure~\ref{img:contrast_fact_ex} for more details). In both the middle and bottom panels, dashed lines indicate the position of the spectral lines determined using a simple Gaussian fit. The continuous vertical lines indicate the positions after the automatic recentering process (in order of increasing $\lambda$, \hb\ in yellow, O\thinspace{\sc iii} 4959 in blue, O\thinspace{\sc iii} 5007 in orange, \ha\ in red, and \sii\ in green). }
 \label{img:single_spec_ex}
 \end{figure*}

The three panels of Figure~\ref{img:single_spec_ex} show the steps of the processing that precede the stacking phase. Initially, the contamination is removed from the original spectra\footnote{The contamination as a function of $\lambda$ is obtained directly from the original WISP data}. Then, we estimate the continuum at each $\lambda_{i}$ (central red curve in the top panel of Figure~\ref{img:single_spec_ex}). for this purpose, we linearly fit the observed flux measured in the wavelength intervals immediately preceding ($\Delta\lambda^{-}_{i}$) and following ($\Delta\lambda^{+}_{i}$) the wavelength considered. 
We extrapolate the values of the continuum expected at $\lambda_{i}$ from both the two sides, considering their average value.
In this process, we mask both the main spectral lines and the noisy central wavelength range, where the sensitivity of the two grisms is lower. After this initial computation, the continuum is further smoothed to obtain a result less dependent on the small-scale variations.

After the continuum subtraction, the local level of noise $\sigma_{F}(\lambda_{i})$ is obtained as the standard deviation of the fluxes $F(\lambda)$ measured in the same $\lambda$ intervals considered when computing the continuum ($\Delta\lambda^{-}_{i}$ and $\Delta\lambda^{+}_{i}$). Also, the noise level is subsequently smoothed. Values of $F(\lambda)$ exceeding a local 6$\sigma$ threshold (highest and lowest red curves in the two upper panels of Figure~\ref{img:single_spec_ex}) are masked and not used in the successive steps (unless when the excess is due to the presence of an actual spectral line).

In order to precisely align all of the single spectra to the common rest frame, every ``\emph{single--line}'' spectrum is automatically finely recentered following the line identification performed by the algorithm.
In principle, the recentering could be automatically performed by fitting Gaussian functions to all of the main emission lines and using the weighted average of the peaks' positions (in $\lambda$) to precisely estimate the redshift. This approach works well when the spectral lines considered are detected well above the level of the noise.
In practice, the same solution is not well suited for determining the position of weak lines ($F\lesssim\sigma_{F}$). 
In these cases, the result is that the alignment to a common rest frame relies only on the (poor) fit of the strongest emission line, detected just above the noise threshold, while the fit to the other lines is completely unreliable and often fails.

To overcome the problems of the technique described, we elaborated a wavelength-dependent \emph{contrast factor} ($CF$), which we compute for each of the main emission lines. We use this function to compute the most probable position of each spectral line. For example, $CF_{H\alpha}(\lambda_{1})$ quantifies the relative probability associated with the \ha\ emission being centered at $\lambda_{\mathrm{obs}}^{H\alpha}=\lambda_{1}$. This can be immediately translated into the relative probability for the source to be located at $z=\lambda_{1}/\lambda_{\mathrm{int}}^{H\alpha}-1$. The sum of the contrast factors computed for each spectral line, weighted for the local noise, can be used to determine the most probable redshift of the source.
The bottom panel of Figure~\ref{img:single_spec_ex} shows the behavior of $CF(z, \sigma_{\lambda})$ for each of the main lines in the visibility range, and the (not-normalized) PDF resulting from their sum.

\begin{figure}[!ht]
 \centering
 \includegraphics[width=7.5cm]{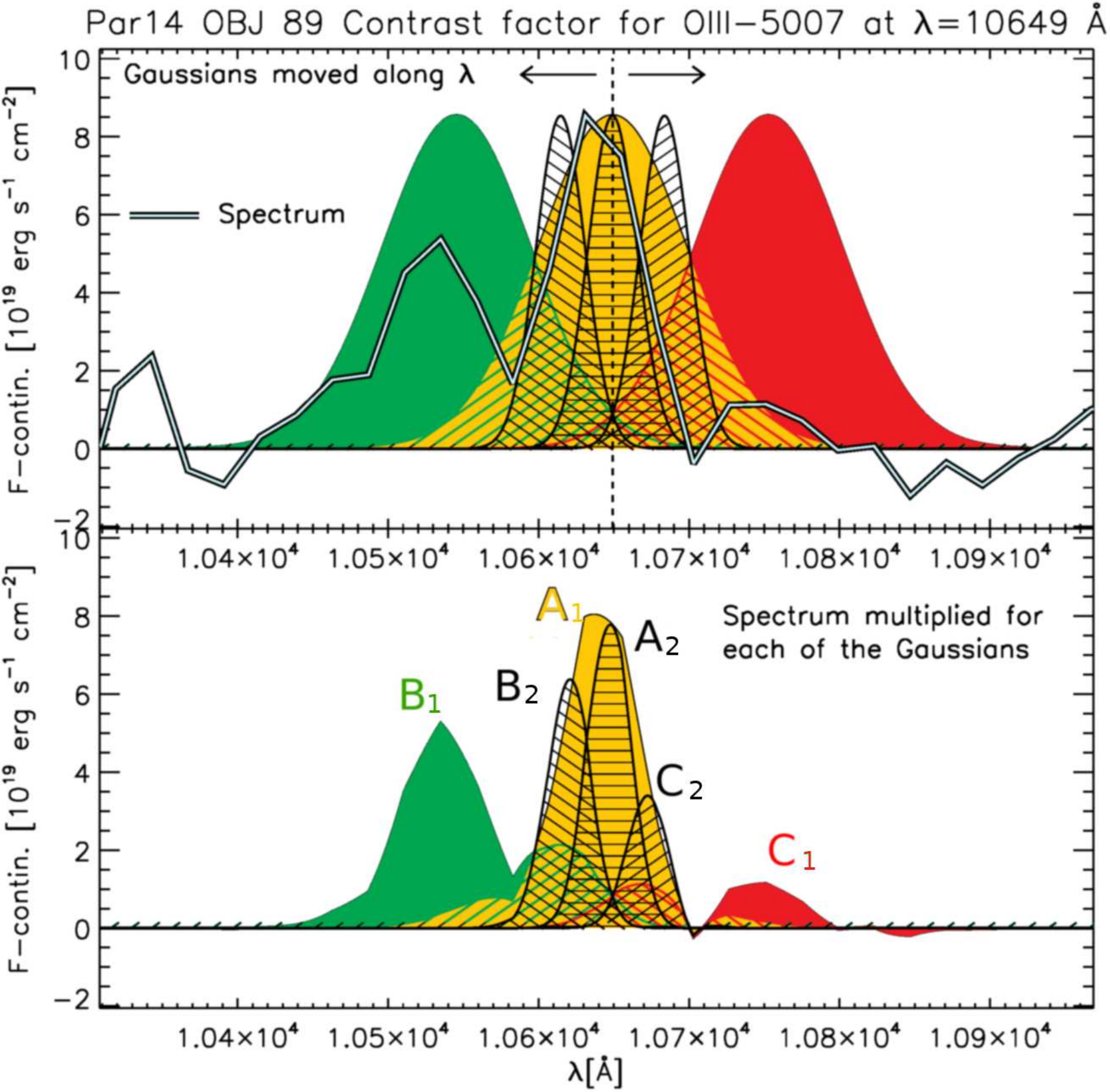}
 \caption{ Before the stacking, every spectrum is finely recentered to the common rest frame using a contrast factor $FC(\lambda)$ that we compute for each of the main emission lines.
While a simple Gaussian fit is sensitive to the noise at all scales, our contrast factor acts as a combination of two bandpass filters calibrated on the expected width of the spectral line. This approach guarantees the limited influence of the noise at all the other scales.
The {\bf top panel} shows the same spectrum of Figure~\ref{img:single_spec_ex}, corresponding to the O\thinspace{\sc iii} 5007 emission (following the identification performed by our algorithm).
   Two sets of Gaussians functions, each of which is characterized by a specific value of $\sigma$, are shown in the same plot. In both the two sets, three Gaussians are centered at $\lambda_{i}$ and at $\lambda_{i}\pm 2\sigma$, where $\lambda_{i}$ is the value of the wavelength considered during one generic step of the process (at each step, the Gaussian functions are shifted along $\lambda$, as explained below).
   For the three Gaussian functions in the first set (green, yellow and red areas), $\sigma$ is similar to the width of the emission line as expected from the apparent size of the source. In the second set (black shaded areas), $\sigma$ corresponds to 1/3 of the same value. At every $\lambda$, the spectrum $F(\lambda)$ is multiplied for the value assumed at that wavelength by each of the Gaussian functions ({\bf bottom panel}).  For a given spectral line, and at each wavelength $\lambda_{i}$, the contrast factor $FC(\lambda_{i})$ is obtained by combining the integral of these functions as described in Equation~\ref{EQN:CF2}. Then, all of the Gaussian functions are coherently shifted and the process is repeated in order to obtain the value of $FC(\lambda)$ at different wavelengths.
   A similar contrast factor is computed for all the spectral lines considered and the precise redshift of the source is obtained as the value of $z$ maximizing the weighted sum of the resulting functions ($FC_{\mathrm{TOT}}(z)$: black curve in the bottom panel of Figure~\ref{img:single_spec_ex}). }
 \label{img:contrast_fact_ex}
 \end{figure}

Figure~\ref{img:contrast_fact_ex} schematically describes how the contrast factor is computed for the OIII 5007 line that is visible in the spectrum already shown in Figure~\ref{img:single_spec_ex}, but the same description can be extended to all the emission lines considered. We initially compute two sets of normalized Gaussian functions $G(\lambda)$.
In the first set, the value of $\sigma$ is similar to the width of the emission lines (as expected from the size of each source), while in the second set $\sigma$ corresponds to $1/3$ of this value.
Each of the two sets is made up of three functions, $G^{0}_{\lambda_{i}}$, $G^{-}_{\lambda_{i}}$ and $G^{+}_{\lambda_{i}}$, characterized by identical shapes and normalizations, but centered at $\lambda_{i}$ (the wavelength considered) and at $\lambda_{i} \pm 2\sigma$. Every function $G(\lambda)$ is multiplied, at each $\lambda$, for the spectrum $F(\lambda)$. In our computation of $CF$, we consider the integrals of the six Gaussian functions, normalized for the local level of noise $\sigma_{F}(\lambda)$ measured from the spectrum. For each of the two sets, we have
\begin{equation}
A(\lambda_{i})=\int_{0}^{\infty} G^{0}_{\lambda_{i}}(\lambda)\frac{F(\lambda)}{\sigma_{F}(\lambda)} \ \mathrm{d}\lambda
\end{equation}
\begin{equation}
B(\lambda_{i})=\int_{0}^{\infty} G^{-}_{\lambda_{i}}(\lambda)\frac{F(\lambda)}{\sigma_{F}(\lambda)} \ \mathrm{d}\lambda
\end{equation}
\begin{equation}
C(\lambda_{i})=\int_{0}^{\infty} G^{+}_{\lambda_{i}}(\lambda)\frac{F(\lambda)}{\sigma_{F}(\lambda)} \ \mathrm{d}\lambda
\end{equation}
At this point, the six integrals ($A_{1}$ and $A_{2}$ corresponding with the central Gaussians, $B_{1}$ and $B_{2}$ corresponding with the Gaussians centered at $-2\sigma$, $C_{1}$ and $C_{2}$ corresponding with the Gaussians centered at $+2\sigma$) are combined to compute the contrast factor:
\begin{eqnarray}
CF(\lambda_{i})=A_{1}(\lambda_{i})-\frac{1}{2}[B_{1}(\lambda_{i})+C_{1}(\lambda_{i})]+ \nonumber \\
+A_{2}(\lambda_{i})-\frac{1}{2}[B_{2}(\lambda_{i})+C_{2}(\lambda_{i})]
 \label{EQN:CF2}
\end{eqnarray}
These contrast factors, computed for each spectral line and for each value of $\lambda_{i}$ (i.e. for each value of $z_{i}$), are finally summed together, obtaining a unique function $CF_{\mathrm{TOT}}(z)$.
The spectrum is recentered considering the value of $z$  that maximizes the function $CF_{\mathrm{TOT}}(z)$.

The contrast factor described works essentially as a combination of two bandpass filters. The band of the first filter (i.e. the $\sigma$ of the first three Gaussian functions) is tuned on the line width expected from the observed size. This approach allows us to mitigate the effects of the noise at different spatial ($\lambda$) scales, focusing on the possible presence of flux excesses at these scales. An additional filter (the other three Gaussians) allows us to refine the estimation of the central position of the line by considering the flux peaks, observable at higher spatial frequencies.

In our analysis, we use the contrast factor described only to recenter the spectra. However, we notice that in future analyses the same approach could be used to improve the accuracy of the line identification process. In fact, the contrast factor is to all effects a redshift probability distribution function that could be integrated into our computations\footnote{The same can be said for the $\chi^{2}$ resulting from the best-fitting Gaussian functions, but with the limitations previously described, concerning line fluxes are at the level of the noise}.

After recentering each single spectrum, we compute, at each $\lambda$, the weighted average of $F(\lambda)$, normalized for the total flux of the detected emission line. As a weight, we use the local value of $\sigma_{F}(\lambda)$. The mean is computed after excluding all the values of $F(\lambda)$ exceeding the 16th and 84th percentiles from the median value ($\pm 1\sigma$).



 \bibliographystyle{apj}
\bibliography{biblio3}{}

\end{CJK*} 
\end{document}